\documentclass[journal]{IEEEtran}

\ifCLASSOPTIONcompsoc
  \usepackage[caption=false,font=footnotesize,labelfont=sf,textfont=sf]{subfig}
\else
  \usepackage[caption=false,font=footnotesize]{subfig}
\fi

\usepackage{afterpage}
\usepackage{rotating}
\usepackage[english]{babel}
\usepackage{multirow}
\PassOptionsToPackage{hyphens}{url}
\usepackage[numbers,square, comma, sort&compress]{natbib}
\usepackage[hyphens]{url}
\usepackage{hyperref}
\hypersetup{breaklinks=true}
\usepackage[normalem]{ulem}
\usepackage{enumerate, enumitem}
\usepackage[table]{xcolor}
\usepackage{amssymb, amsmath, amsthm}
\usepackage{tabularx, tabulary}
\usepackage{array}
\usepackage{booktabs}
\usepackage{pifont}
\usepackage{lscape,threeparttable,pdflscape}
\usepackage{tikz}
\usetikzlibrary{shapes,arrows.meta, calc, shadings, shadows, shapes.arrows, chains}
\usepackage{xtab}
\usepackage{wasysym}
\usepackage{comment}
\usepackage{xspace}
\usepackage{balance}
\newcommand{\subparagraph}{}

\usepackage{titlesec}

\titlespacing*{\subsubsection}
{0pt}{2.25ex plus 1ex minus 0.2ex}{1.5ex plus 0.2ex}
\renewcommand\thesubsubsection{%\hspace{-1em}
\Alph{subsection}.\arabic{subsubsection}}

\titleclass{\subsubsubsection}{straight}[\subsection]

\newcounter{subsubsubsection}[subsubsection]
\renewcommand\thesubsubsubsection{\hspace{-0.2em}
\arabic{subsubsection}.\arabic{subsubsubsection})}

\titleformat{\subsubsubsection}
{\vspace{1em}\normalfont\normalsize\itshape}{\thesubsubsubsection}{0.5em}{}
\titlespacing*{\subsubsubsection}
{0pt}{3.25ex plus 1ex minus .2ex}{1.5ex plus .2ex}

\makeatletter
\renewcommand\paragraph{\@startsection{paragraph}{5}{\z@}%
  {3.25ex \@plus1ex \@minus.2ex}%
  {-1em}%
  {\normalfont\normalsize\bfseries}}%%%
\renewcommand\subparagraph{\@startsection{subparagraph}{6}{\parindent}%
  {3.25ex \@plus1ex \@minus .2ex}%
  {-1em}%
  {\normalfont\normalsize\bfseries}}
\def\toclevel@subsubsubsection{4}
\def\toclevel@paragraph{5}
\def\toclevel@paragraph{6}
\def\l@subsubsubsection{\@dottedtocline{4}{7em}{4em}}
\def\l@paragraph{\@dottedtocline{5}{10em}{5em}}
\def\l@subparagraph{\@dottedtocline{6}{14em}{6em}}
\makeatother

\newcommand{\BAstar}{BA$\star$\xspace}

\newcommand{\ghost}{\textsc{Ghost}\xspace}

\tikzstyle{block} = [draw, rectangle, minimum width=2cm, minimum height=2cm, fill=white]
\tikzstyle{keyblock} = [draw, rectangle, minimum width=2cm, minimum height=2cm, fill=blue!10!white]
\tikzstyle{microblock} = [draw, circle, minimum size=1.5cm, fill=black!10!white]
\tikzstyle{comchainconfig} = [draw, rectangle, minimum width=1.5cm, minimum height=2cm, fill=purple!10!white]
\tikzstyle{comchaintx} = [draw, rectangle, minimum width=1cm, minimum height=1cm, fill=white]
\tikzstyle{fruit} = [draw, circle, minimum size=1.5cm, fill=orange!10!white]
\tikzstyle{lineconnect} = [draw, circle, minimum size=0.2cm, fill=black]
\tikzstyle{hgwitness} = [draw, circle, minimum size=0.8cm, fill=white]
\tikzstyle{hgevent} = [draw, circle, minimum size=0.5cm, fill=black!5!white]
\tikzstyle{tangletx} = [draw, rectangle, minimum width=0.5pt, minimum height=0.5pt, fill=white]
\tikzstyle{tangletip} = [draw, rectangle, minimum width=0.5pt, minimum height=0.5pt, fill=yellow!10!white]
\tikzstyle{blsend} = [draw, rectangle, minimum width=0.5pt, minimum height=0.5pt, fill=blue!10!white]
\tikzstyle{blrec} = [draw, rectangle, minimum width=0.5pt, minimum height=0.5pt, fill=red!10!white]
\tikzstyle{beaconsideblocka} = [draw, rectangle, minimum width=0.5pt, minimum height=0.5pt, fill=red!10!white]
\tikzstyle{beaconsideblockb} = [draw, rectangle, minimum width=0.5pt, minimum height=0.5pt, fill=blue!10!white]
\tikzstyle{beaconblock} = [draw, rectangle, minimum width=0.6cm, minimum height=0.6cm, fill=white]

\xdefinecolor{DarkGreen}{cmyk}{0.8,0,0.8,0.3}
\definecolor{brown}{RGB}{225,128,128}

\newcolumntype{x}[1]{>{\centering\arraybackslash\hspace{0pt}}p{#1}}

\newcommand{\cmark}{\cellcolor{green!25}\ding{51}}

\newcommand{\xmark}{\cellcolor{red!25}\ding{55}}

\newcommand{\mmark}{\cellcolor{orange!25}\ding{51}/\ding{55}}

\newcommand{\ycell}{\cellcolor{orange!25}}
\newcommand{\gcell}{\cellcolor{green!25}}
\newcommand{\rcell}{\cellcolor{red!25}}

\newtheorem{theorem}{Theorem}
\newtheorem{definition}[theorem]{Definition}

\begin{document}
\title{
Deconstructing Blockchains:\\ A Comprehensive Survey on Consensus, Membership and Structure
}

\author{ \IEEEauthorblockN{Christopher Natoli,
    \ Jiangshan Yu\IEEEauthorrefmark{1}\thanks{\IEEEauthorrefmark{1}
    Corresponding author.}
    ,\ Vincent Gramoli
    ,\ and Paulo Esteves-Verissimo
  }
  \IEEEcompsocitemizethanks{
    \IEEEcompsocthanksitem Christopher Natoli is with University of Sydney, Australia.
    \IEEEcompsocthanksitem Vincent Gramoli is with University of Sydney and CSIRO Data61, Australia.%
    \IEEEcompsocthanksitem Jiangshan Yu is with Monash University,
    Australia. E-mail: J.Yu.Research@gmail.com%
    \IEEEcompsocthanksitem Paulo Esteves-Verissimo is with the
    Interdisciplinary Centre for Security, Reliability and Trust,
    University of Luxembourg, Luxembourg.}}

    \date{}

\maketitle

\begin{abstract}

  It is no exaggeration to say that since the introduction of Bitcoin,
  blockchains have become a disruptive technology that has shaken the
  world.
  % Since Bitcoin's release in 2008,
  % blockchains have played a critical role in financial technology and
  % innovation, resulting in an increase of interest and
  % attention. 
  However, the rising popularity of the paradigm has led to a flurry
  of proposals addressing variations and/or trying to solve problems
  stemming from the initial specification. This added considerable
  complexity to the current blockchain ecosystems, amplified by the
  absence of detail in many accompanying blockchain whitepapers.

  Through this paper, we set out to explain blockchains in a simple
  way, taming that complexity through the deconstruction of the
  blockchain into three simple, critical components common to all
  known systems: \emph{membership selection}, \emph{consensus
    mechanism} and \emph{structure}. We propose an evaluation
  framework with insight into system models, desired properties and
  analysis criteria, using the decoupled components as criteria. We
  use this framework to provide clear and intuitive overviews of the
  design principles behind the analyzed systems and the properties
  achieved. We hope our effort will help clarifying the current state
  of blockchain proposals and provide directions to the analysis of
  future proposals.
\end{abstract}

%% Sections
\section{Introduction}

\begin{figure*}[t]
    \begin{center}
        \includegraphics[width=\textwidth]{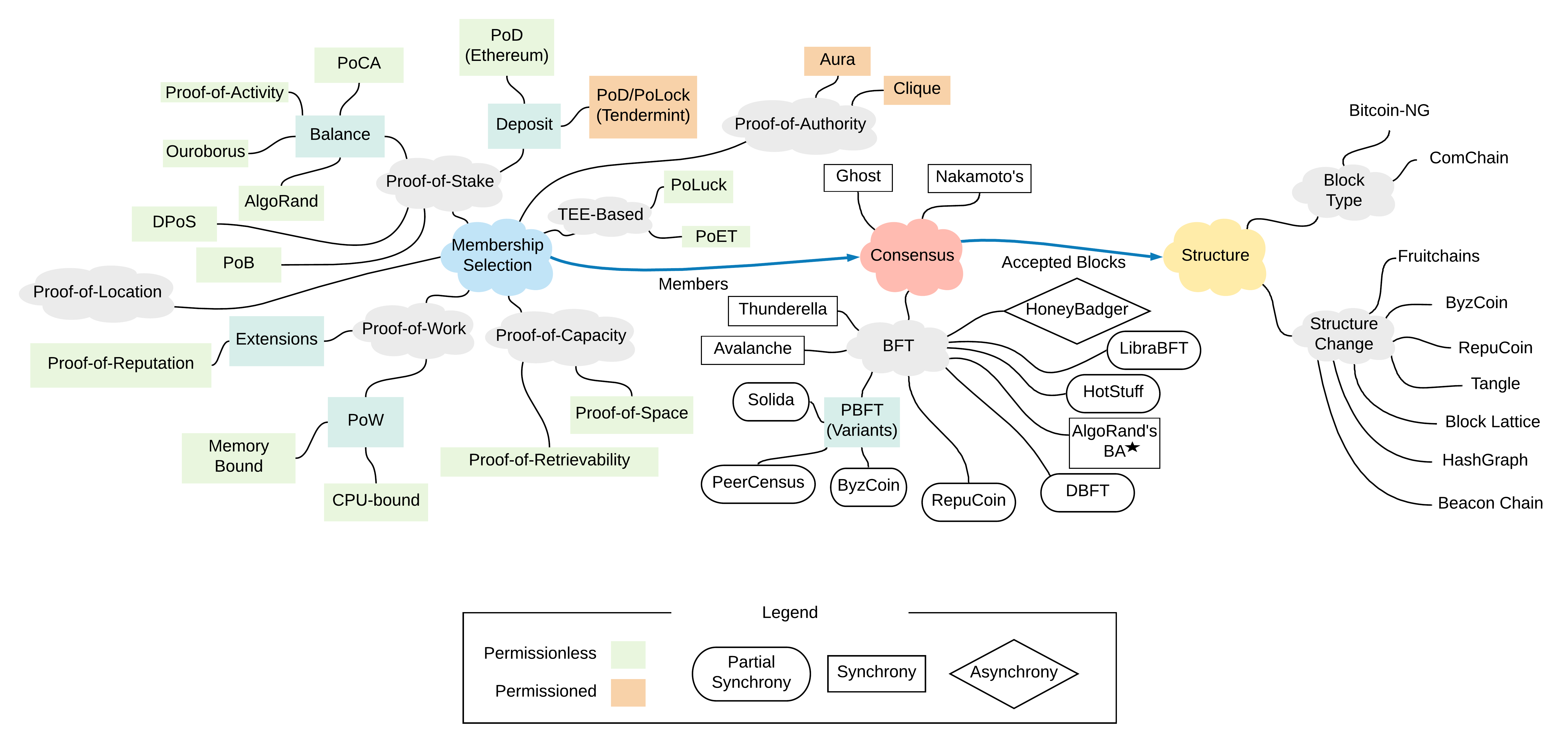}
        \caption{The blockchain landscape.\label{fig:overview}}
    \end{center}
\end{figure*}

\IEEEPARstart{I}{n} 2008 Satoshi Nakamoto released
Bitcoin~\cite{Nakamoto08-Bitcoin}, highlighting the power and importance of
distributed systems. The use of distributed systems in our lives has been
transparently controlled by significant actors. The introduction of Bitcoin
provided the public with insight into ungoverned distributed systems and
sparked a movement towards decentralization.  Initially, the blockchain
technology went vastly unnoticed and was heavily integrated with the deep web
due to its pseudonymity properties~\cite{guardian-silkroad, silkroad-CN13}.
However, the concept of the blockchain quickly gained interest and grew to what
it is today. The growing interest promoted the introduction of new chains and
the growth of distributed ledger technologies~\cite{ethereum, Corda, ripple,
stellar, waves, dash, miers2013zerocoin}.  However, the rising popularity of
the paradigm has led to a flurry of proposals addressing variations and/or
trying to solve problems arising from the initial specification.

The complexity introduced through popularity was amplified by the absence of
detail in the proposals. As an example, the vast majority were presented
through white papers~\cite{kwon2014tendermint, iota-whitepaper, cosmos, nxt,
slimcoin,cryptonote, Nakamoto08-Bitcoin, rippleoriginal, ripplewhitepaper},
wiki documentations~\cite{parityPoA, EIP225, aura-POA, parityaura} and
websites~\cite{steemio, dogecoin, litecoin, loibl2014namecoin}. This is in
contrast with the traditional academic research that conveys results through
scientific publications typically peer-reviewed by specialists of particular
domains, like distributed systems, cryptography, networking or game
theory~\cite{Kokoris-KogiasJ16-ByzCoin,EyalGencer-etc-16-BitcoinNG,
GHMVZ-algorand, ouroboros-praos}. The lack of detail has left room for
interpretation that has sometimes led researchers to different
conclusions~\cite{armknecht2015ripple, heilman2019cryptanalysis,
ekparinya2019clones}.

Through this paper, we set out to explain blockchains in a simple way,
taming that complexity through the deconstruction of the blockchain
into three simple, critical components common to most known systems:
\emph{membership selection}, \emph{consensus mechanism}, and
\emph{structure}.  The \emph{membership selection} component
determines a committee of nodes that participate in the consensus; the
\emph{consensus mechanism} component is responsible for deciding on
the next block, run by the selected committee of nodes; and the
\emph{structure} component represents how the data is organized in the
blockchain.  Thanks to our deconstruction, we are able to provide a
clear and unique landscape of blockchains, as depicted in
Figure~\ref{fig:overview}. We then propose an evaluation framework
with insights into different key system models, desired properties, and analysis
criteria, using the decoupled components as parameters.
We use this framework to provide overviews of the design
principles behind the analyzed systems and the achieved properties.

There are notable works aggregating and analyzing blockchains, each
providing unique evaluation but primarily focusing on one aspect and
constructing specific frameworks to comparatively analyze. Cachin and
Vukoli\'c present a comparison of consensus protocols in permissioned
blockchains~\cite{CachinV17}.  Abraham and Malkhi also present the
idea of deconstructing the blockchain into
layers~\cite{abraham2017blockchain} but with a focus on relating
Nakamoto's consensus versus Byzantine fault tolerant (BFT)
protocols% ,
% under the perspective of traditional distributed systems
.  Similarly, Vukoli\'c~\cite{Vukolic15} explores the contrast between
blockchains and BFT replicated state machines, pointing to scalability
problems faced by both and provide insight into upcoming
proposals. Gramoli~\cite{gramoli2017blockchain} investigates
mainstream blockchains and discusses classical Byzantine consensus in
the context of the blockchain and discusses some of the dangers of
misunderstanding guarantees, problems that we address and systematize
in our paper.  Wang et al.~\cite{wang-abs-1805-02707} approach the
blockchain consensus mechanisms in the perspective of game theory and
the strategies of adoption of nodes. Bano et
al.~\cite{DBLP:journals/corr/abs-1711-03936} provide an overview on
how different blockchain consensus work, and compare existing
Proof-of-* against other proposed consensus mechanisms based on their
properties.
However, no other work has deconstructed the blockchain into these unique
components and analysed all three.
% In this paper, We provide a richer deconstruction of the
% blockchain components, directly analyzing membership selection (e.g.,
% Proof-of-* and others), consensus mechanisms against the several
% categories, as well as the different structure alternatives.

While existing work provides good analysis of specific sets of the properties of
blockchains, it remained challenging, even for the educated but non-expert readers to
get a thorough and comparative picture of the design
principles of different complicated systems. To our knowledge, we are
the first to classify blockchain based on a decomposition into
membership selection, consensus mechanism and structure, categorizing
and providing an analysis of leading proposals. As an addition to our
analysis, we provide an overview of the attacks and threats related to
membership selection, consensus goals, and structural assumptions. We
hope that our analysis will clarify future design directions
and inspire new designs based on innovative coherent combinations
revealed by our decomposition and categorization.

Thus, this paper presents the following contributions:

\begin{itemize}
  \item It innovatively deconstructs the blockchain into three simple, critical
    components: membership selection, consensus mechanism and
    structure.
  \item It provides an evaluation framework with insight into system
models, desired properties and analysis criteria, using the decoupled
components as parameters.
  \item It provides a clear and intuitive overview of the design
    principles behind analyzed systems and their properties, in terms
    of the parameters above.
  \item It proposes, based on the analyzed state of the art, a categorization
    of membership selection and consensus approaches.
In addition, it provides generic
    charts representing the design principles for each category of membership
    selection, to simplify future designs.
\end{itemize}

The remainder of the paper is as follows. Section~\ref{sec:background}
provides background information about blockchains and the core
properties relating to our decomposition. Section~\ref{sec:criteria}
outlines the criteria used for comparison and
analysis. Section~\ref{sec:membership} discusses and analyzes the
membership selection and proposals. Section~\ref{sec:cons} provides an
analysis of consensus mechanisms. Section~\ref{sec:structure}
overviews structure proposals for the
blockchain. Section~\ref{sec:attack} highlights proposed attacks
relating to the membership selection and consensus.
Section~\ref{sec:discussion} discusses other insights and aspects% provides a discussion of the systems and
% areas details that are not covered
. We conclude with a summary of our analysis and discuss future
directions.

\section{Background\label{sec:background}}

In this section, we explore the foundational concepts of the blockchain and
discuss the initial specification of Bitcoin. We then discuss the applicability
of blockchains and conclude by highlighting the known impediments.

\subsection{Blockchain}

Blockchain is an append-only distributed ledger of transactions. First
introduced with Bitcoin~\cite{Nakamoto08-Bitcoin}, the blockchain
originated as a decentralized electronic payment system, which removed
the need for any third-party involvement for payment transfers.
%
% Data
% in blockchain is initially organised a chain of blocks directed using
% the hash value of blocks, thus the system is named blockchain. Later
% proposals may use paralleled chains or graphs to organise data,
% however, they are still commonly referred as blockchain, where the
% term ``blockchain'' becomes a misnamed concept of distributed ledgers.
The original Bitcoin blockchain organized data as a chain of blocks directed by
utilizing the block hash values, hence the name ``blockchain''. Later proposals
expanded from a single chain to parallel
chains~\cite{Kokoris-KogiasJ16-ByzCoin, comchain, Yu2018Repucoin} and
graphs~\cite{iota-whitepaper, baird2016swirlds, raiblocks-whitepaper}. This has
led to the term ``blockchain'' becoming a misnamed concept for distributed
ledgers.

The decentralized nature of blockchains means that all nodes verify and store
the transactions that have taken place in the system, and propose new blocks to
append to the chain.  The blockchain structure can be seen as a linearly
increasing linked list of transactions batched into blocks.  The chain begins
with a \textit{genesis} block at index 0 and each block appended links to its
direct predecessor forming the chain. This, however, is the combination of
pre-existing ideas constructed together to form what the blockchain is today.
Haber et al.~\cite{Haber1991} first introduced the idea of timestamping to
digitally verify a document, which is paralleled by the timestamped block being
appended on the chain. Similarly, the distributed ledger of
cryptographically-linked blocks can be represented as a fully replicated state
machine~\cite{schneider-replicatedstatemachine}, where the state machine is a
hashchain~\cite{lamport1981password}.

From the initial specification detailed by Bitcoin, various
blockchains have been proposed; improving and extending the original work by
adding new features and functionality on top of the decentralized payment.
An example is Ethereum~\cite{wood2014ethereum}, which introduced the ability to
execute Turing complete code on the blockchain and perform conditional payments
based on actions through \textit{smart contracts}. A number of new
blockchains~\cite{cardano, dash, verge, stellar, neo, waves}, and distributed
ledger technologies~\cite{ripple, Corda}, have been created to fulfill new
purposes and help integrate the blockchain into critical infrastructure today.

Such extensions have provided ways for which blockchains can be applied to
existing systems. Transport~\cite{arcade-city, chasyr},
Healthcare~\cite{yue2016healthcare, peterson2016blockchain,
mettler2016blockchain}, and Finance~\cite{Corda,hyperledger-fabric, ripple}
have shown potential for the use of blockchain technology and a number of
applications are being used in systems today~\cite{ibm-abudhabi, walmart-food,
blockchain-finance-ibm, forbes-blockchain-use}.

\subsection{Bitcoin}

% Describes the inherent ways in which Bitcoin works - more detailed depiction
% of Bitcoin and the motivation behind it.

The seminal paper on Bitcoin~\cite{Nakamoto08-Bitcoin} was the first
to introduce blockchain as the completely decentralized electronic
cash system on a peer-to-peer network. It facilitates pseudonymous
payment between two parties without the requirement of a third party.

Each account on the Bitcoin blockchain is composed of a public and
private key pair. The hash of a public key identifies the account of
the key owner by forming an \textit{address}. The public key address
is used to accept coins, and the private key of an account is used to
authorize a spending. Payment occurs with a ``payer'' (or ``payers'') signing a
transaction using the private key, transferring assets to the address
of the ``payee'' (or ``payees''). Once signed, the ``payer'' broadcasts the
transaction to the network, where the transaction is then mined into a
block. The ``payee(s)'' can use its private key of the address to claim
its ownership.

A transaction contains three data fields, namely metadata, inputs, and
outputs. The metadata field records the unique ID of the transaction
(i.e. the hash of the entire transaction), the size of the
transaction, the number of inputs, the number of outputs, and a
\texttt{lock\_time} field defining the time period that one have to wait to
validate this transaction. An input specifies a previous transaction
using its hash value. The input also contains the index of the
previous transaction's outputs that is being claimed (as there may be
more than one output). A valid signature for the current transaction
is also required to prove the ownership of the claimed output (of the
previous transaction), by using the signing key associated to the
address of the claimed output. An output has two fields, namely value
and address. They define the value to be transferred to the
address. Bitcoins are just transaction outputs with an arbitrary value
with 8 decimal places of precision. The smallest possible value is
$10^{-8}$ BTC which is called 1 \emph{Satoshi}. For the transaction to
be valid, every input must be an unspent output (i.e. coin) of a
previous transaction. Every input must be digitally signed. The total
value in the input field must be no smaller than the total value in
the output field. If the total value in the input is greater than the
total value in the output, then the difference between the two total
values is called a transaction fee.

Transactions are broadcast in the network. If a node receives the
same transaction multiple times, it only broadcasts it once. So that
transactions will not be broadcast in the network forever. For a
received valid transaction, a node will use it as a part of the input
in its ``mining'' process, and the transaction is accepted if it is
included in the blockchain. All nodes in the network have the option
to participate in the mining process, computing a valid hash as the
Proof-of-Work (PoW). We will detail different Proof-of-Work systems in
\S~\ref{subsec:pow}. The block containing a set of verified
transactions is then propagated through the network, and if valid and
has been decided as the correct extension of the chain, the block
containing transactions will be accepted.

The creator of a block that is included in the chain obtains bitcoins
as reward of its work. A reward has two parts, namely a pre-defined
amount of coins as mining reward, and transaction fees of contained
transactions.

% The Bitcoin system has a finite supply of bitcoins --- 21
% million in total. The reward for a winner is 50 bitcoins at the
% beginning of the system, and halves every 210,000 blocks
% (approximately every four years provided a block is created every 10
% minutes). So coins will run out in 2040, and no new bitcoins will be
% created unless rules change.

% Once the
%transaction has been mined, the transaction has been appended in the ledger and
%is distributed to all nodes in the network.
The seminal work introduced by Bitcoin was a scalable membership
selection and consensus, dubbed Nakamoto's Consensus, discussed in
\S~\ref{subsec:nakamotocons}, as well as the application of a
distributed timestamping service for decentralized validity on a
ledger. This assembly of ideas presented new technology, causing a
paradigm shift in digital payments and providing a number of new
possibilities for this technology to evolve.

% \subsection{User Accounts}
%
% % All users create their account with a private key
% % They then derive a public key, have some hash functions performed and they
% % Are allocated an address
%
%
% \begin{figure}[h!]
%     \begin{center}
%       \begin{tabular}{p{0.2\columnwidth}|p{0.65\columnwidth}}
%           Private Key &  {\tt b1d0a747204fbd8cd649cfc223\newline 27caa585491740b33dac28c345\newline 6befbb6862f} \\
%           \midrule
%           Public Key &   {\tt 0x45a9cb0ddb3a1888c56eef23\newline c1179adf237c3f7c3ee6ba4564\newline 9f26e7d29a7e4cddb54019420d\newline 1d58f1d18b21d0525f21ee45e9\newline bca9e70a9648370e786733aa5} \\
%           \midrule
%           Address &      {\tt 0x9183C012BC7174CE4800ecc8\newline Ee8950577f4312f}
%         \end{tabular}
%         \caption{User Account Example.\label{fig:account}}
%     \end{center}
% \end{figure}
%
% User accounts on the blockchain are managed through the public-private
% key infrastructure. Each user creates a private key locally, which is
% then used to derive the associated public key. To obtain a user
% address, the public key is put through a series of hash
% functions. This address is the unique identifier for each user and is
% what transactions are addressed to with asset transfer.
%

\subsection{Smart Contracts}

As the blockchain continues to evolve, new functionalities emerge, allowing them
to be applied in numerous use cases and further integrated into daily use. One
major evolutionary impact was the introduction of \textit{Smart Contracts} in
Ethereum~\cite{ethereum, wood2014ethereum}. The Ethereum blockchain runs a
virtual machine in the lower layer, known as the EVM, holding the blockchain
state and executing all blockchain commands. Through this, Ethereum introduced
the functionality to run \textit{bytecode} on the EVM of all nodes and interact
directly with the blockchain. This provides mechanisms not only for conditional
payment, but programmable applications that interact directly with the
blockchain verified by all nodes.

A smart contract is typically written in a high-level language, such as
Solidity~\cite{solidity-readthedocs}, and compiled down to bytecode. A
transaction is then composed to deploy the contract on the blockchain. The
transaction data contains the contract code and any constructor arguments to
instantiate the contract. Once deployed, the contract address is returned as
part of the transaction receipt and can be interacted with. If a contract
function modifies the blockchain state, it must be invoked through a
transaction, which does not require any asset transfer but must have enough
``Gas''\footnote{Ethereum employs the concept of Gas to prevent indefinite code
execution. Each instruction on the Ethereum Virtual Machine has an associated
cost, so function invocations would have a cost to run.}. However, any
functions that do not require state modification, or, viewing any public
variables can be invoked without a required transaction.

Since the blockchain is seen as ``immutable'', once a contract has been
deployed, it cannot be modified. This has been revealed as a severe weakness
for many applications, as any vulnerable code is stuck on the blockchain and
cannot be updated or patched. Current techniques often require ``proxy''
contracts which allow for contacts to be ``upgraded'' with minimal
impact\footnote{A proxy contract provides an address which will always point to
the ``current'' version of a contract, so if a new upgrade is deployed on the
chain, people will still use the same contract address to access the upgraded
contract.}.

Most blockchain applications, known as Distributed Applications, or DApps,
interact with contracts through libraries, written in languages such as
JavaScript, Python, Go or Java. These are often seen as mobile or web
applications and are similar to those used commonly today.

Other blockchain systems, such as HyperLedger~\cite{hyperledger} and
EoS~\cite{eos} also provide smart contract capabilities. However, blockchains
such as Bitcoin allow for ``scripts'' to be written into the data field of
their transactions and executed on the blockchain directly. Overall, smart
contracts allow for the blockchain to be integrated with business logic and
allow conditional payment through contract invocation.

\subsection{Known Causes of Blockchain Varieties}

The original Bitcoin blockchain provided the foundation for blockchain systems
today by introducing the decentralized peer-to-peer transaction system.
However, aspects of Bitcoin served as focal points of evolution to new
blockchain varieties. We list below the main triggers for these varieties.

% The blockchain faces a number of identified impediments that impact on the
% applicability in a number of areas. However, these impediments are the focus
% for research and improvement.

\subsubsection{Transient Forks}

With multiple nodes working to propose new blocks, there is a high
chance of concurrent proposals. With structures that follow a single,
canonical chain, the concurrent proposals create a transient branch in
the chain, known as a \textit{fork}, in which the proposed blocks
share a common predecessor and can continue on their own separated
path. At this point, the nodes of the network must decide which is the
correct chain to extend through consensus. We defer the explanation of
the different consensus approaches to Section~\ref{sec:cons}. The
occurrence of forks impact transaction finality and can lead to a case
of double-spending, where the same coin is spent multiple times on
different branches. However, forks can also be used for protocol upgrades,
known as \textit{hard forks}, where the system diverges from a given block
height and all nodes running the latest version will be on the same branch.

%As all nodes are working to propose new blocks, there is a high chance of
%concurrent proposals. In some blockchains, namely those following the chain
%structure followed by Bitcoin, the concurrent proposals from nodes create a
%transient branch in the chain, known as a \textit{fork}, in which two blocks
%have the same predecessor. At this point, the nodes must decide which is the
%correct block to extend, through Nakamoto's consensus or Ghost, further
%explained in sections~\S~\ref{subsec:nakamotocons} and~\S~\ref{subsec:ghost}
%respectively. This presents an effect on the finality of transactions as the
%risk of a transaction being included in a fork can lead to double spending.
%Other blockchains that deploy different consensus or structures may not
%experience forks, or utilize the forks alternatively.

%At the point of concurrent proposals, a transient fork
%appears, changing the single chain into a tree.
%At this point, only one block
%can be accepted as the correct head of the chain, invoking the consensus as
%the conflict, or fork, resolution.

\subsubsection{Storage Size}

The blockchain is designed as an append only ledger, resulting in an linearly
increasing structure. This requires storage space that is growing at dynamic
rates. Bitcoin, currently
$182.206$GB\footnote{\url{https://www.blockchain.com/charts/blocks-size}.
  Accessed 09 Sep 2018.} and Ethereum is $94$GB (with Geth's Fast
Sync)\footnote{\url{https://etherscan.io/chart2/chaindatasizefast}. Accessed 13
Sep 2018.} and $600+$GB (full
node)\footnote{\url{https://bitinfocharts.com/ethereum/} Accessed on 13 Sep
2018.} reveal the size requirements for a node on the blockchain.
The growing storage requirements present limitations for certain devices to
participate, such as mobile or IoT, as well as long-term problems in
maintaining a node. Current efforts, such as pruning and light clients, focus
on this problem and aim to provide low space requirements for nodes.

\subsubsection{Scalability}

Blockchains provide a seemingly scalable service to growing numbers of users
and nodes. However, this can be contrasted by the increasing energy consumption
of Proof-of-Work blockchains and, in some cases, increased transaction latency.
Scalability of decentralized distributed systems is a major problem area for a
number of systems with a numerous proposals for improvements. Another important
aspect of scalability is the system throughput, defined as the number of
transactions per second included into a block, which is one of the major
hindrances to blockchain deployment. When compared to current centralized
systems, blockchain systems fail to meet throughput requirements to handle
daily spending. The cost of true decentralization is observed through the lower
throughput. However, many efforts are being made to improve the blockchain
throughput while still offering the same guarantees~\cite{Vukolic15}. We divert
further discussions of scalability to Section~\ref{subsec:scalability}.

\subsubsection{Clique Formation and Centralization}

% Bitcoin's main selling message was decentralization
Bitcoin's primary attraction was the decentralized distributed system, without
a required presence of a controlling party. Although the ideal blockchain would
have homogeneous nodes distributed and owned by unique users, the incentive of
block reward proved to be more profitable if miners formed cliques to produce
blocks. Due to these cliques, seen as mining pools in current PoW blockchains,
the decentralization of the system has been heavily impacted. For example, in
Bitcoin's blockchain, as of February 2019, 4 pools can be seen to control
larger than 50\%, and 15 pools control more than
80\% of the total reported hashrate~\cite{blockchaininfo-pools}.

% No middle or controlling party to trust
% The block reward was given in such a way that clique formation proved to be
%   a more profitable way to participate.
% Due to this, the decentralization aspect is slowly forming into large
%   cliques, or pools.
% Bitcoin as an example, 5 pools > 50%. Large amount hosted in china.

\section{Analysis Criteria\label{sec:criteria}}

In this section, we introduce the criteria used in evaluation and comparison of
the different membership selection and consensus mechanisms. We begin by
describing the common assumptions made and follow by describing the properties
of membership selection and consensus.

\subsection{Common Assumptions}

Both membership selection and consensus require assumptions to achieve their
goals, often encapsulating system behavior (e.g.\ network delay, node numbers,
etc) as well as a threat model (e.g.\ threshold of faults, adversary behavior).

\subsubsection{Network}

The network assumptions detail the bound of time for messages to be delivered
and can be categorized into three distinct classifications:

\begin{itemize}
  \item \textit{Synchronous}: there is a known, fixed upper bound on the time
    required for a message to be sent from one processor to another.
  \item \textit{Asynchronous}: there is no upper bound on the time taken for a
    message to be delivered. It is proven by Fischer et al.~\cite{FischerLP85-FLP}
    that it is impossible to have deterministic
    consensus schemes in an asynchronous model.
    \item \textit{Partially Synchronous}: the upper bound for message delivery
      exists but is unknown a-priori~\cite{DLS1985}.
\end{itemize}

\subsubsection{Online Presence}

The online presence considers whether all nodes are online at any given point
in time, or only a subset of nodes.

\begin{itemize}
  \item \textit{All Online}: The model in which all nodes are online and have network
    connections to other nodes.
  \item \textit{Sleepy}: The sleepy model considers crash fault tolerance in
    the BFT context. This concept was initially introduced for classical state
    machine replication~\cite{liu2016xft}, and later introduced for
    blockchains~\cite{BentovPS16-sleepyModel, pass-shi-sleepy}.  Nodes are
    classified as ``alert'' or ``sleepy''.  An alert node is one that is
    currently online and has predictable network connections to all other alert
    nodes. A sleepy node is one that is currently offline, but may become alert
    later, and vice versa.

\end{itemize}

\subsubsection{Adversary Threat Model}

The adversary threat model captures the threshold of Byzantine behavior that a
system can tolerate. These threat models have two distinct properties, type and
threshold, where the type defines what adversary is referenced, and the
threshold is the upper bound on Byzantine power specific to the type~\cite{abraham2017blockchain}. Traditionally, Byzantine Fault
Tolerant (BFT) systems put an assumption on the maximum number of committee
members and clients that an attacker can control. This is often referred to as
the maximum number $f$ of Byzantine nodes. In our generic analysis, in order to
capture all situations, we define $N$ as the total voting power of the system,
and $f$ as the maximum voting power controlled by all attacker nodes (as a
fraction of $N$).  The $f-to-N$ relation becomes then a fraction defining the
maximum percentage voting power that can be controlled by the adversary without
the system failing to reach its goals.

\begin{itemize}
  \item \textit{Traditional}: The number of votes a Byzantine actor can
    control, often represented as the number of Byzantine votes $f$ as
    a proportion of $N$, the total number of votes.
  \item \textit{Computational}: The adversary is defined by the amount of
    computing power they control. The adversary is expressed as
    the amount of computing power $f$ with respect to the total computing
    power of the system, $N$.
  \item \textit{Stake}: For stake-based systems, the adversary is defined
    by the amount of asset the adversary has to stake. This is defined as
    the amount of a resource in possession of the adversary $f$ in regards
    to total resources possessed by all users in the system $N$.
  \item \textit{Space}: The amount of storage space readily available
    to an adversary with low-latency reads. This adversary, similar to
    the computational adversary, controls a percentage of the total storage
    space in the system, represented respectively as $f$ and $N$.
\end{itemize}

\subsubsection{Trust}

The trust considers whether the proposal assumes a trusted presence to interact
and participate.

\begin{itemize}
  \item \textit{Hardware}: The proposed mechanism assumes that trusted hardware
    components will be used in the operation and all output is taken as
    correct.
  \item \textit{Participants}: This assumes that trusted authorities or
    participants, through policy or otherwise, are operational in this
    mechanism. This can be seen as an oracle or trusted entities in
    the system.
  \item \textit{None}: This means there is no assumption on the trust of the
    hardware or participants. All aspects assume that there is no trust
    required in the system to operate.
\end{itemize}

% \begin{enumerate}[label=$\bullet$, listparindent=1.5em]

%    \item \textbf{Trust} 
% \end{enumerate}

% Removed under Vincent's suggestion due to the paper depth/length being too
% Great. This can be a follow-on project?
% \subsection{Features}
%
% The features that are provided as a combination of the goals and the
% assumptions if they are met.
%
% \begin{itemize}
%     \item \textbf{Consensus Finality} is achieved if both \emph{Agreement} and
%         \emph{Total Order} are satisfied~\cite{Vukolic15, vitalikfinality}.
%       \item \textbf{Responsiveness}\todo{Define}\JS{I'm not sure if we
%           want to include this -- I'm not really convinced by the
%           definition they give, and there is no other literature
%           discussing it. In an updated version, they even removed the
%           Attiya-Dwork-Lynch-and-Stockmeyer work that was previously
%           cited as ``the source'' for this definition. Maybe, to help
%           us decide whether to include this, we should see first, what
%           is the real benefits of having this property; and second,
%           how should we analyse this against other systems.}
%     \item \textbf{Non-Forkable} is the feature in which at any point in the
%         execution, no two blocks are agreed upon. It is achieved if both
%         \emph{Agreement} and \emph{Termination} are satisfied.
% \end{itemize}

\subsection{Membership Properties}

The membership selection provides the subset of nodes that are then admitted to
participate in the consensus or block proposal. Each mechanism exhibits
properties that allow for classification and categorization.

% Membership selection provides the subset of nodes that are admitted to
% participate to the consensus, that is, become members of the consensus
% group.

\subsubsection{Openness}

The openness represents the rules on how nodes participate in the system and how
committees are formed. This can be seen as a fundamental aspect in the design
of blockchain algorithms and depicts its suitability for certain
environments~\cite{publicprivatechain, Vukolic15}.

\begin{itemize}
  \item \textit{Permissionless}: A permissionless, or public, system has no
    restriction on the set of candidates and their behavior. That is, the size
    of the consensus committee is not pre-defined nor fixed, and a node is able
    to participate or leave at any given time.
  \item \textit{Permissioned}: A permissioned, or private, system has complete
    restriction on the nodes participating. A node must join the system by
    being accepted in and as a known member. The consensus committee is often
    predefined, or, operates under the assumption that all nodes are known to
    all.
  \item \textit{Partially-Permissioned}: A partially-permissioned system is
    often seen in consortium scenarios, where there are certain restrictions
    present on either the consensus members or the node participation. Some
    systems allow nodes to freely join, but only a pre-defined subset of nodes
    form the consensus; other systems require nodes to be accepted in the
    system but allow for any node to become a member of the consensus.
\end{itemize}

\subsubsection{Selection Approach}

The selection approach details the process used to rank and/or select the nodes
which will form the consensus group. These approaches are used to differentiate the
consensus nodes from other nodes according to the way they prove to have met
the pre-defined criteria for some specific parameter(s). In order to increase
the probability of being selected, a node must show that it is better than
other nodes according to the selection criteria. For example:

\begin{itemize}
  \item \textit{Work}: This work parameter measures the ``amount of valid
    work'' a node has produced and can be directly related to the amount of
    computational power used to produce the work.
  \item \textit{Stake}: The amount of resources the node has
    ``staked''. This can be seen as the node's balance, or, the amount of
    asset the node has decided to place as stake. For a node to increase
    probability, they would have to increase the amount of stake.
  \item \textit{Resource Count}: In some mechanisms each individual resource,
    for example CPUs, counts towards the probability of selection. For a node
    to increase the probability of selection, they must possess more of the
    elected resource.
\end{itemize}

\subsubsection{Incentive}

To allow for scalable and robust consensus, some schemes provide a mechanism to
incentivize honest behavior through reward or punishment. We consider the
quality of incentives a major factor for the stability and effectiveness of the
selection mechanisms and consensus. When effective, they introduce
\emph{rationality} and thus influence node participation and heavily constrain
Byzantine behavior.

\begin{itemize}
  \item \textit{Reward}: measures the benefit that one will get as a result of
    its contribution to the system.
  \item \textit{Punishment}: measures the cost that one needs to pay if it
    behaves incorrectly.  It is used to discourage any malicious behavior, and
    plays an important part to the efficiency of the system and may affect the
    overall performance.
\end{itemize}

\subsection{Consensus Properties}

Consensus forms the core property of the blockchain where the valid blocks
are agreed upon to be appended to the chain. The \textit{blockchain
consensus} is composed of consensus instances, where each consensus
instance is the decision for the
block decision at each index of the chain. The valid blocks to be decided contain
valid transactions, defined as the following.

\begin{definition}[Transaction]
  A command that changes the state of the Blockchain.
\end{definition}

\begin{definition}[Valid Transaction]
  A transaction that is consistent with the current blockchain state and
  has correct signatures and structure.
\end{definition}

\subsubsection{Generic Blockchain Consensus Properties}

The consensus instances exhibit properties that define how the decision is made
and what guarantees are in place.

\begin{itemize}
  \item \textit{Agreement}: For a given consensus instance $i$, if a correct
    node decides block $B$, then all correct nodes decide $B$.
    % with probability $P_a$
    % If a correct node decides a state
    % $\sigma$, then all correct nodes will decide $\sigma$ in this
    % consensus instance.~\cite{Guerraoui2000, wood2014ethereum}.
  \item \textit{Termination}: All correct nodes eventually decide~\cite{quorumstorage, MA06}.
    %with probability $P_t$
    % A consensus instance is terminated when
    % all honest nodes eventually decide a block for a given height of
    % the chain~\cite{quorumstorage, MA06}.
  \item \textit{Validity}: If all correct nodes propose a valid transaction
    before starting a consensus instance $i$, then the block decided in $i$ is
    not empty.
\end{itemize}

    % \item \textbf{Validity} is the property that a block appended
    %     to the chain satisfies~\cite{blockchain2byzantine}:
    %         \begin{itemize}
    %             \item Contains no conflicting transactions.
    %             \item Adheres to a validity predicate defined for the
    %               given system.
    %         \end{itemize}
    % Removed under Vincent's suggestion
    % \item \textbf{Liveness} \todo{define}

\subsubsection{Blockchain State Properties}

The blockchain state provides the foundation for the way transaction finality
is interpreted and how state transitions are effected. However, both properties
may not be observed at the same time.

\begin{itemize}
    \item \textit{Total Order Prefix}: A blockchain exhibits a total order
      prefix when consensus instances have terminated with agreement on a
      common prefix of states $\sigma_0, \sigma_1, \dots \sigma_i$, but have
      not yet terminated for the states starting from $\sigma_{i+1}$. This is
      commonly observed with a \textit{fork}~\cite{PSS16, GarayKL17,
      garay2015Bitcoin}.
    \item \textit{Total Order}: A blockchain exhibits total order when
      all consensus instances have terminated and agreement has been
      reached for all states.
\end{itemize}
%
% \CN{ Partial Order: Vincent wants us to present the definition in a more formal
%   way, as we are looking for something stronger than the traditional partial order
%   as explained in order theory/set theory.\\
%
%
%   Let $l_i = \left \langle B_i, P_i \right \rangle$ be the blockchain view at
%   node $p_i$. Let $l_j \left \langle B_j, P_j \right \rangle$ be the blockchain
%   view at node $p_j$. A blockchain is \textbf{partially ordered} if the following
%   properties hold in both $l_i$ and $l_j$:
%
%   \begin{enumerate}
%     \item There exists a set of blocks, $S$, that is a common prefix in both
%       $l_i$ and $l_j$.
%     \item The size of the common prefix, $S$, is at least 1, where the common
%       block is the genesis block. Formally, $\left | S \right | \geq 1$.
%     \item The size of the common prefix is monotonically increasing over time.
%   \end{enumerate}
%
%   Make sure to cite:~\cite{PSS16, GarayKL17, garay2015bitcoin}
% }

\section{Membership Selection\label{sec:membership}}

In this section, we present an overview of the membership selection algorithms
deployed in current blockchain systems. The primary purpose of membership
selection is to determine a committee of nodes to participate in the consensus,
whose primary goal is to propose, or validate, blocks and provide a decision
for the correct block at the next index of the chain. A summary of our overview
is presented in Table~\ref{membershiptable}.

\afterpage{
\onecolumn
  \begin{landscape}
    \vspace*{1.9cm}
  \begin{center}
  \begin{table*}[h]
    \caption{Membership Selection Overview \label{membershiptable}}
        \centering
        \fontsize{6}{8}\selectfont
        \begin{threeparttable}
          %\hspace{-2cm}
          \resizebox{1.3\textwidth}{!}{
        \begin{tabular}{
            p{0.06\textwidth} | % Subheading
            p{0.05\textwidth} | % Goal/Assumption
            x{0.07\textwidth} | % PoW
            x{0.07\textwidth} | % PoR
            x{0.07\textwidth} | % PoCA
            x{0.07\textwidth} | % PoD
            x{0.07\textwidth} | % PoLock
            x{0.07\textwidth} | % PoActivity
            x{0.073\textwidth} | % Ouroboros
            x{0.07\textwidth} | % DPoS
            x{0.07\textwidth} | % AlgoRand
            x{0.07\textwidth} | % PoB
            x{0.07\textwidth} | % PoA
            x{0.07\textwidth} | % PoC
            x{0.07\textwidth} |  % PoL/PoET
            x{0.07\textwidth} % PoLocation
        }
            \toprule
            % Table Header
            \multicolumn{2}{c|}{} &
            %& % Subheading
            %& % Goal/Assumption
            \textbf{PoW} & % PoW
            \textbf{PoR} & % PoR
            \textbf{PoCA} & % PoCA
            \textbf{PoD} & % PoD
            \textbf{PoLock} & % PoD
            \textbf{PoActivity} & % PoActivity
            \textbf{Ouroboros} & % Ouroboros
            \textbf{DPoS} & % DPoS
            \textbf{AlgoRand VRF} & % AlgoRand
            \textbf{PoB} & % Proof-of-Burn
            \textbf{PoA} & % Proof-of-Authority
            \textbf{PoC} & % PoC
            \textbf{PoL/PoET} & % PoL/PoET
            \textbf{PoLocation} % PoLocation

        \\\midrule

            %%% Seminal Chain
            \textbf{Chain} &
            Seminal Chain & % Goal/Assumption
            Bitcoin \vfill \cite{Nakamoto08-Bitcoin} & % PoW
            Repucoin \cite{Yu2018Repucoin} & % PoR
            PPCoin \vfill \cite{king2012ppcoin} & % PoCA
            Ethereum \cite{casperfriendlyghost, ethPoSFaq}& % PoD
            Tendermint \cite{kwon2014tendermint, buchman2016tendermint} & % PoLock
            -- \vfill \cite{bentov2014proof} & % PoActivity
            Ouroboros \cite{KiayiasKRDO16-Ouroboros} & % Ouroboros
            BitShares \vfill \cite{DPoS-bitsharestalk} & % DPoS
            AlgoRand \cite{Micali16-algorand, GHMVZ-algorand} & % AlgoRand
            Slimcoin \vfill \cite{slimcoin} & % Proof-of-Burn
            Ethereum \cite{EIP225, parityPoA} & % Proof-of-Authority
            Permacoin \cite{MillerJSPK14-Permacoin} & % PoC
            Intel Ledger\tnote{*} \cite{Intel-PoET} & % PoL/PoET
            Platin \vfill \cite{platinio-whitepaper} % PoLocation

            \\\midrule

            %%% Assumptions
            \multirow{3}{*}{\raisebox{-\heavyrulewidth}{\bf Assumptions}} &
            Trust & % Goal/Assumption
            \gcell None & % PoW
            \gcell None & % PoR
            \gcell None & % PoCA
            \gcell None & % PoD
            \gcell None & % PoLock
            \gcell None & % PoActivity
            \gcell None & % Ouroboros
            \gcell None & % DPoS
            \gcell None & % AlgoRand
            \gcell None & % Proof-of-Burn
            \ycell Participants (Authorities) & % Proof-of-Authority
            \gcell None & % PoC
            \rcell Hardware & % PoL/PoET
            \gcell None % PoLocation
            \\\cmidrule{2-16}

            & % Subheading
            \multirow{2}{*}{\raisebox{-\heavyrulewidth}{Adversary}} & % Goal/Assumption
            Computational & % PoW
            Computational & % PoR
            Stake & % PoCA
            Stake & % PoD
            Stake & % PoLock
            Stake / Computational & % PoActivity
            Stake & % Ouroboros
            Stake & % AlgoRand
            Stake & % DPoS
            Stake & % Proof-of-Burn
            Traditional & % Proof-of-Authority
            Space & % PoC
            Traditional & % PoL/PoET
            Traditional % PoLocation

            \\\midrule

            %%% GOALS
            \multirow{3}{*}{\raisebox{-\heavyrulewidth}{\bf Properties}} &
            Openness & % Goal/Assumption
            Permissionless & % PoW
            Permissionless & % PoR
            Permissionless & % PoCA
            Permissionless & % PoD
            Permissioned & % PoLock
            Permissionless & % PoActivity
            Permissionless & % Ouroboros
            Permissionless & % DPoS
            Permissionless & % AlgoRand
            Permissionless & % Proof-of-Burn
            Permissioned & % Proof-of-Authority
            Permissionless & % PoC
            Permissionless  & % PoL/PoET
            Permissionless % PoLocation

            \\\cmidrule{2-16}

            & % Subheading
            Selection Approach & % Goal/Assumption
            Memory / Hash Power & % PoW
            Past Contribution & % PoR
            Coin Age & % PoCA
            Coins Deposited & % PoD
            Coins Locked & % PoLock
            Coins Held, Online & % PoActivity
            Balance & % Ouroboros
            Votes & % DPoS
            Balance & % AlgoRand
            Coins Burnt & % Proof-of-Burn
            Elected Authority & % Proof-of-Authority
            Storage & % PoC
            Number of CPUs  & % PoL/PoET
            GPS Location, Nearby Peers % PoLocation

            \\\cmidrule{2-16}

            & % Subheading
            Incentive Reward & % Goal/Assumption
            \cmark & % PoW
            \cmark & % PoR
            \cmark & % PoCA
            \cmark & % PoD
            \xmark & % PoLock
            \cmark & % PoActivity
            \cmark & % Ouroboros
            \cmark & % DPoS
            \xmark~\tnote{$\dagger$} & % AlgoRand
            \cmark & % Proof-of-Burn
            \xmark & % Proof-of-Authority
            \cmark & % PoC
            \cmark & % PoL/PoET
            \cmark % PoLocation

            \\\cmidrule{2-16}

            & %
            Incentive Punishment & % Goal/Assumption
            \xmark & % PoW
            \cmark & % PoR
            \xmark & % PoCA
            \cmark & % PoD
            \cmark & % PoLock
            \cmark & % PoActivity
            \xmark & % Ouroboros
            \cmark\tnote{$\ddagger$} & % DPoS
            \cmark & % AlgoRand
            \xmark & % Proof-of-Burn
            \xmark~\tnote{$\#$} & % Proof-of-Authority
            \cmark & % PoC
            \xmark & % PoL/PoET
            \cmark % PoLocation

        \\\bottomrule

        \end{tabular}
      }
        \begin{tablenotes}
        \item[*] Intel's Ledger was later HyperLedger Sawtooth~\cite{sawtoothpoet}
        \item[$\dagger$] AlgoRand reward based incentive mentioned as future
          work in the paper~\cite{GHMVZ-algorand}.
        \item[$\ddagger$] DPoS punishment is that the malicious node will lose
          trust in the community and be voted out.
        \item[$\#$] There is no incentive mechanism paired with PoA, but an intrinsic reputation
          to uphold.
        \end{tablenotes}
        \end{threeparttable}
\end{table*}
\end{center}
\end{landscape}
\twocolumn
}

\subsection{Proof-of-Work (PoW)\label{subsec:pow}}

\begin{figure}[t]
    \begin{center}
        \includegraphics[width=\columnwidth]{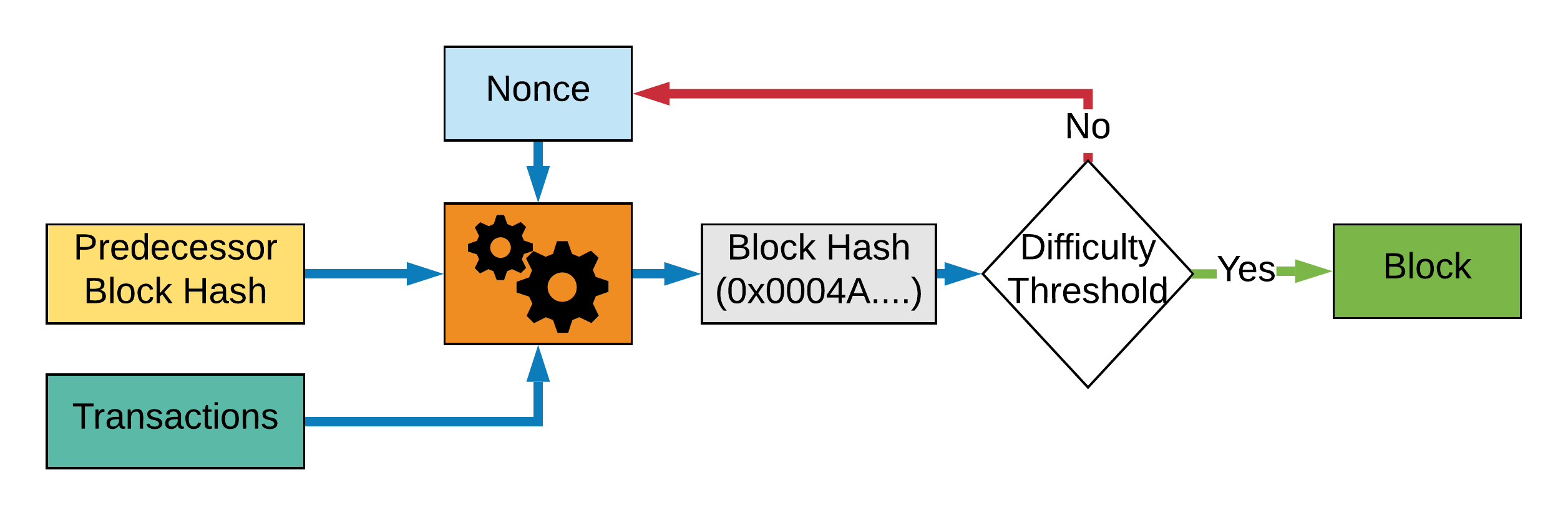}
        \caption{Proof-of-Work flow.\label{fig:pow}}
    \end{center}
\end{figure}

Proof-of-Work (PoW) is the mechanism used by Bitcoin and most mainstream
blockchain systems today. It was proposed in this context to address the Sybil
Attacks~\cite{douceur-sybil}, where an attacker creates multiple entities and dominates the
consensus committee. In particular, PoW addresses the Sybil attack by requiring
nodes to prove that they have performed some work at a non-negligible cost,
such as consumed electricity, time and hardware.
The ``work'' is to produce a solution to a \textit{puzzle}, where the
solution is relatively difficult to find, but can be easily verified by others.
This, in turn, acts as a disincentive for attackers due to the involved cost.
In Bitcoin and its variants, the process of working to obtain a solution is
called \textit{mining}, where the nodes performing the work are called
\textit{miners}.

Proof-of-Work in Bitcoin was heavily derived from the initial proposal by Dwork
et al.\ to combat junk mail~\cite{dwork1992pricing}, which was later
adopted by HashCash~\cite{Bla02}. The technique has two variants of
implementations, the CPU-bound approach, as seen in
Bitcoin~\cite{Nakamoto08-Bitcoin}, and the memory-bound
approach~\cite{AbadiBW03, DworkGN03}.

Proof-of-Work's strength lies in its ability to provide a mechanism for
validity to a seemingly unlimited, unknown set of nodes which is necessary
for this blockchain context. This quality has seen Proof-of-Work integrated as the
backbone for other hybrid blockchains~\cite{Kokoris-KogiasJ16-ByzCoin,
AbrahamMNRS16, pass2017thunderella, Yu2018Repucoin}.

\paragraph{\noindent\bf Key Concept}

The core concept of Proof-of-Work is proving validity through work. This allows
for a distributed system to provide a validity check, as an aspect of
scalability, to a large number of nodes whilst also adding resilience against
Sybil attacks. In the blockchain context, a valid block proposal from a node
must encapsulate the proof showing they have worked for some cost.

Proof-of-Work utilizes the process of solving the puzzle to express membership
selection. If a node wishes to participate in the proposal of a block in the
consensus, it must find a solution prior to learning of any other blocks that
have been proposed for that given index. The membership selection accepts any
number of nodes as long as they have provided a correct solution.

To produce a valid block with Proof-of-Work, as depicted in
Figure~\ref{fig:pow}, a miner repeatedly submits a chosen nonce along with the
predecessor block hash, the transaction root hash, and other metadata to a
hashing function. The output produced is then compared against the difficulty
threshold to determine its validity.  The difficulty threshold, dependent upon
the blockchain, is directly proportional to the number of hashes in expectation
to be performed to find a valid solution, as well as acting as a delay between
proposals.

% For example, in Bitcoin the difficulty threshold is adjusted to
% ensure one block is proposed every 10 minutes on average.  This provides the
% time for messages to be propagated across the network, while also reducing the
% number of concurrent proposals.

\paragraph{Weaknesses}

Although Proof-of-Work achieves its goal for blockchain membership selection,
it is hindered by a number of limitations. The process of mining required
to solve the Proof-of-Work puzzle requires large amounts of computation,
resulting in high resource and energy consumption. The rising difficulty to
maintain a constant block proposal time results in increasing costs for running
a miner, heavily impacting the scalability and future operation of the
blockchain.

The cost of computation is directly proportional to the increasing difficulty.
%result in throughput issues as the speed of block production. 
Finding a valid solution to the Proof-of-Work puzzles is a lengthy process,
both allowing for fairness in computational cost as well as giving the network
the chance to disseminate the information of the latest block before a forked
block is found. The inherent delays in proposals relate directly to a weakness
of Proof-of-Work in mainstream blockchains, as it must cater for a seemingly
unknown network size as well as large amounts of hashing power.

Proof-of-Work assumes the distribution of the hashing power, where the honest
nodes control the majority. However, there are many plausible attacks proposed
against Proof-of-Work where an adversary can gain control of hashing power, or
manipulate the node's connections to perform double spending.

\paragraph{Goals and Assumptions}

Proof-of-Work strives to provide membership selection in an open,
permissionless environment, where all nodes are freely able to join or
participate. Proof-of-Work ties the probability of membership
selection to the hash power of the node, assuming that an
increase in hash power requires an increase of cost. The computational
adversary assumptions made by Proof-of-Work follow this, where the
majority of the computational power of the network will be owned by
honest nodes, where the probability of selecting an adversary multiple
times in a row is low.

For nodes to be incentivized, Proof-of-Work is often accompanied by a reward
structure that rewards the nodes for their work, as seen in
Bitcoin~\cite{Nakamoto08-Bitcoin}, or Ethereum~\cite{ethereum} with the block
rewards.

\begin{table*}[t]
  \begin{center}
    \caption{The relation between difficulty and the expected number of hashes. As
    the difficulty increases, there is an observable increase in the number of
  leading $0$'s in the target, so a valid block must have a value lower than the
target.}
  \label{tab:mining_difficulty}
  \resizebox{0.8\textwidth}{!}{
    \begin{tabular}{r|l|l}
        \textbf{Difficulty} & \textbf{Expected Number of Hashes} & \textbf{Approximate Target} \\\midrule
    1 & 4.295e+09 & {\tt 0x00000000FFFF00000000000000000000\ldots}\\
    100 & 4.295e+11 & {\tt 0x00000000028F5999999999A000000000\ldots}\\
    1000 & 4.295e+12 & {\tt 0x00000000004188F5C28F5C2800000000\ldots}\\
    10000 & 4.295e+13 & {\tt 0x0000000000068DB22D0E560400000000\ldots}\\
    1000000 & 4.295e+15 & {\tt 0x00000000000010C6E6D9BE4CD7000000\ldots}\\
    100000000 & 4.295e+17 & {\tt 0x000000000000002AF2F2D14354120000\ldots}\\
    1000000000000 & 4.295e+21 & {\tt 0x00000000000000000119787E99468E30\ldots}\\
    \end{tabular}
    }
    \vspace{2pt}
  \end{center}
\end{table*}

\subsubsection{Proof-of-Work Variants}

\subsubsubsection{CPU-bound Proof-of-Work}
Bitcoin~\cite{Nakamoto08-Bitcoin} was the first mainstream blockchain,
which introduced the Proof-of-Work algorithm for cryptocurrency. The
design in bitcoin was heavily influenced by HashCash~\cite{Bla02}
together with Dai's B-Money~\cite{wdai-bmoney}.  The
HashCash cryptographic puzzle requires a node to find the solution by
repeatedly putting a \emph{nonce} through a pseudorandom function with
a SHA-1 hash. The Bitcoin Proof-of-Work deviates by requiring the
pseudorandom function to produce a SHA-256 hash, the input is the
combination of the nonce and the new block hash. The CPU-bound
functions are directly related to the calculation speed of the
hardware. Modern ``Application-Specific Integrated Circuit'' (ASIC)
devices, such as the {\em Antminer R4}~\cite{asicminer-antminer-r4}
%\footnote{\url{https://asicminermarket.com/product/antminer-r4-mute-8ths/}}
were designed to surpass ordinary computer hashes per unit of
money. This provides a major advantage to the owners that performed
the Proof-of-Work CPU-bound mining. The introduction of the ASIC
miners in the Bitcoin blockchain have heavily influenced the market,
and since have dominated the mining due to the advantages
provided. From the introduction of CPU-bound Proof-of-Work in Bitcoin,
it has been implemented in most major mainstream blockchains, such as
Ethereum and Litecoin.

To create a solution for the CPU-bound Proof-of-work, the miner
batches a number of transactions and creates a Merkle
tree~\cite{merkle1987digital} from the transaction hashes. The miner
knows the global {\em threshold}, or difficulty, the latest block and
the transactions. They then select a nonce and apply a pseudo-random
function to the new block of transactions.
%until they retrieve a hash value lower than the known difficulty.
The higher the difficulty is,
the more leading zero is required in the hash value of a valid block.
Table~\ref{tab:mining_difficulty} depicts the relationship between difficulty,
target and the number of expected hashes. The target is the number that the
block hash must be lower than, and as shown the number of leading zeroes
increases with increasing difficulty, which requires more expected effort to
solve. We refer readers to~\cite{BTCDifficulty} for a detailed explanation.

\subsubsubsection{Memory-bound Proof-of-Work}

In 2005, Dwork et al.~\cite{DworkGN03} further studied the concept of
memory-bound proof of work for preventing email spam. To cope with the influx
of ASIC miners that dominate the node hash power, memory-bound Proof-of-Work
algorithms~\cite{AbadiBW03,DworkGN03,BiryukovK16} are adapted by blockchains.
In particular, the memory-bound Proof-of-Work, also known as {\em egalitarian
Proof-of-Work}, relies on random access to slow memory rather than
computational hashing power, emphasizes a dependency on latency. This makes the
performance bound by memory-access speed rather than hashing power. Therefore,
this provides the system with resilience to ASIC miners and fast memory-on-chip
devices.

Following the success of Proof-of-Work in cryptocurrencies, but seeing the
disadvantage of ASIC miners, the EquiHash~\cite{BiryukovK16} algorithm was
adopted to memory-bound Proof-of-Work. The CryptoNote~\cite{cryptonote} as well
as ZeroCash~\cite{sasson2014zerocash} both adopt the memory bound proof-of-work
in the concept of a cryptocurrency.

\subsubsection{Proof-of-Reputation\label{subsubsec:repu}}

Proof-of-Reputation~\cite{Yu2018Repucoin} was proposed to extend Proof-of-Work
to mitigate the risks deriving from the ability to quickly gain computational
power, which are at the root of bribery, or flash
attacks~\cite{Bonneau16-bribery-attacks}. Rather than using instantaneous
mining power to select members, Proof-of-Reputation
%restricts the ability to quickly gain computational power by considering the
considers node's \textit{integrated power}.
%The integrated power utilizes the miner's historical performance and honesty.
In fact, it is calculated using
the total amount of valid \textit{work} a miner has contributed to the system,
over the regularity of that work over the entire period of \textit{time} during
which the system has been active. Hence the name, and the physical bound on
power growth rate. When a miner deviates from the system specifications,
the miner's reputation, and hence its voting power, will lower
in consequence of this negative contribution. This prevents a powerful
malicious miner from attacking the system repeatedly without significant
consequences, as can occur in the PoW-based systems.

In this way, an adversary can only be successful after gaining
sufficient reputation, requiring costly investment over time. Thus, it
can tolerate an attacker who controls a majority of mining power for a
short time period. In addition, the node's reputation drops to zero as
soon as an attack is detected, meaning an adversary can only launch
this attack at one time before having to rebuild reputation.
Therefore, any bribes or rented computational power require costly
time investments, lowering the economical incentive for the adversary
to launch the attack, further mitigating the risk of this attack
occurring.

\subsection{Proof-of-Stake (PoS)}

Proof-of-Work's major impediments include the extremely high resource
consumption to produce a block as well as the low throughput of
transactions per second~\cite{GervaisKWGRC16, CachinV17}. Efforts have
been made to pursue new mechanisms that can provide similar guarantees
for a permissionless environment whilst improving on the fall-backs of
Proof-of-Work. The first noted proposal on a stake-based voting system
was through Wei Dai's b-money~\cite{wdai-bmoney} in 1998, where it
mentions users' voting on included transactions, staking money for
dispute resolution. The concept of Proof-of-Stake in blockchain was
first proposed in a Bitcoin community forum~\cite{posbitcoincommunity}
in 2011 to provide quicker and more definite transaction confirmation
through a virtual mining mechanism. The main idea of Proof-of-Stake is
that the consensus participants are required to deposit something of
value at stake, and this deposit will be taken away if the node is
found to be acting incorrectly.

The virtual mining not only provides an environmentally friendlier blockchain
system, by saving the overall energy consumption when compared to PoW, but also
significantly improves the throughput as blocks can be appended and committed
to the chain faster.

The concept of Proof-of-Stake provides the flexibility to be implemented in a
variety of ways. Each implementation calculates the user's stake differently
and imposes different incentive mechanisms to promote node behavior.

\paragraph{\bf Key Concept}

\begin{figure}[t]
    \begin{center}
        \includegraphics[width=\columnwidth]{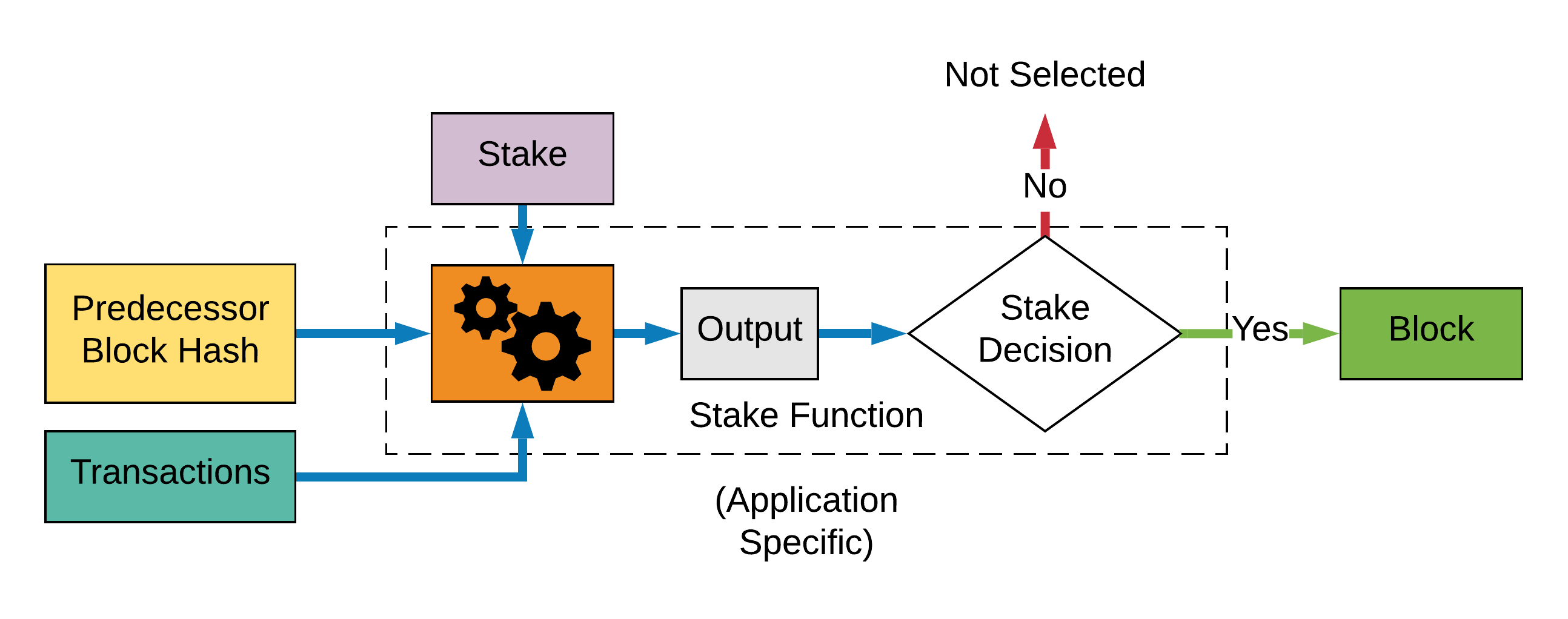}
        \caption{Proof-of-Stake flow.\label{fig:pos}}
    \end{center}
\end{figure}

The primary motivation behind Proof-of-Stake is to mitigate the energy waste
caused by the block production process in Proof-of-Work. The core concept lies
in a node proving validity by staking assets, replacing the hashing to solve a
cryptographic puzzle with a stake-based selection, while still preserving the
permissionless nature of the blockchain. The stake is either taken from their
current balance, or, is locked or deposited by the shareholder. The voting
power of the node can be proportionally mapped to the stake they have issued.
Figure~\ref{fig:pos} presents the Proof-of-Stake calculations.

``Nothing at Stake'' is a well known attack on Proof-of-Stake systems, where
stakeholders vote for all possible proposals by using the same stake. Since the
voting will lead to an eventual block decision, the stakeholders will not lose
any assets but gain profit, acting as an incentive to behave in this way. We
will discuss more about such attacks in Section~\ref{sec:attack}. To
prevent such an attack, designs include punishments that result in asset loss
upon detection of a stakeholder's malicious behavior.

The Proof-of-Stake core concept has been adapted into a number of different
membership selection attributes. The goal of the membership selection is to
filter the nodes to form a committee to participate in consensus, and can
assign a weight to the node's vote proportional to the amount of stake held.

\paragraph{Weaknesses}

Whilst Proof-of-Stake mitigates the energy usage tied to Proof-of-Work, it is
prone to a number of inherent weaknesses. The outstanding issue with
Proof-of-Stake is the increased potential of centralization and the concerns of
governance. The distribution of assets, as well as people's willingness to
stake their assets, greatly influence the membership selection and could lead
to a small subset of nodes being in a position of power for the entire system
operation. To solve this problem, some Proof-of-Stake implementations utilize
frequent membership changes and integrate randomness so the committee changes
and a larger pool of nodes is used. However, this is still greatly impacted by
the use of incentive mechanisms~\cite{ethproblemswiki}.

Proof-of-Stake, while satisfying the requirements of membership selection,
still suffers from a number of attacks. We defer the discussion of attacks
to Section~\ref{sec:attack}.

\paragraph{Goals and Assumptions}

All Proof-of-Stake implementations share some common goals and assumptions.
Members are selected through stake calculations using balances, deposits or
votes. To increase the probability of selection, nodes obtain or deposit more
stake. However, in some cases such as Delegated
Proof-of-Stake~\cite{DPoS-bitsharestalk}, the nodes are selected by being voted
in, where the votes are weighted proportionally to balance. Proof-of-Stake
assumes that there is a limited amount of resources for a node to stake, and
gaining the resource requires time or significant cost.

The various implementations provide incentives by rewarding, or punishing,
behavior and participation. Nodes that are selected are often rewarded through
transaction fees or block rewards. However, Proof-of-Stake is often accompanied
by punishment; when a node is found guilty of misbehavior it loses the stake
that it has deposited, or owns.

Similar to Proof-of-Work, Proof-of-Stake assumes that an adversary does not
control large portions of the stake, that the stake follows a distribution that
allows for the honest nodes to be selected with higher probability than the
adversary.

\subsubsection{Deposit based}

Deposit Based implementations of Proof-of-Stake allow any node to
deposit stake to have the chance to be selected as a member. The nodes
have a choice to increase their probability of selection by depositing
more stake, and vise-versa. Existing deposit based implementations
include proof-of-lock and proof-of-deposit, as follows.

\subsubsubsection{Proof-of-Lock}

Tendermint's~\cite{kwon2014tendermint} implementation of Proof-of-Stake
requires nodes to ``lock'' or ``freeze'' coins as stake and if found to act
maliciously, the frozen assets are removed. The membership selection selects
the nodes that have frozen coins to be part of the consensus group, where their
voting weight is proportional to the amount of coins they have frozen as a
ratio to total amount of frozen coins. This is effective against the ``nothing
at stake'' attack, as the malicious behavior will result in direct loss of
deposited coins.

\subsubsubsection{Proof-of-Deposit}

Similar to Proof-of-Lock, the Proof-of-Deposit as implemented in Ethereum's
original Proof-of-Stake~\cite{ethPoSFaq} proposal, requires nodes to deposit a
selected amount of coins. The blockchain keeps track of a set of nodes that
have deposited their coins and randomly selects members from this set to form
groups for the consensus. Dependent upon the
implementation~\cite{casperfriendlyghost}, the number of selected nodes is
different, i.e., a single node is selected if the chain-based implementation is
chosen, or, several nodes are selected if the BFT-style implementation is used.

\subsubsection{Balance Based}

Alternative to the deposit-based implementations, balance-based
implementations calculate stake using the node's current account balance, giving no
choice in the amount of asset staked by a node. However, some implementations
allow nodes to present their stake when they wish to be selected, whereas 
others assume the entire set of nodes are willing to participate in the
membership selection.

\subsubsubsection{Proof-of-Coin-Age}

Proof-of-Stake allows for anyone to quickly gain assets and utilize them as
stake, which could have adverse effects on the system stability. To prevent
instantaneous stake gain, PeerCoin, or PPCoin~\cite{king2012ppcoin}, implements
Proof-of-Stake through Proof-of-Coin-Age, which calculates the stake based on
balance multiplied by the length of \textit{time} a node has held that coin
for. If the coin is ever traded, the elapsed time is reset to 0 and the stake
is lost. This provides a disincentive for adversarial behavior as there is now
a required time investment, proving to be costly.

\subsubsubsection{Proof-of-Activity \label{subsubsubsec:poactivity}}

Although instantaneous stake gain is a challenge faced by Proof-of-Stake, many
implementations do not require explicit safeguards to provide resilience
against powerful actors. Proof-of-Activity~\cite{bentov2014proof} selects
``lucky'' nodes which must be online to participate and get rewarded. This
approach utilizes Proof-of-Work where a miner creates an empty block header to
randomly derive $N$ \textit{Satoshi's}\footnote{A ``Satoshi'' is the lowest
value of currency in the Bitcoin blockchain.} amongst all coins in existence.
The Proof-of-Activity's \textit{Follow-the-Satoshi} is then invoked to select
the owners of the selected Satoshi. The owner of this Satoshi is a stakeholder
chosen to propose the next block. All stakeholders verify the empty block
header to verify if they are the chosen stakeholder and sign the empty block.
Once a stakeholder can verify that $N-1$ signatures have signed the empty block
header, the stakeholder creates a wrapped block that extends the empty block
header, adding the transactions and $N$ valid signatures. Once the block has
been accepted, the transaction fees are shared between the original miner of
the empty block and the $N$ stakeholders that signed.

\subsubsubsection{AlgoRand's VRF}

Previous approaches to Proof-of-Stake may be vulnerable to an adversary
determining consensus members prior to the group creation, which could lead to
a Denial-of-Service attack against the consensus and prevent block signing. To
prevent this, AlgoRand's~\cite{Micali16-algorand, GHMVZ-algorand}
Proof-of-Stake implementation randomizes a new subset of members for each step
of block commitment using a Verifiable Random Function (VRF). This
implementation provides a unique feature, where only the nodes who are selected
for the next consensus step know that they are selected, while other nodes can
only verify this membership selection result afterwards. The use of VRF
prevents an adversary from predicting the set of members chosen for the next
step of committing a block. AlgoRand measures stake based on the amount of
currency held by an entity. Each member is randomly selected, with voting power
proportional to the amount of stake owned.

\subsubsubsection{Ouroboros}

Similar to AlgoRand, Ouroboros~\cite{KiayiasKRDO16-Ouroboros}\footnote{A
version of Ouroboros is presented in~\cite{ouroboros-praos} which applies the
``semi-synchronous'' network model.} applies a lottery-like method for
membership selection. In particular, the set of current members for the
consensus run a multi-party coin-flip to generate a uniformly random string.
This string is then used to randomly derive the leader for the current epoch as
well as the set of members for the next epoch.

\subsubsection{Delegated Proof-of-Stake}

Delegated Proof-of-Stake~\cite{bitshares, DPoS-Bitshares-overview, DPoS-Bitshares-docs,
lisk-DPoS} (DPoS) provides a democratic, vote-based membership selection. The
\textit{Stakeholders}, nodes that hold coins, vote on nodes to become
\textit{Delegates} for specified intervals. In some implementations, such as
Lisk~\cite{lisk-DPoS}, the nodes are assigned a voting weight proportional to
the amount of coins held. The Delegates are responsible for block production in
the entire system, forming the consensus committee and signing blocks to be
appended to the chain. To incentivize participation, Delegates are rewarded
with block rewards or transaction fees.

BitShares~\cite{bitshares} was the original Delegated Proof-of-Stake system to
be implemented. The concept of ``voting through transactions'' was first
discussed in the forums~\cite{DPoS-bitsharestalk} and was later developed. Since
BitShares, a number of DPoS systems have been implemented; Lisk~\cite{liskio},
Steem~\cite{steemio}, and EOS~\cite{eos} are major examples.

\subsubsection{Proof-of-Burn (PoB)}

Proof-of-Burn~\cite{stewart2012proof} acts as an alternative to Proof-of-Work
and is built upon the primitives of Proof-of-Stake. Initially discussed in the
BitcoinTalk forum~\cite{proofofburn-bitcointalk}, the concept was proposed to
rival Proof-of-Stake to provide something ``difficult'' for all to do. At the
time it was proposed, ``staking'' in Proof-of-Stake only locked currency, in
which a node could withdraw and their currency would still circulate. To
prevent ``burning coins'' being effected by potential chain reorganization, the
Proof-of-Burn specification outlines that the burning of coins should happen a
number of blocks prior. The initial concept of burning coins was proposed as a
way to transition between
cryptocurrencies~\cite{bitcointalk-btcsecondwhitepaper,secondwhitepaperlink}.

Slimcoin~\cite{slimcoin}, influenced by Bitcoin~\cite{Nakamoto08-Bitcoin} and
PPCoin~\cite{king2012ppcoin}, implemented Proof-of-Burn alongside Proof-of-Work
as a mechanism to mitigate some of the requirements of powerful hardware.
Proof-of-Work blocks are produced as in Bitcoin, whereas the Proof-of-Burn
blocks are created by burning a threshold of coins, where the threshold is a
function of the difficulty of the Proof-of-Work block as well as the number of
coins burnt by the network. Proof-of-Burn blocks are only able to appear after
a Proof-of-Work block has been mined. This makes Proof-of-Burn applicable to
permissionless systems, as it inherits the same guarantees and properties as
Proof-of-Work with the added benefits of combined Proof-of-Stake and
Proof-of-Burn.

\subsection{Proof-of-Authority (PoA)}

\begin{figure}[t]
    \begin{center}
        \includegraphics[width=\columnwidth]{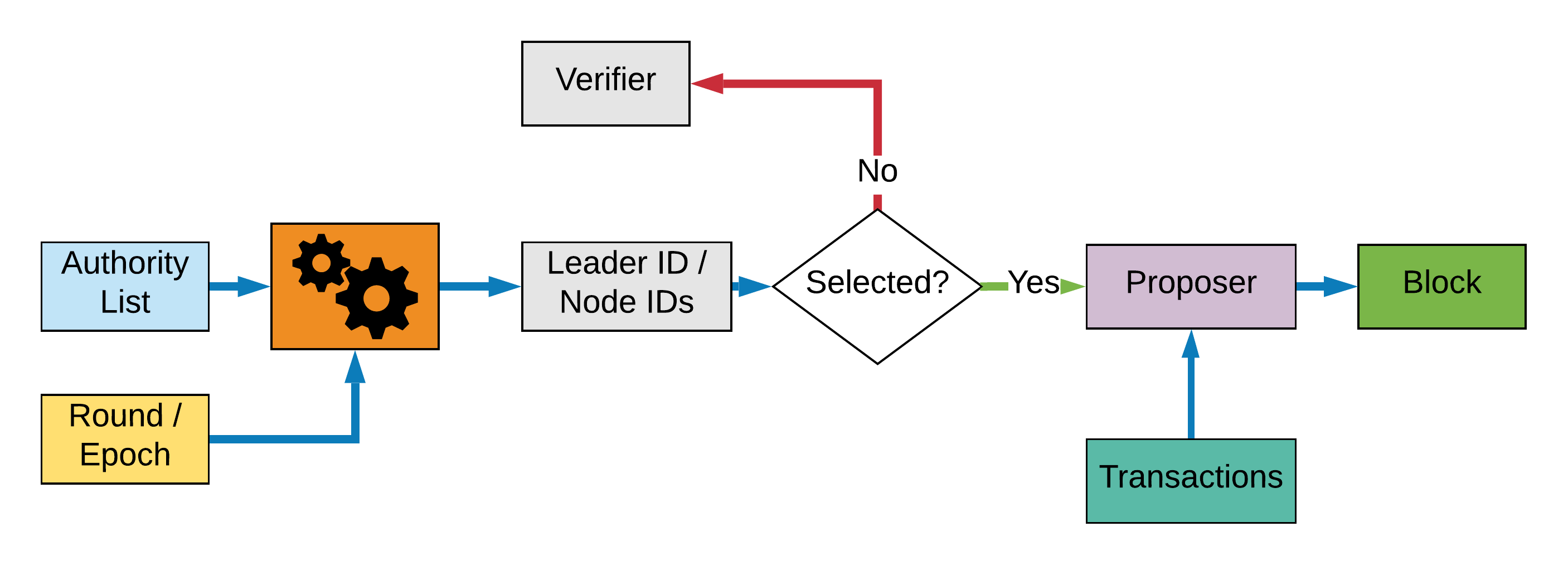}
        \caption{Proof-of-Authority flow.\label{fig:poa}}
    \end{center}
\end{figure}

Proof-of-Authority was initially proposed as an addition to the Ethereum
blockchain~\cite{parityPoA,EIP225} for testnet\footnote{A testnet is a
  blockchain network that runs the blockchain but is used primarily for testing
  and has no financial value. It reflects the same characteristics as the main
blockchain network, but can often be seen with changed consensus, smaller
difficulty or quicker block times.} usage, which was a response to the risks with
Proof-of-Work in this context. The core of Proof-of-Authority is that only
chosen, trusted nodes are able to mint new blocks~\cite{de2018pbft}. Although
this centralizes the overall membership of the blockchain, the usage is
suitable for small networks and testnets.

\paragraph{Key Concept}

Proof-of-Authority was proposed to maintain a blockchain with minimal waste of
energy and primarily focused for the testing network context. The core require
Authorities that are chosen to verify and produce blocks for the blockchain.
For each round, or epoch, a leader or subset of the authorities are chosen to
produce the blocks that are proposed to extend the chain. The authority nodes
are agreed upon in the configuration of the blockchain. A depiction of this
can be found in Figure~\ref{fig:poa}.

\paragraph{Weaknesses}

Although Proof-of-Authority requires Authorities to be chosen as part of the
blockchain configuration, there is a selected leader for each round. If the
leader exhibits malicious behavior, there are cases where they are able to
censor and produce invalid blocks. However, in some implementations the leader
can be effectively removed as an authority if this behavior is observed and
voted upon by the other authorities.

\paragraph{Goals and Assumptions}

To provide membership selection in this environment, Proof-of-Authority
variants require pre-selected nodes to form the authority committee, and thus,
the consensus committee. From this committee, a leader is selected based on
calculations of time~\cite{parityPoA} or blocks~\cite{EIP225}. The network is
therefore placing trust in the authorities to create valid blocks and progress
the blockchain. Although this membership selection has a committee of
``chosen'' members, there is no explicit incentive reward or punishment, rather
an implied reputation for the nodes running as the authorities to maintain
their authority position as well as their status in the community.

\subsubsection{Aura}

Aura~\cite{parityPoA} is the Proof-of-Authority implemented by the Parity
Ethereum client~\cite{parity-ethereum} and progresses in rounds. Out of the
trusted entities, one is the chosen leader of the round who proposes, or
``seals'' by signing, blocks to the other members. Each round a leader is
expected to propose a block, if there is no transaction available then an empty
block is proposed. In the case that the block proposed is invalid, the trusted
entities form a vote to kick the malicious leader.

\subsubsection{Clique}

Alternatively, Clique~\cite{EIP225} is the Proof-of-Authority implemented by
the Go Ethereum client~\cite{geth-client} (Geth). Similar to Aura, Clique
progresses in epochs and has a leader for an epoch. However, to minimize the
threat of a malicious leader, Clique PoA allows more than one Authority to
propose a block, specifically $A - (A/2 + 1)$ where $A$ is the number of
Authorities. Each epoch elects a new leader, and each Authority is limited to
proposing every $A/2 + 1$ blocks.

\begin{figure}[h]
    \begin{center}
      \includegraphics[width=0.8\columnwidth]{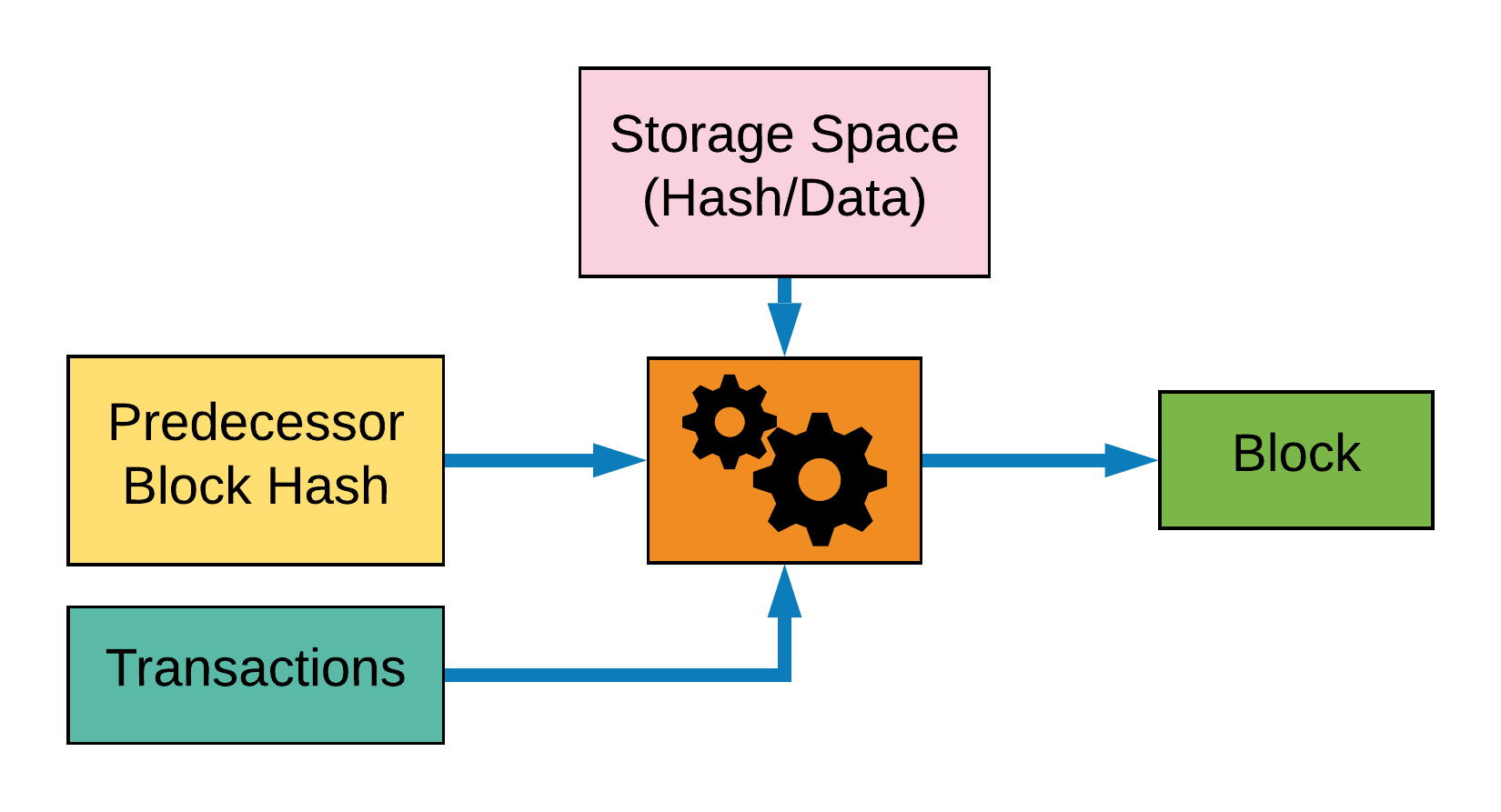}
        \caption{Proof-of-Capacity flow.\label{fig:poc}}
    \end{center}
\end{figure}

\subsection{Proof-of-Capacity (PoC)}

Similar to the motivation behind Proof-of-Stake, Proof-of-Capacity aims to
reduce the amount of wasted resources on the blockchain by utilizing the
resources that would be used for block production.

As mentioned previously, Proof-of-Work has resulted in an observable increase
in the investment of computing hardware, through the renting of cloud
computing, or, the purchasing of specialized hardware for mining. As a
consequence, a drastic change to the mining process may cause the hardware to
rapidly depreciate in value. The repercussions of such depreciation could see
the loss of adoption in the new membership selection.

Proof-of-Capacity, otherwise known as Proof-of-Retrievability~\cite{juels2007pors, MillerJSPK14-Permacoin} or
Proof-of-Space~\cite{dziembowski2015proofs, ParkPAFG15-Spacemint}, provides a use for the hardware running the blockchain so that
the block production process is not wasted. The process for block production
utilizes hardware storage which is then used to maintain the blockchain, rather
than wasting resources on useless computing work. The proposed membership selection
requires a node to prove validity through storing and retrieving shards of a
given file. Once the node has successfully accessed the storage and delivered
the file, the node has not only proven validity, but also aided in the
operation of the blockchain as the stored files are part of the blockchain
itself. This process allows for the blockchain to be stored and sharded amongst
the entire network, reducing the overall replication required, but maintaining
enough redundancy to be fully decentralized. By doing so, information can be
accessed at any given time while minimizing the storage costs for the
blockchain. The implementation has been adapted to a number of blockchains,
such as SpaceMint~\cite{ParkPAFG15-Spacemint},
Permacoin~\cite{MillerJSPK14-Permacoin} and
BurstCoin~\cite{burstcoinwhitepaper}.

However, a major issue with proving validity through capacity is the impact of
cloud infrastructure and the ease of providing largely scalable storage. To
oppose this, implementations were designed to require fast access to storage,
where the cost of outsourcing impacts the node's ability to participate. This
mitigates the risk of an adversary purchasing large amounts of cloud storage to
overcome the network for a short period of time.

\paragraph{Key Concept}

The core focus of Proof-of-Capacity was to re-purpose hardware that was used
for Proof-of-Work, avoiding the waste of computational and electrical resources.
The main concept for this membership selection is that a node proves validity
by storing important information and showing they can retrieve it at a later
point in time. This provides motivation for a number of implementations that
use the block creation process to aid the operation of the blockchain. By
sharding information amongst nodes successfully, the blockchain storage can be
decentralized whilst minimizing the number of full replications. A figure
presentation of Proof-of-Capability is shown in Figure~\ref{fig:poc}. Similar
to the computation of Proof-of-Work, storage space is used to construct a proof
of validity allowing a node to propose a block to the consensus.

\paragraph{Weaknesses}

The variants of Proof-of-Capacity suffer from the advances of technology. The
foundation is that space is utilized as an aspect of stake, a node provides
storage space for the power to create blocks. However, the advances in cloud
providers and large companies reveal an increasing possibility of
centralization and monopolization in the mining process.

\paragraph{Goals and Assumptions}

To provide membership selection in a permissionless environment, the
variants of Proof-of-Capacity utilize storage space which also acts as
a proof that the node is valid. For a node to be selected, it must perform a
function on it's storage space, committing space or retrieving
information, which generates the proof similar to that of
Proof-of-Work. To increase the probability of selection, a node can
increase its storage space which has an assumed cost. However, to act
as a disincentive for cloud-based solutions, the proofs employ
mechanisms to ensure nodes can only have local storage. This is
enforced by taking data access time into consideration, and nodes
access to remote cloud storage will lose its advantage in the mining
process. Nodes are rewarded for successful selection and
proposal with freshly minted coins, which also acts as an incentive to continue
to commit storage space. In SpaceMint~\cite{ParkPAFG15-Spacemint}, ``punishment
transactions'' allow misbehaving nodes to be punished and their rewarded coins
revoked.

\subsubsection{Proof-of-Retrievability}

Proof-of-Retrievability was first proposed for storing large files in a
distributed archive in 2007~\cite{juels2007pors}, which was later adapted into
the blockchain context through Permacoin~\cite{MillerJSPK14-Permacoin} in 2014.
Generating a proof of validity requires a node to prove it has access to
selected files. A valid proof provided by a node shows that the node has a
copy of the associated file. To achieve a successful distributed archive, the
system needs to incentivize storing files locally, away from cloud-based
solutions.  Proof-of-Retrievability employs this by requiring the consensus
member to create a block through local random access to segments of files. In
PermaCoin, the payment private key is integrated into the puzzle solution, so
the solution must be created with access to this file. In addition to this, and
as a further disincentive for cloud-based storage, the solution requires
frequent random and non-precomputable access to storage. Thus, in comparison to
storing files locally, a node outsourcing storage will have a delay in creating
a valid proof.

\subsubsection{Proof-of-Space}

Similarly, nodes in Proof-of-Space~\cite{dziembowski2015proofs} prove
that they have committed dedicated storage space to the system. This
concept was later adapted into
SpaceMint~\cite{ParkPAFG15-Spacemint}. To create a block and to be
selected in the SpaceMint, each node creates a transaction committing
dedicated storage space, and ties the commitment to its public
key. Once the transaction is formed, the node computes a proof using
the Proof-of-Space algorithm by taking the hash value of the last
block as input and outputting not only the proof of space, but also a
``quality'' of the proof. The quality of the proof determines whether
the node will be selected to propose a new block.

\subsection{TEE-Based}

\begin{figure}[t]
    \begin{center}
        \includegraphics[width=\columnwidth]{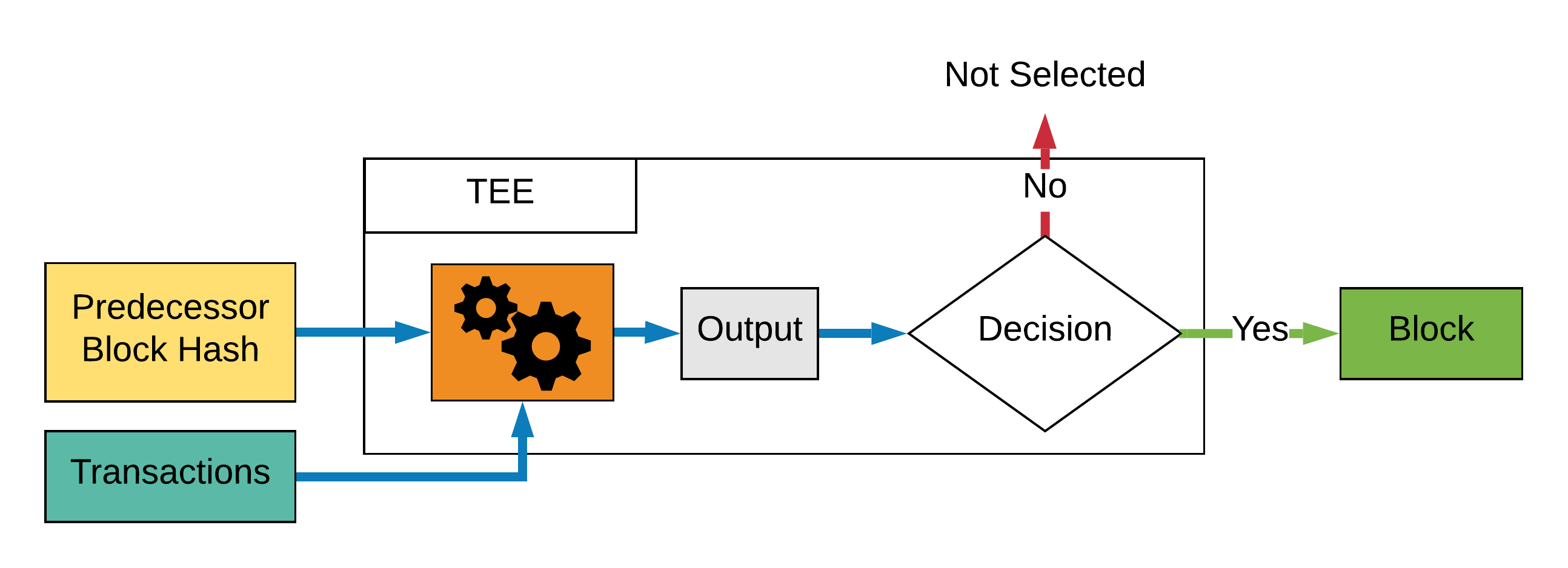}
        \caption{TEE-based flow.\label{fig:TEE}}
    \end{center}
\end{figure}

With technological advancements in hardware, new alternatives were
made possible to be used in blockchain systems. \textit{Trusted
  Execution Environment} (TEE) is a secure space of a trusted
hardware. It guarantees that code running inside will honestly follow
the pre-defined policies and algorithms. The concepts of Trusted
Execution Environments were originally heavily influenced by the ARM
TrustZone~\cite{arm-trustzone}. From this, many major hardware
manufacturers, such as AMD~\cite{amd-psp} and Intel~\cite{intel-sgx}
provide Trusted Execution Environments (TEEs).

TEEs allow for trust to be placed on the hardware running on a node,
allowing algorithms that use randomness and waiting time to be
utilized. This places stronger assumptions on the system, but allows
for a range of different techniques to be explored, resulting in
overall improved performance. However, although the technology
provides great guarantees for running secure code, it is slowly being
integrated with average consumer hardware. In addition, it requires
users to trust the manufacturers of the trusted components.

\paragraph{\textbf{Key Concept}}

Members can be selected by using a random that utilizes the trusted execution
environments on hardware. Rather than wasting work or requiring nodes to
sacrifice funds, TEE-based membership selection selects nodes with a
probability proportional to the number of TEEs that node controls. This means a
node is able to increase the probability of selection by purchasing more
hardware devices. One critical aspect of TEE hardware is that each hardware
device has a unique identification, meaning each node's hardware device can
only be used once. This prevents a node from spoofing the number of hardware
devices they are in control of.

\paragraph{Weaknesses}

The prominent weakness in TEE-based membership selection is the reliance on
specialized hardware. Although Trusted Execution Environments are manufactured
to provide guarantees, other factors may influence the trust placed in the
hardware. Trusting hardware requires placing trust in the vendors that supply
the hardware, which inevitably can lead to centralization or monopolization
through vendors.

However, even if monopolization or centralization provide negligible impact,
the inherent trust in the hardware may prove fatal. TEEs, although manufactured
and developed correctly, may suffer from a number of flaws or
attacks~\cite{moghimi2017cachezoom, wang2017leaky, schwarz2017malware,
jang2017sgx, Gotzfried2017sgx}, where an adversary can bias the operation of
the blockchain.

\paragraph{Goals and Assumptions}

TEE-based membership selection techniques, such as Proof-of-Elapsed-Time and
Proof-of-Luck, place trust into hardware to aid with the selection of members.
This allows for lottery-like algorithms to be used to randomly generate data,
which is then used to determine whether a node is selected. The function to
generate the randomization is run in the trusted environment, so the algorithm
can assume that the output will be random. To increase the selection probability,
the node will need to purchase more specialized hardware with TEEs.
Figure~\ref{fig:TEE} presents the abstract idea of this class of membership
selection.

Proof-of-Elapsed-Time and Proof-of-Luck are often paired with reward-based
incentive mechanisms, where winning proposals are rewarded with freshly
minted coins or transaction fees.

\subsubsection{Proof-of-Elapsed-Time (PoET)}

Intel ltd.\ proposes Proof-of-Elapsed-Time~\cite{Intel-PoET,
  chen2017security}, which relies on secure instruction execution to
achieve the features provided by PoW-based systems, while not
consuming as much energy as PoW. This project was later implemented as
a HyperLedger project, called ``HyperLedger Sawtooth'' \cite{sawtoothpoet}.

Proof-of-Elapsed-Time is performed by each node in the blockchain network
executing a `\textit{sleep}' function for a random amount of time. The first
node to wake from the sleep gets elected as the leader to propose the new
block. The time to sleep is a random number generated using Intel's Software
Guard Extensions (SGX)~\cite{intel-sgx}, which allows applications to run
trusted code in a protected environment, this in turn ensures that the
randomness of the sleep time generated for each machine is trusted randomness.

% The sleep time is generated using Intel's Software Guard Extensions
% (SGX)~\cite{intel-sgx}, which allows applications to run trusted code in a
% protected environment, such as ensuring the randomness of the sleeping time for
% each machine.

\subsubsection{Proof-of-Luck}

As an alternative method, Proof-of-Luck~\cite{MilutinovicHWK16-PoLuck}
proposes a similar concept in which nodes perform computation inside a Trusted
Execution Environment (TEE), such as Intel's SGX, to generate a random winner for block
proposal. The generation of the random winner occurs as all nodes
execute the block proposal code inside the TEE producing a number that
is compared against all other proposed numbers. The core element of
this mechanism lies in the trust of the TEE to guarantee that all code
run will produce valid expected output. The Proof-of-Luck also
utilizes sub-proofs to prove that a node has provided the correct
solution.

\subsection{Proof-of-Location}

\begin{figure}[t]
    \begin{center}
      \includegraphics[width=\columnwidth]{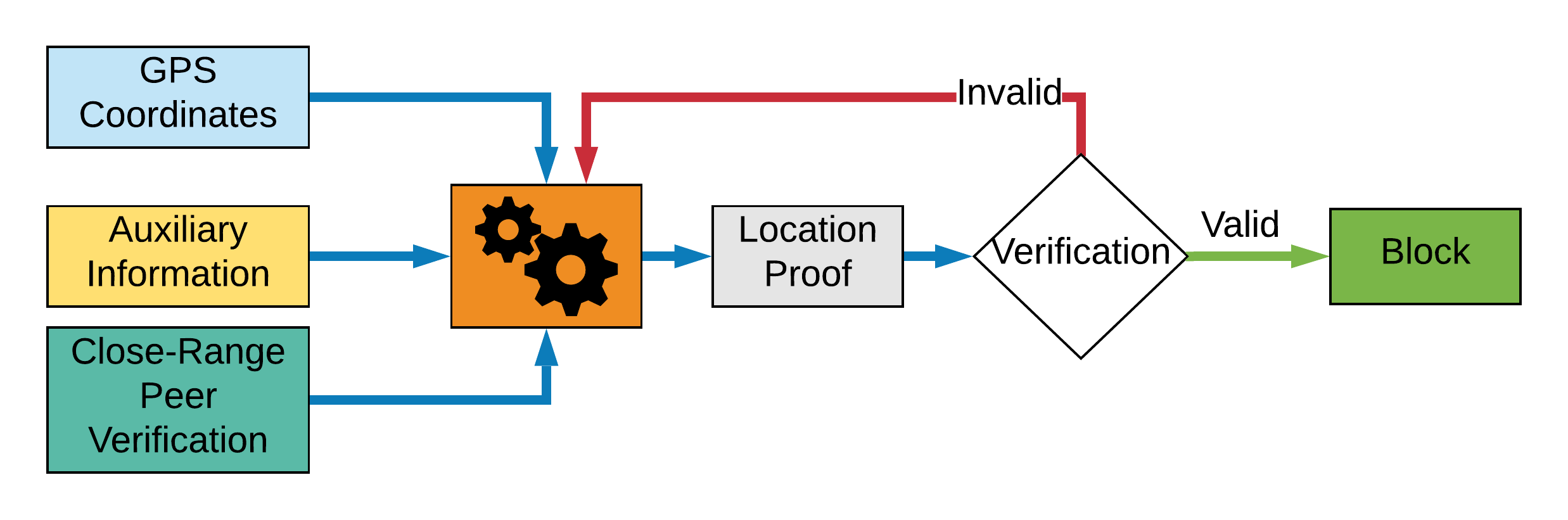}
        \caption{Proof-of-Location flow.\label{fig:polocation}}
    \end{center}
\end{figure}

The growing use of mobile and IoT devices has influenced the introduction of
the blockchain to these platforms~\cite{kshetri2017can, huh2017managing,
dorri2017blockchain, dorri2017towards, reyna2018blockchain}. However, the major
blockchains utilizing Proof-of-Work based membership requires computationally
intensive tasks, impractical for the low-powered devices. The
Proof-of-Location~\cite{polocation-peer2peer, platinio-whitepaper} aims to
utilize the dynamic location of mobile devices to create and verify proofs of a
given device's location. This can be used for signing transactions, interacting
with blockchain contracts or for pure verification of location.

Proof-of-Location~\cite{polocation-peer2peer} heavily relies on close peers to
validate the location of another. Originally proposed for collusion
resistance~\cite{zhu2011toward}, the Proof-of-Location utilizes the geographic
location of the node that is then signed by other nodes to verify that it is
correct, as depicted in Figure~\ref{fig:polocation}. The location verification
is required to stop geographic location spoofing, which is done through
location requests as well as utilizing low-range communication channels.

\paragraph{Key Concept}

The Proof-of-Location's core concept is to utilize the dynamic geographic
location of mobile devices to integrate into the interaction with the
blockchain. To mitigate the risk of spoofed location, some proposals for
Proof-of-Location utilize a range of communication
channels~\cite{polocation-peer2peer} for validation and verification of another
node's location. Dependent upon implementation, the selection will work via
verification of location, either a node is in a certain
location~\cite{platinio-whitepaper}, or, are able to verify their location with
closest peers~\cite{polocation-peer2peer}. To ensure good behavior, the
Proof-of-Location provides incentives to act good, through reward, as well as
punishment if they are found to falsify the location.

\paragraph{Weaknesses}

Although there are countermeasures to location spoofing, collusion amongst
players may lead to spoofed locations being accepted as valid.

\paragraph{Goals and Assumptions}

Proof-of-Location aims to provide a membership selection
mechanism\footnote{Proof-of-Location is used in a variety of ways. Some
  location proofs are utilized for interaction, others are utilized to share
  location in a decentralized context depending on the
usage.~\cite{dorri2017towards, platinio-whitepaper, polocation-peer2peer}}
based of a device's geographical location. To do so, the location must be
verified by other nearby devices. This will enable the membership selection to
be applied to a permissionless environment with no inherent trust required. The
combination of both rewards and punishment act as incentives for the nodes
to behave correctly.

\section{Consensus \label{sec:cons}}

In this section we present an overview of the consensus mechanisms that have
been applied to blockchain systems today. A summary of the consensus covered
in this section can be found in Table~\ref{table:cons}.

The consensus in the blockchain holds the responsibility of deciding the next
chosen block to be appended to the chain. This step is vital to the operation
of the blockchain as it produces the blocks that contain the transactions and
state transfers. Byzantine Fault-Tolerant (BFT) protocols, also known as Byzantine Agreement
(BA) protocols, were proposed to achieve consensus in the presence of
distributed participants where some may be potentially malicious. This problem
was originally detailed as the \emph{Byzantine Generals
Problem}~\cite{LamportSP82} in distributed computing systems and has been
thoroughly studied for over 40 years.

\paragraph{The Byzantine Generals Problem}

The Byzantine Generals Problem depicts a situation where the Byzantine army is
separated into different units, each commanded by a general. The army wishes to
make an attack on the city, however, the attack will only succeed if the units
are coordinated to attack at the same time. To prepare a coordinated attack,
the generals must communicate through messengers and decide whether to attack
or retreat. However, one or more generals may be traitors who will try to
intentionally subvert the process so the loyal generals cannot arrive at a
unified plan. The goal is to have every honest general agreeing on the same decision
even in the presence of traitors.

To illustrate the problem, let the army have $2N+1$ generals in total, and one
of them is a traitor. If half of the loyal generals want to attack, and half
want to retreat, the traitor has the ability to misguide the set of generals to
attack or retreat. This can lead to the situation where $N$ generals attack and
$N$ retreat, fatal for the Byzantine army. This problem becomes worse as
messengers are used for communication and the traitor may prevent messages from
being delivered to a loyal general, or, by forging a falsified message.

% \paragraph{The Byzantine Generals Problem}
% Consensus has been a thoroughly studied problem in distributed systems for over
% 40 years. The concept of reaching agreement in the presence of faults was
% formally introduced with the Byzantine Generals Problem~\cite{LamportSP82}.
% The Byzantine Generals Problem depicts a situation where the Byzantine army is
% separated into different units, each commanded by a general. The army wishes to
% make an attack on the city, however, the attack will only succeed if the units
% are coordinated to attack at the same time. To prepare a coordinated attack,
% the generals must communicate through messengers and decide whether to attack
% or retreat. However, one or more generals may be traitors who will try to
% intentionally subvert the process so the loyal generals cannot arrive at a
% unified plan. The goal is to have every honest general agreeing on the same decision
% even in the presence of traitors.

% Even though how to reach consensus in the presence of
% malicious has been a research focus for the last 40
% years, how to design a secure and scalable consensus for
% permissionless blockchains are still an open challenge.

%%%%%%%%%%%%%%%%%%%%%%%%%%%%%%%%%%%%%%%%%%%%%%%%%%%%%%%%%%%%%%%%%%%%%%%%%%%%%%%%
%% Nakamoto's Consensus
%%%%%%%%%%%%%%%%%%%%%%%%%%%%%%%%%%%%%%%%%%%%%%%%%%%%%%%%%%%%%%%%%%%%%%%%%%%%%%%%
\subsection{Nakamoto's Consensus\label{subsec:nakamotocons}}

Nakamoto's Consensus, introduced with Bitcoin, was the first consensus to be
applied in the permissionless blockchain context, and has been adopted by many
other blockchains~\cite{litecoin,wood2014ethereum,sasson2014zerocash}. The core
aspect of Nakamoto's consensus is the \textit{Longest Chain} rule, which is
used in the event that two, or more, blocks had been proposed at the same
height and the chains have progressed in a forked state. The longest chain
states that nodes must select the chain with the highest block number as the
single, canonical chain. The longest chain is believed to have the most work
performed, adhering to the principle that the strict majority of the network,
and therefore computational power, is owned by honest participants. In the case
two chains have the same length, the nodes perform work on the first valid
block they receive, while rejecting other conflicting blocks until one branch
becomes the longest branch.

In the context of a permissionless blockchain, Nakamoto's consensus only
guarantees eventual consistency~\cite{vogels2009eventually}, that is, if there
are no new blocks proposed, all nodes will eventually see the latest block at
the height of the chain.  During the time the blockchain experiences a
transient fork and consistency is not met, Nakamoto's consensus will provide
the fork resolution eventually reaching a consistent state on all nodes.

Nakamoto's consensus provides security under the assumption that the strict
majority of the network is honest, and when blocks are propagated through the
network prior to any new proposals. With high block creation speed, more nodes
will propose blocks based on stale information, as the information has not
fully propagated before their proposal is created. This results in an
increasing number of blocks not extending the ``longest chain'', which may have
a number of implications for performance or potential attacks.

The rule of selecting the longest chain as a fork resolution protocol causes a
number of proposals to be wasted. If paired with Proof-of-Work-based membership
selection, then this can lead to a waste of mining power and, at large, energy,
as all blocks that are not selected on the main branch are considered ``not
accepted''~\cite{decker2013information, PSS16, GKK16,
NayakKMS16stubbornmining}.

One hindrance to Nakamoto's Consensus is the reduced security with frequent
block proposals. As the voting power in the network increases, through node
number or otherwise, the number of possible proposals increases. This resulted
in a requirement in Nakamoto's consensus to have a limited number of proposals
with satisfactory delay between proposals\footnote{Typically, Nakamoto's
consensus is coupled with Proof-of-Work, as in Bitcoin, and the delay of
proposals are adjusted through a dynamically changed difficulty parameter.}.

The increased frequency of block proposals presents a problem for Nakamoto's
consensus, as it is heavily reliant on the dissemination of blocks prior to any
mining. Once the block creation speed exceeds the propagation time, the
probability of concurrent proposals increases and the honest nodes will have
more conflicting blocks, wasting overall mining power. At this point,
Nakamoto's consensus will select the longest chain and as a consequence a
number of attacks, such as double spending, may have a higher success rate
without the required majority voting power.

% 
% The increased frequency of block proposals presents a problem for Nakamoto's
% consensus, as it is heavily reliant upon the block proposals disseminating
% prior to any new proposals being created. Once the block creation speed becomes
% faster than the propagation time, the probability of adversarial chains being
% the longest chain grows due to the number of proposals using stale blocks. In
% this case, members with the majority of the voting power will see many blocks
% created at the same depth and the adversarial chain having the highest number
% of blocks. At this point, Nakamoto's consensus will select the longest chain and
% as a consequence a number of attacks, such as double spending, may have a higher
% success rate without the required majority voting power.

\paragraph{Goals and Assumptions}

Nakamoto's consensus will operate correctly assuming a synchronous network,
as it requires all nodes to have knowledge of the latest blocks in order to
correctly extend the main chain. Due to this, the consensus will only
terminate given that all blocks for a given height are seen by all the nodes.
In practice, it provides probabilistic consensus when its time bounds are exceeded
as
% periods of asynchrony are 
experienced by the Internet. The probabilistic consensus can be due to either
termination or agreement not being reached as there is a possibility of unseen
chains after sufficient time has elapsed. The Nakamoto Consensus model can be
seen to provide partial order during time before agreement is met or before
termination is achieved. Once the fork has been resolved, the common prefix
increases. Nakamoto's consensus fails to ensure validity, where all correct
nodes proposing a block with a transaction may still decide on an empty block
upon termination of the consensus instance for the given height.

%The validation of the proposals is
%linked to the chain structure, as long as the proposing block, or
%chain, leads to the correct genesis and has valid paths of blocks,
%it is seen as valid.
% \JS{Why the validations is linked to the chain
%   structure? I thought that miners only verify whether this extends
%   the longest chain or not. Of course that they also need to verify
%   the validaty of the blocks including its contents.
%   \CN{Rewrote, how's this?}
% }

The adversary model employed by Nakamoto's Consensus places heavy
assumptions on the block proposals, where the honest nodes produce new
proposals quicker than the adversarial nodes, such that the honest nodes
will always be proposing a longer, valid chain. This can be achieved
through the majority of the voting power resource being owned by the honest
nodes.

%%%%%%%%%%%%%%%%%%%%%%%%%%%%%%%%%%%%%%%%%%%%%%%%%%%%%%%%%%%%%%%%%%%%%%%%%%%%%%%%
%% Ghost
%%%%%%%%%%%%%%%%%%%%%%%%%%%%%%%%%%%%%%%%%%%%%%%%%%%%%%%%%%%%%%%%%%%%%%%%%%%%%%%%

\subsection{Ghost \label{subsec:ghost}}

Inherent impediments caused by the longest chain consensus prompted a
proposal for a new consensus mechanism. The \textit{Greedy Heaviest Observed
Sub-Tree}, or \ghost, protocol~\cite{sompolinsky2015secure} was
proposed as a new consensus mechanism for fork resolution in the blockchain,
attempting to alleviate the problems discovered in the longest chain approach.
The conflict-resolution mechanism employed by \ghost\ utilizes blocks
that are not included on the canonical chain, providing a mechanism to consider all
efforts on the blocks. The chain selection algorithm iteratively selects the
heaviest sub-tree, rooted at the genesis block, block 0, and building up the
correct chain to the current block.

By including the number of proposed blocks that are not included in the canonical
branch, as a calculation of a sub-tree's weight, \ghost\ provides some
resilience to forks occurring in high throughput environments. The main
advantage of \ghost\ is that it maintains the 50\% threshold for attacks at
higher throughput and larger block sizes. The key concept of \ghost\ is that
the blocks not accepted on the canonical chain still impact the weight and therefore
contribute to the fork-resolution protocol, providing resilience to some
adversary activity in the case of high throughput.

\paragraph{Goals and Assumptions}

\ghost, similar to Nakamoto's consensus, requires synchrony to achieve a
consensus agreement. Under the conditions where synchrony is not achieved, it
will fail to terminate or reach agreement, as nodes will be observing stale
branches of the chain. During the point at which no branch is heaviest, \ghost
provides a partial order, as the prefix of blocks is common to all forked
branches until the fork has been resolved. However, \ghost does not achieve
validity in proposals, as empty blocks can still be considered even though all
valid nodes propose blocks with valid transactions being contained.

The adversary model follows Nakamoto's consensus, where the strict majority of
the voting power belongs to honest nodes. This ensures that the heaviest
sub-tree has the highest probability of being created by honest nodes.

\paragraph{\bf Ethereum and \ghost}

Computing the entire sub-tree requires a substantial amount of time and
resources, impeding the progress of the blockchain. Ethereum utilizes a variant
of \ghost, a simplified \ghost\ protocol for reward\footnote{Details can be
  found at
\url{https://github.com/ethereum/wiki/wiki/White-Paper\#modified-ghost-implementation}},
where each block specifies an \textit{uncle} block if it has competing
predecessor blocks of the same height. The simplification comes with the depth,
as the ancestors considered are only in the order of $7$ generations, whereas
the full version would go the full depth.

%%%%%%%%%%%%%%%%%%%%%%%%%%%%%%%%%%%%%%%%%%%%%%%%%%%%%%%%%%%%%%%%%%%%%%%%%%%%%%%%
%% BFT
%%%%%%%%%%%%%%%%%%%%%%%%%%%%%%%%%%%%%%%%%%%%%%%%%%%%%%%%%%%%%%%%%%%%%%%%%%%%%%%%

\subsection{BFT Protocols \label{subsec:bft}}

Traditional BFT protocols provide solutions to this problem through techniques
such as the use of signatures, being able to detect the absence of messages, or
otherwise. Given the context of the blockchain, where a number of nodes may be
acting malicious or experiencing unusual network traffic, BFT protocols were a
suitable selection to achieve consensus.

To utilize BFT protocols in a Bitcoin-like context, three main problems must be
addressed.  First, the original membership selection has been designed for
Nakamoto/Ghost-like consensus, and it cannot be directly applied to the
traditional BFT protocols.  If not paired with a cost, this will allow an
attacker to launch Sybil attacks~\cite{douceur-sybil}
and disrupt the BFT consensus from reaching agreement or terminating. Secondly,
the set of participants in the Blockchain eligible to participate in the
consensus is generally not fixed, nor predefined. Thirdly, the participants may
join or leave the system at arbitrary times, therefore the quorum
size~\cite{MalkhiR97} required for agreement is not constant. These questions
are generally addressed by applying a membership selection technique.

\afterpage{
\onecolumn
  \begin{landscape}
    \vspace*{0.5cm}
  \begin{center}
  \begin{table*}[h]
        \caption{Consensus Mechanisms \label{table:cons}}
        \centering
        \fontsize{6}{8}\selectfont
        \begin{threeparttable}
          \hspace{-7cm}
          \resizebox{0.86\textwidth}{!}{
            %\hspace{-3.8cm}
            \begin{tabularx}{\textheight}{
            p{0.048\textwidth}| % Subheading
            p{0.052\textwidth} | % Goal/Assumption
            x{0.078\textwidth} | % Nakamoto
            x{0.078\textwidth} | % Ghost
            x{0.07\textwidth} | % AlgoRand
            x{0.07\textwidth} | % HoneyBadger
            x{0.075\textwidth} | % RepuCoin
            x{0.07\textwidth} | % DBFT
            x{0.07\textwidth} | % Tendermint
            x{0.07\textwidth} | % PeerCensus
            x{0.07\textwidth} | % Solida
            x{0.07\textwidth} | % ByzCoin
            x{0.074\textwidth}| % Thunderella
            x{0.07\textwidth} | % Avalanche
            x{0.07\textwidth} | % HotStuff
            x{0.07\textwidth} % Libra
        }
        \cmidrule[.2em]{1-16}
            % Table Header
            \multicolumn{2}{c|}{} &
            %& % Subheading
            %& % Goal/Assumption
            \textbf{Nakamoto} & % Nakamoto
            \textbf{Ghost} & % Ghost
            \textbf{\BAstar} & % AlgoRand
            \textbf{HoneyBadger} & % HoneyBadger
            \textbf{RepuCoin} & % RepuCoin
            \textbf{DBFT} & % DBFT
            \textbf{Tendermint} & % Tendermint
            \textbf{PeerCensus} & % PeerCensus
            \textbf{Solida} & % Solida
            \textbf{ByzCoin} & % ByzCoin
            \textbf{Thunderella} & % Thunderella
            \textbf{Avalanche} & % Avalanche
            \textbf{HotStuff} & % HotStuff
            \textbf{LibraBFT} % Libra

            \\\cmidrule{1-16}

            %%% Assumptions
            \textbf{Chain} &
              % Subheading
            Seminal Chain & % Goal/Assumption
            Bitcoin \vfill \cite{Nakamoto08-Bitcoin} & % Nakamoto
            - \vfill \cite{sompolinsky2015secure} & % Ghost
            AlgoRand \vfill \cite{Micali16-algorand, GHMVZ-algorand} & % AlgoRand
            - \vfill \cite{MillerXCSS16-Honey-Badger-BFT} & % HoneyBadger
            RepuCoin \vfill \cite{Yu2018Repucoin} & % RepuCoin
            RedBelly \vfill \cite{CGLRCons,crain2018evaluating} & % DBFT
            Tendermint \vfill \cite{kwon2014tendermint, buchman2016tendermint} & % Tendermint
            PeerCensus \vfill \cite{decker2016bitcoin} & % PeerCensus
            Solida \vfill \cite{AbrahamMNRS16} & % Solida
            ByzCoin \vfill \cite{Kokoris-KogiasJ16-ByzCoin} & % ByzCoin
            Thunderella \vfill \cite{pass2017thunderella} & % Thunderella
            AVA \vfill \cite{avalanchewhitepaper} & % Avalanche
            - \vfill \cite{yin2018hotstuff} & % HotStuff
            Libra \vfill \cite{librabft, librawhitepaper} % Libra

         \\\cmidrule{1-16}

            %%% Assumptions
        \multirow{3}{*}{\raisebox{-\heavyrulewidth}{\hspace{-0.2cm}\bf Assumptions}} &
              % Subheading
            Network & % Goal/Assumption
            \ycell Sync & % Nakamoto
            \ycell Sync & % Ghost
            \ycell Sync & % AlgoRand
            \gcell Async & % HoneyBadger
            \gcell Partial-Sync & % RepuCoin
            \gcell Partial-Sync & % DBFT
            \gcell Partial-Sync & % Tendermint
            \gcell Partial-Sync & % PeerCensus
            \ycell Sync & % Solida
            \gcell Partial-Sync & % ByzCoin
            \gcell Partial-Sync & % Thunderella
            \ycell Sync\tnote{\S} & % Avalanche
            \gcell Partial-Sync   & % HotStuff
            \gcell Partial-Sync % Libra

            % \\\cmidrule{2-16}

            % & % Subheading
            % Trust & % Goal/Assumption
            % \gxmark & % Nakamoto
            % \gxmark & % Ghost
            % \gxmark & % AlgoRand
            % \gxmark & % HoneyBadger
            % \gxmark & % RepuCoin
            % \gxmark & % DBFT
            % \gxmark & % Tendermint
            % \gxmark & % PeerCensus
            % \gxmark & % Solida
            % \gxmark & % ByzCoin
            % \gxmark   % Thunderella

            \\\cmidrule{2-16}

            & % Subheading
            Online Presence & % Goal/Assumption
            Online & % Nakamoto
            Online & % Ghost
            Online & % AlgoRand
            Online & % HoneyBadger
            Online & % RepuCoin
            Online & % DBFT
            Online & % Tendermint
            Online & % PeerCensus
            Online & % Solida
            Online & % ByzCoin
            Sleepy & % Thunderella
            Online & % Avalanche
            Online & % HotStuff
            Online % Libra

            \\\cmidrule{2-16}

            & % Subheading
            Adversary & % Goal/Assumption
            %$N > 2f$
            % {\tiny VP(Adv) $< \frac{\textup{VP(Total)}}{2}$}
            $f' < \left \lfloor \frac{N'}{2}\right \rfloor$
            %{\tiny P\textsubscript{hon} $>$ 2P\textsubscript{adv}}
            \tnote{*}& % Nakamoto
            %$N > 2f$
            % {\tiny VP(Adv) $< \frac{\textup{VP(Total)}}{2}$}
            $f' < \left \lfloor \frac{N'}{2}\right \rfloor $
            %{\tiny P\textsubscript{hon} $>$ 2P\textsubscript{adv}}
            \tnote{*}& % Ghost
            $f' < \left \lfloor \frac{N'}{3}\right \rfloor $
            & % AlgoRand
            $f \leq \left \lfloor \frac{N}{3}\right \rfloor $
            & % HoneyBadger
            $f' \leq \left \lfloor \frac{N'}{3}\right \rfloor$,
            $f \leq \left \lfloor \frac{N}{3}\right \rfloor $
            \tnote{$\star$} & % RepuCoin
            $f \leq \left \lfloor \frac{N}{3}\right \rfloor $ & % DBFT
            $f \leq \left \lfloor \frac{N}{3}\right \rfloor $
            & % Tendermint
            $f \leq \left \lfloor \frac{N}{3}\right \rfloor $
            & % PeerCensus
            $f \leq \left \lfloor \frac{N}{3}\right \rfloor $ & % Solida
            $f \leq \left \lfloor \frac{N}{3}\right \rfloor $ & % ByzCoin
            $f \leq \left \lfloor \frac{N}{3}\right \rfloor$ & %\tnote{$\dagger$}% Thunderella
            $f < \frac{N}{5}$\tnote{\P} & % Avalanche
            $f \leq \left \lfloor \frac{N}{3}\right \rfloor$  & % HotStuff
            $f \leq \left \lfloor \frac{N}{3}\right \rfloor$ % Libra

            \\\cmidrule{1-16}

            %%% GOALS
            \multirow{4}{*}{\raisebox{-\heavyrulewidth}{\bf Goals}} &
            Agreement & % Goal/Assumption
            \ycell Probabilistic \tnote{\#} & % Nakamoto
            \ycell Probabilistic \tnote{\#} & % Ghost
            Common Coin & % AlgoRand
            Common Coin & % HoneyBadger
            \gcell Deterministic & % RepuCoin
            \gcell Deterministic & % DBFT
            \gcell Deterministic& % Tendermint
            \gcell Deterministic & % PeerCensus
            \gcell Deterministic & % Solida
            \gcell Deterministic & % ByzCoin
            \gcell Deterministic & % Thunderella
            \ycell Probabilistic \tnote{\#} & % Avalanche
            \gcell Deterministic & % HotStuff
            \gcell Deterministic % Libra

            \\\cmidrule{2-16}

            & % Subheading
            Termination & % Goal/Assumption
            \ycell Probabilistic \tnote{\#} & % Nakamoto
            \ycell Probabilistic \tnote{\#} & % Ghost
            \gcell Deterministic & % AlgoRand
            \ycell Probabilistic & % HoneyBadger
            \gcell Deterministic & % RepuCoin
            \gcell Deterministic & % DBFT
            \gcell Deterministic & % Tendermint
            \gcell Deterministic & % PeerCensus
            \gcell Deterministic & % Solida
            \gcell Deterministic & % ByzCoin
            \gcell Deterministic \tnote{$\dagger$} & % Thunderella
            \ycell Probabilistic \tnote{\#} & % Avalanche
            \gcell Deterministic & % HotStuff
            \gcell Deterministic % Libra

            \\\cmidrule{2-16}

            & % Subheading
            Validity & % Goal/Assumption
            \xmark & % Nakamoto
            \xmark & % Ghost
            \xmark & % AlgoRand
            \xmark & % HoneyBadger
            \xmark & % RepuCoin
            \cmark & % DBFT
            \xmark & % Tendermint
            \xmark & % PeerCensus
            \xmark & % Solida
            \xmark & % ByzCoin
            \xmark & % Thunderella
            \xmark & % Avalanche
            \xmark & % HotStuff
            \xmark % Libra

            \\\cmidrule{2-16}

            % & % Subheading
            % Validity & % Goal/Assumption
            % \gcell Hash, Predecessor & % Nakamoto
            % \gcell Hash, Predecessor & % Ghost
            % \gcell Metadata, Predecessor & % AlgoRand
            % \gcell Signatures, Merkle Tree & % HoneyBadger
            % \gcell \todo{}& % RepuCoin
            % \gcell Application Validity Predicate & % DBFT
            % \gcell Hash, Signatures & % Tendermint
            % & % PeerCensus
            % & % Solida
            % & % ByzCoin
            % \gcell Signature % Thunderella

            % \\\cmidrule{2-16}

            & % \textbf{Blockchain State} & % Subheading
            Total \vfill Order & % Goal/Assumption
            \mmark & % Nakamoto
            \mmark & % Ghost
            \cmark & % AlgoRand
            \cmark & % HoneyBadger
            \cmark & % RepuCoin
            \cmark & % DBFT
            \cmark & % Tendermint
            \cmark & % PeerCensus
            \cmark & % Solida
            \cmark & % ByzCoin
            \cmark & % Thunderella
            \mmark & % Avalanche
            \cmark & % HotStuff
            \cmark % Libra

            \\\cmidrule{1-16}

            \multirow{3}{*}{\raisebox{-\heavyrulewidth}{\bf Partition}} &
              % Subheading
            Consistency & % Goal/Assumption
            \xmark & % Nakamoto
            \xmark & % Ghost
            \cmark & % AlgoRand
            \cmark & % HoneyBadger
            \cmark & % RepuCoin
            \cmark & % DBFT
            \cmark & % Tendermint
            \cmark & % PeerCensus
            \cmark & % Solida
            \cmark & % ByzCoin
            \xmark\tnote{$\ddagger$} & % Thunderella
            \xmark & % Avalanche
            \cmark & % HotStuff
            \cmark % Libra

            \\\cmidrule{2-16}

            & % Subheading
            Availability & % Goal/Assumption
            \cmark & % Nakamoto
            \cmark & % Ghost
            \xmark & % AlgoRand
            \xmark & % HoneyBadger
            \xmark & % RepuCoin
            \xmark & % DBFT
            \xmark & % Tendermint
            \xmark & % PeerCensus
            \xmark & % Solida
            \xmark & % ByzCoin
            \cmark & % Thunderella
            \cmark & % Avalanche
            \xmark & % HotStuff
            \xmark % Libra

            \\\cmidrule[.2em]{1-16}

        \end{tabularx}
      }
        \begin{tablenotes}
          \item[$f$] Denotes the number of Byzantine nodes.
          \item[$f'$] Denotes voting power of the adversary
          \item[$N$] Denotes the total number of nodes.
          \item[$N'$] Denotes Total voting power.
          \item[\cmark] The system provides the associated property.
          \item[\xmark] The system does not provide the associated
            property.
          \item[\mmark] The system experiences forks resulting in
            states of total order prefix with tails of unconfirmed blocks.
          %\item[\mmark] The system \textbf{partially provides} the associated property.
         %   With high probability, a prefix of the
         %   blockchain achieves total ordering. \CN{I think we need to explain
         %   this clearly in the section 3. We should define this as ``partial
         %   order'' and say what you said here}
          %\item[$\dagger$] Achieves responsiveness under: $\left \lfloor
          %  \frac{n+f}{2} + 1 \right \rfloor $
          \item[\S] The paper assumes a synchronous network, but conjecture
            that the results hold in a partially synchronous network. However, the proof
            is left to future work in the paper~\cite{avalanchewhitepaper}.
          \item[\P] The authors in \cite{avalanchewhitepaper} recommend that
            system designers select their value of Byzantine threhsold. The
            value here was an example provided in the paper.
          \item[*] The adversary model follows that the majority of the voting
            power is owned by the honest majority. This can translate into ``the
            honest majority produces blocks faster than the adversary''.
          \item[\#] The termination OR agreement can be seen as probabilistic.
            Discussed in Section~\ref{sec:discussion}.
          \item[$\star$] With RepuCoin, only the top reputed miners in
            the committee have voting power, other participants have a
            reputation score, but have zero voting power before they
            become a top reputed miner.
          \item[$\dagger$] Under optimistic conditions with an honest leader.
            Under non optimistic conditions and a faulty leader, falls back to
            Proof-of-Work for the cooldown period.
          \item[$\ddagger$] The Thunderella blockchain utilizes a Hybrid
            mechanism, which in non-optimal cases can fall back to availability
            in the presence of a partition.
          % \item \JS{I tried to use f' and N' to replace VP(adv) and
          %     VP(Total).Is all f are about the number of nodes in this
          %     row, and all f' are the voting power?}
        \end{tablenotes}
        \end{threeparttable}
\end{table*}
\end{center}
\end{landscape}
\twocolumn
}

\subsubsection{PBFT-based solutions}

The Practical Byzantine Fault Tolerance (PBFT)~\cite{CastroL02-PBFT} algorithm
achieves consensus through leader-based communication with three phases. The
first phase, \textit{pre-prepare}, begins with the leader assigning a sequence
number and multicasting to all other nodes. Once a node receives the
pre-prepare, it enters the \textit{prepare} phase and muilticasts to all other
nodes. If a node receives $2f$ \textit{prepare} messages that match the
\textit{pre-prepare} message, where $f$ is the number of potentially faulty
nodes, it enters the commit stage and multicasts a \textit{commit} message. A
message is then \textit{committed} if a node had received enough commit
messages from $2f+1$ replicas that match the pre-prepare for the message. The
\textit{pre-prepare} and \textit{prepare} phase are used to totally order the
messages, whereas the prepare and commit phases are used to ensure requests
that commit are totally ordered across all nodes.

In the case that the primary exhibits faulty behavior, or the replica
``timers'' expire, PBFT offers a \textit{view-change} mechanism that
elects a new primary. To perform the view-change, all replicas will
send the \texttt{VIEW-CHANGE} message to all replicas, which contains
the current state. The new primary then sends the \texttt{NEW-VIEW}
message to all replicas to initiate the new view.

To provide BFT to the blockchain, systems such as
PeerCensus~\cite{decker2016bitcoin},
ByzCoin~\cite{Kokoris-KogiasJ16-ByzCoin}, and
Solida~\cite{AbrahamMNRS16} utilize variants of the PBFT algorithm
tailored to their requirements.

\paragraph{Goals and Assumptions}

The variants of PBFT explored in this survey, PeerCensus~\cite{decker2016bitcoin} and
ByzCoin~\cite{Kokoris-KogiasJ16-ByzCoin}, exhibit similar goals and
assumptions. Consensus can be reached in a partially-synchronous environment,
where there are strictly less than a third of the nodes exhibiting Byzantine
behavior. However, Solida~\cite{AbrahamMNRS16} assumes a synchronous
environment to ensure that a delta message delay is known a-priori.
Agreement is reached, through the use of signed messages in a leader-based
consensus protocol, however, ByzCoin provides a variant that uses a two-phase
signing scheme to reach agreement. The termination, that of PBFT, is achieved
when the nodes vote on the proposed block.

\paragraph{PeerCensus' Consensus}

The consensus mechanism employed by the PeerCensus~\cite{decker2016bitcoin}
blockchain is a variant of the PBFT leader-based consensus mechanism. The
ability for a member to join and leave are managed through a secure group
membership protocol (SGMP)~\cite{Reiter96}, which allows for the members in the
consensus committee to observe any changes in membership. This provides the
knowledge of all members in the committee, allowing communication between known
nodes and thus can perform the required communication.

To commit a transaction, the transaction must be sent to the current leader.
The leader then performs the PBFT-based consensus with the current committee.
Similarly, blocks are committed through the committee performing PBFT with the
leader proposing any conflicting blocks and obtaining a decision from the
members on which block is to be decided. This solves the second and third
problems mentioned above.

\paragraph{ByzCoin's Consensus}

ByzCoin~\cite{Kokoris-KogiasJ16-ByzCoin} improves upon PeerCensus by altering
the consensus and adding scalability through the use of a collective signing
scheme, CoSi~\cite{SytaTVWJGGKF16-CoSi}. In particular, the consensus committee
has a fixed size\footnote{The size of the consensus committee is implemented as
the size of ``window'' in~\cite{Kokoris-KogiasJ16-ByzCoin}}, and any newly
elected members replace the oldest member. For each consensus execution, the
committee executes the PBFT view-change protocol and elects a new leader.

The consensus protocol is a two-phase PBFT protocol which replaces the PBFT MAC
authentication with the collective signing scheme presented in
CoSi~\cite{SytaTVWJGGKF16-CoSi}. The signing scheme allows a large set of
selected members to efficiently and collectively commit and produce a
signature. This allows for larger committee sizes to run the consensus, thus
improving scalability in respect to the committee size.

\paragraph{Solida's Consensus}

In Solida~\cite{AbrahamMNRS16}\footnote{This work was previously known as
``Solidus: An Incentive-compatible Cryptocurrency Based on Permissionless
Byzantine Consensus'', which was first available one arXiv.org at Dec 2016.},
the consensus is a variant of PBFT, which is performed by a dynamically
configured group of members. Similar to ByzCoin, newly selected members replace
the oldest members in the committee, but also become the leader until a new
leader is elected. A leader that is unable to make progress is replaced through
a similar mechanism to the PBFT view-change.

Members are selected by ranking non-member miners based on their solution to
the Proof-of-Work puzzle. If two solutions are found at the same time, the
miners are ranked using the hash value of their solution. However, the
selection progress has preference to select non-member miners.

\subsubsection{RepuCoin}

RepuCoin's consensus, integrated with the Proof-of-Reputation as
discussed in \S~\ref{subsubsec:repu}, provides a weighted vote-based
consensus. In particular, each member in the consensus committee is
assigned a weight associated with that member's reputation. To reach
agreement, RepuCoin requires both sufficient number of votes, and,
majority of collective weight of the votes.

The main novelty of this weighted consensus comes from the decoupling
of mining power and voting power. In particular, the voting power in
RepuCoin is ``integrated power'' (i.e. reputation) which represents
the total contribution of a node in the system, rather than the node's
``instantaneous mining power'' (e.g. hashing power in PoW) that can be
gained quickly. So, a newly joined miner may have a lot of computing
power, but it will not have any voting power before it achieves a
contribution threshold that is relative to the contributions of other
miners. In this way, RepuCoin is secure against an attacker who can
even obtain a majority of mining power in a short time. Also, when
detecting malicious behavior of a node, the reputation of this node
will be set to 0. This not only reduces the incentive of a node
attacking the system, but also prevents a malicious node from
repeating the attacks without any consequence.

RepuCoin did not provide a detailed consensus algorithm, rather it
provides ways to adapt existing secure BFT protocols, making the
design more generic. In particular, one modification to make for
adapting existing BFT protocols is the change of views and update of
consensus committee. In RepuCoin, at the end of each epoch, the system
enforces view-change and membership updates. Also, upon detecting
crashed committee member, the system also updates the membership.

\paragraph{Goals and Assumptions}

Currently, all proof-of-work based systems assume that no single
attacker can control more than 50\% of the mining power at any time, as
otherwise the attacker can launch 51\% attacks to double spend
coins. For example, Ethereum classic has been attacked by 51\% attack,
and lost millions of dollars \cite{ETC-news-1,ETC-news-2}.

The main goal of RepuCoin is to tolerate a malicious miner who may
even control a majority of mining power for a time period. This goal
is achieved by the separation of mining power and voting power, as
previously mentioned. In addition, RepuCoin provides a higher
throughput (about 10K TPS) by using a parallel chain structure, which
we will detail in \S\ref{sec:structure}.

RepuCoin's leader-based consensus operates using a selection of miners
who hold the highest reputation, which is then translated into
weighted votes in the consensus. The consensus operates under the
assumption of a partially-synchronous network and can be achieved given
that (a) no more than $\frac{1}{3}$ of selected nodes for running
consensus are malicious; and (b) the honest nodes collectively hold
more than $\frac{2}{3}$ of the voting power in the consensus
committee.

\subsubsection{Thunderella}

Thunderella~\cite{pass2017thunderella} provides a unique consensus for
blockchains in two conditions, the optimistic conditions and the worst case
conditions. In optimal conditions, the leader is honest and the adversary
controls strictly less than $\frac{1}{4}$ of the online voting power in each
round and can only corrupt a target node after a certain time delay. In
non-optimal conditions, where the adversary controls strictly less than
$\frac{1}{2}$ of the total voting power, Thunderella resorts to the underlying
blockchain to satisfy the liveness and safety guarantees of their consensus
model.

During optimistic conditions, the consensus is to simply collect $\frac{3}{4}$
votes from the consensus committee. This provides a fast, two-round
communication consensus for a transaction to be committed. If the leader is
seen as dishonest, and the optimistic conditions are not met, Thunderella falls
back to Nakamoto's consensus. In particular, miners create a blockchain with
uncommitted transactions. If the leader is honest, they should commit these
transactions within a defined time period through the optimistic consensus. If
this fails to occur, it provides evidence towards an incorrect leader and
Thunderella falls back for a cool-down period. The
Nakamoto's consensus ensures consistency of the miners' views of the
blockchain.

\paragraph{Goals and Assumptions}

Thunderella aims at eliminating the complex process of view-change in
PBFT, by replacing it with the Nakamoto consensus. In this way, when
the attacker is less powerful, Thunderella provides a high
throughput; when the attacker is more powerful, then it falls back to
the guarantees of normal Nakamoto consensus.

It assumes a partially-synchronous environment, and provides
deterministic leader-based consensus in the optimal case. If there is
a faulty leader, Thunderella falls back to Nakamoto's consensus. The
optimistic case of Thunderella requires $\frac{3}{4}$ of the votes to
be collected for a single block, in which only two communication
rounds are needed to commit the block.  Else, similar to the
traditional blockchain consensus, $\frac{1}{2}$ of votes must be from
honest members.

The consensus operates like a classical leader-based BFT consensus. The leader
receives transactions and signs with a sequence number, which is then broadcast
to the committee. Once the committee has received the transaction, they
acknowledge and vote on the transaction, but only one per sequence number. Once
the transaction has gained $\frac{3}{4}$ votes, it is considered committed. In
the case it cannot be committed, the cool-down period will come into effect and
Nakamoto's consensus will be utilized.

\subsubsection{AlgoRand's BA$\star$}

AlgoRand's~\cite{Micali16-algorand, GHMVZ-algorand} BA$\star$ provides a
vote-based Byzantine Agreement aimed at providing efficient probabilistic
consensus. The members of the consensus committee are chosen through
randomization provided by a blockchain-based common coin, which is also used in
the consensus rounds. The selection occurs at each new block height, selecting
both members and a leader for the committee.

The common coin is implemented by using the last block of the blockchain. In
particular, leaders and consensus committees are selected based on the hash
value of their signature on some commonly shared information, including the
last block, the current consensus step number and consensus round number. The
unique element in AlgoRand is that nodes are only able to see if they are
selected and cannot see, or predict, other node's participation.

BA$\star$ is expected to terminate within 6 rounds. Each BA$\star$ execution
consists of two phases. The first reduction phase, where a block decision is
reduced to a binary value, and the second phase where voting on the binary
value is run to reach consensus, or, reaches a decision on an empty block. Each
phase consists of a number of steps to reach a final agreement. At each step
the node performs local computation to produce a vote. The votes are then
counted and steps progress only if the threshold of votes is met.

\paragraph{Goals and Assumptions}

AlgoRand aims to provide a balance between scalability,
decentralization, and security. In particular, it aims to prevent an
attacker from predicting consensus committee members. This eliminates
potential targeted attacks.

AlgoRand provides blockchain consensus in a synchronous network. The
agreement of the \BAstar algorithm utilizes a common coin paired with
cryptographic signatures to reach agreement. The Byzantine Agreement is
reached in 6 rounds, 13 worst case, but utilizes a common coin for
randomization if a decision could not be made. The validity of the proposals
requires the message to have a valid signature from the node in the current
sortition, but also ensure that the block \textit{seed}, a random seed for a
round, is correct for the given round.

Although AlgoRand requires synchrony to reach agreement, it can achieve safety
in periods of weak synchrony where the network is asynchronous for a bound
period of time. The AlgoRand \BAstar can provide consensus given that there are
more than two thirds honest nodes.

\subsubsection{Tendermint}

Tendermint~\cite{kwon2014tendermint, buchman2016tendermint} provides a
leader-based BFT consensus for the blockchain. The selected nodes form a
committee and take turns proposing a new block for the given height in rounds. Each
round consists of a proposal of a valid block, voting to decide upon accepting
the block, and finally, if more than two-thirds of the committee vote to
commit the block, the block gets appended to the blockchain and committed.

The Tendermint consensus requires a locking mechanism to ensure that a
selected node does not double vote for different blocks at the same height. To
ensure this, the selected node is seemingly locked if they had pre-voted and
pre-committed for a specified block, however, they are able to unlock their
vote if they have witnessed more than two-thirds of the votes for a block in a
given round. The switching of a vote is permitted to protect liveness, but must
be done in a way that it does not compromise safety, therefore only after
two-thirds of the committee have voted to commit a block.

\paragraph{Goals and Assumptions}

Tendermint provides consensus under the assumption of partial-synchrony, where
the honest nodes control more than $\frac{2}{3}$ of the voting power. The
consensus is deterministic, utilizing increasing timeouts for the proposal
round, but allowing for asynchrony during vote stages. During the event that
the network is partitioned, or more than a third of the nodes exhibit Byzantine
behavior, the network may halt all-together.

\subsubsection{HoneyBadger BFT}

The HoneyBadger BFT~\cite{MillerXCSS16-Honey-Badger-BFT}, aims at providing
blockchain consensus for permissioned blockchains, where the network is
asynchronous but reliable channels exist between nodes that guarantee message
delivery once a message is placed into the channel. The execution of the
HoneyBadger BFT utilizes a reduction from multi-value consensus to a binary
consensus, utilizing erasure codes to further improve the efficiency. To cope
with the complete asynchrony, the implementation uses a common coin for
randomization in the consensus.

For each consensus execution, each node of a pre-defined consensus committee
(of size up to 104) randomly selects a set of uncommitted valid transactions,
encrypts it through a threshold-based encryption scheme, and disseminates it to
all nodes through a reliable broadcast protocol. Each node produces its part of
the decryption share for each received message, broadcasts it, and perform the
threshold-based decryption upon receiving $f+1$ shares. The successfully
decrypted transactions are considered to be accepted by the consensus
committee, and are sorted in canonical order to place into the block. Following
this, the adversarial model adopted is static faults, where an adversary is
able to control up to $f$ nodes, where $3f+1 \leq N$.

\paragraph{Goals and Assumptions}
HoneyBadger BFT provides blockchain consensus in a completely asynchronous
network through the use of a common coin for randomization. Although the
network is completely asynchronous, with no bound on the message delay, there
is an assumption that there is a reliable channel between nodes to guarantee
the message delivery. The common coin provides a high probability that agreement
will be reached. Due to this, the termination is probabilistic, as the common
coin can lead to disagreement.  If less than a third of the members are
exhibiting malicious behavior, then consensus can be achieved.

\subsubsection{Democratic Byzantine Fault Tolerant (DBFT) \label{subsubsec:cglrcons}}

% The goal of Blockchain consensus is to agree on a set of transactions to commit
% to a ledger. The current Proof-of-Work model allows for all nodes in the
% network to propose a list of transactions, but only after proving validity by
% solving a cryptographic puzzle. Due to this, the blockchain suffers from
% throughput degradation. Other blockchain consensus models require randomization
% to reach consensus with high probability.

The consensus mechanism proposed by Crain et al.\ in~\cite{CGLRCons} provides a
deterministic algorithm for Blockchain consensus. The unique properties of this
consensus is that it requires no leaders, signatures or randomization. However,
by sacrificing liveness, it ensures safety even in the event of arbitrary
delays. Another unique aspect of the Democratic Byzantine Fault Tolerant
(DBFT) consensus is that it performs a union on the proposed values, allowing
for the maximum set of values to be agreed upon and committed by the consensus
committee.

\paragraph{Goals and Assumptions}

The DBFT consensus provides blockchain consensus in a partial synchronous
network. It can provide consensus given the Byzantine nodes are less than a
third of the total consensus nodes, so a decision can be made with adequate
votes. By removing the randomization, the agreement is deterministic based on
the votes received by the nodes, in turn providing deterministic termination.
DBFT provides validity by providing a union of proposals, such that if all
honest nodes propose valid transactions, the block agreed in the consensus
instance will not be empty.

%%%%%%%%%%%%%%%%%%%%%%%%%%%%%%%%%%%%%%%%%%%%%%%%%%%%%%%%%%%%%%%%%%%%%%%%%%%%%%%%
%% Avalanche
%%%%%%%%%%%%%%%%%%%%%%%%%%%%%%%%%%%%%%%%%%%%%%%%%%%%%%%%%%%%%%%%%%%%%%%%%%%%%%%%

\subsubsection{Avalanche \label{subsubsec:avalanche}}

Avalanche~\cite{avalanchewhitepaper} introduces a class of leaderless Byzantine
Fault Tolerant consensus protocols. The protocols utilize gossip communication
and select subsets of nodes in the network to converge to a decision. A node uniformly selects
nodes in the network, interacting with the subset of nodes and gossiping
their decision. Once the node has interacted with the entire network, they can
determine the decision based on the given threshold.  Avalanche is built up
from \textit{Slush}, a non-Byzantine consensus core for metastability,
\textit{Snowflake}, a Byzantine Fault Tolerant adaptation of Slush, and
\textit{Snowball}, an adaptation of Snowflake with confidence for switching
decisions. Inspired by Nakamoto Consensus~\cite{Nakamoto08-Bitcoin}, Avalanche
provides probabilistic guarantees with probability $1 - \epsilon$, where
$\epsilon$ is a security parameter chosen by system. Avalanche utilizes a
\textit{DAG} structure for transactions and confidence.

\paragraph{Goals and Assumptions}

Avalanche is modeled in a synchronous environment, but is conjectured to work
in a partial synchronous environment. %deferred to future work.
The leaderless consensus exhibits probabilistic guarantees, inspired by
Nakamoto's consensus in Bitcoin. With these probabilistic guarantees, the
consensus has a preference for availability over consistency, as it will allow
applications to continue to be used and transactions committed once they reach
a certain confidence threshold.

\subsubsection{HotStuff \label{subsubsec:hotstuff}}

HotStuff~\cite{yin2018hotstuff} is a leader-based Byzantine Fault-Tolerant
consensus mechanism that builds upon concepts from PBFT~\cite{CastroL02-PBFT},
Tendermint~\cite{buchman2016tendermint, kwon2014tendermint} and Ethereum's
Casper~\cite{casperfriendlyghost, cbccasperresources, casperffg-onemessage}.
HotStuff presents a three phase consensus, with \textit{\textsc{prepare}},
\textit{\textsc{pre-commit}} and \textit{\textsc{commit}} phases. The three
phases are required for liveness as it allows nodes to change their initial
decision and not lock~\cite{livelock-tendermint}. In each phase, the leader
collects messages, waiting for $N-f$ votes, where $N$ is the number of nodes
participating in the consensus and $f$ is the number of Byzantine faults,
before moving to the next phase. By moving to the next phase as soon as the
leader collected enough messages, it allows for progress to follow the delay of
the network rather than arbitrary waits. This is done by using \textit{Quorum
Certificates} ($QC$), a concept of signed data containing the $N-f$ votes to
show proof of message reception as well as progression.

HotStuff also presents a linear ($O\left( N \right )$) view-change mechanism,
used to elect new leaders~\cite{yin2018hotstuff}. The reduction in complexity
allows the view-change to be integrated as part of the normal operation with
minimal impact to the overall performance.

\paragraph{Chained HotStuff}

Vitalik Buterin proposed a Casper~\cite{casperffg-onemessage} message reduction
to improve efficiency by changing prepare and commit to simple votes.
Similarly, HotStuff proposes an improvement by allowing a view-change to happen
on \textit{every} \textsc{prepare} phase, reducing the number of messages
required for a single round ($r$), as well as pipelining the commit process. When
collecting the votes for the \textsc{prepare} phase, the Quorum Certificate
($N-f$ \textit{votes}) formed is passed to the next round ($r + 1$) leader. The
next leader initiates the \textsc{prepare} for their view, but also includes
the \textsc{pre-commit} for the previous round $r - 1$ as well as the
\textsc{commit} phase for round $r - 2$.

\paragraph{Goals and Assumptions}

HotStuff presents consensus with partial synchrony. The consensus tolerates
strictly less than a third of the participants exhibiting Byzantine behavior.
The protocol utilizes a core of three-phases to come to agreement, not
requiring any randomization or coin flipping and thus provides deterministic
agreement and termination. HotStuff utilizes timeouts to progress, however, if
the threshold of messages arrive earlier than the timeout, it will interrupt
and progress without having to wait for the timeout to be fulfilled.

\paragraph{LibraBFT\label{pg:libra}}

LibraBFT~\cite{librabft} is a blockchain-based Byzantine Fault Tolerant
consensus which builds upon HotStuff~\cite{yin2018hotstuff} and is at the heart
of Facebook's Libra blockchain~\cite{librawhitepaper}. LibraBFT extends
HotStuff by requiring a new leader for each block proposal. The leader proposes
a block, which is then voted upon by the remainder of the committee.  Once the
leader has collected enough votes ($N-f$), they form a Quorum Certificate with
the votes and append the block and certificate to the blockchain. A new leader
is then chosen for the next round.

The LibraBFT also extends HotStuff through a reconfigurable consensus committee
per epoch, where membership selection can be utilized, such as a
VRF~\cite{Micali16-algorand} as mentioned in the proposal.

\section{Structure \label{sec:structure}}

In this section we present overviews and graphical representations of proposed
structural changes to the blockchain. The Bitcoin blockchain originally
featured a single, canonical chain with homogeneous block types. New proposals,
however, provide a variety of new block types and structures that provide
improvements and alternative ways that the blockchain can operate.

\subsection{New block types}

Some proposals focus improvements primarily on block types. By moving away
from homogeneous blocks and adding new block functionality, new opportunities
arise for improvements to the blockchain.

\subsubsection{Bitcoin-NG}

\begin{figure}[h]
  \begin{center}
    \resizebox{.7\columnwidth}{!}{
      \begin{tikzpicture}
      \node[keyblock] (G) at (0,0) {Keyblock$_{i}$};
      \node[left = of G] (d1) {\dots};
      \node[microblock, right= of G] (m1) {Microblock};
      \node[microblock, right= of m1] (m2) {Microblock};
      \node[keyblock, right = of m2] (K1) {Keyblock$_{i+1}$};
      \node[right= of K1] (d2) {\dots};
      %\node[microblock, right= of m2] (m3) {Microblock};
      %\node[microblock, right= of m3] (m4) {Microblock};
      %\node[microblock, right= of m4] (m5) {Microblock};

      \path[<-, >={Latex[scale=2.0]}, thick]
        (d1) edge (G)
        (G) edge (m1)
        (m1) edge (m2)
        (m2) edge (K1)
        (K1) edge (d2);
    \end{tikzpicture}
  }
  \end{center}
  \caption{Bitcoin-NG Block Overview \label{fig:bitcoin-ng}}
\end{figure}
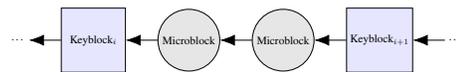

Bitcoin-NG~\cite{EyalGencer-etc-16-BitcoinNG} aims at improving upon Bitcoin's
throughput by utilizing heterogeneous block types. In particular, Bitcoin-NG
introduces two types of blocks, namely \textit{Keyblocks} and
\textit{Microblocks}. Keyblocks are produced through Bitcoin's Proof-of-Work,
taking several minutes to produce a block. The keyblocks contain no
transactions and each time they are created, the miner is chosen as the leader
to produce microblocks, so in this sense the microblocks are linked directly to
the keyblock. Microblocks contain transactions and are produced without
Proof-of-Work, so transactions are batched and can be produced quicker.
Figure~\ref{fig:bitcoin-ng} presents an overview of this structure.

% The Bitcoin-NG~\cite{EyalGencer-etc-16-BitcoinNG} proposal to upgrade the
% Bitcoin blockchain proposed a change to the structure of the blockchain. To
% increase the throughput of Bitcoin, the new structure involves different block
% types that allow for transactions to be committed to the chain without the
% requirement of a hard computation for each block. The Bitcoin-NG proposal was
% to add two block types, \textit{Microblocks} which are blocks to append
% transactions to the chain and are added every few seconds through the selected
% leader; \textit{Keyblocks} which are responsible for determining a new leader.

\subsubsection{ComChain}

\begin{figure}[h]
  \centering
  \resizebox{\columnwidth}{!}{
    \begin{tikzpicture}
      \node[comchainconfig] (1) at (0,0) {Config 1};
      \node[left=of 1] (dle) {\ldots};
      \node[comchaintx, below right= -1cm and 1cm of 1] (2) {txblock$_i$};
      \node[comchaintx, right=of 2] (3) {txblock$_j$};
      \node[comchaintx, right=of 3] (4) {txblock$_k$};
      \node[comchainconfig, above right= -1cm and 1cm of 4] (5) {Config 2};
      \node[comchaintx, below right= -1cm and 1cm of 5] (6) {txblock$_l$};
      \node[comchaintx, right=of 6] (7) {txblock$_m$};
      \node[rectangle, above right= -0.5cm and 1cm of 7, minimum height=1.5cm] (dri) {\ldots};

      \path[<-, >={Latex[scale=2.0]}, thick]
      (dle) edge (1)
      (1.325) edge (2)
      (2) edge (3)
      (3) edge (4)
      (4) edge (5.215)
      (5.325) edge (6)
      (6) edge (7)
      (1.35) edge (5.145)
      (5.35) edge (dri.145)
      (7) edge (dri.250);

    \end{tikzpicture}
  }
\caption{Community Blockchain (ComChain) structure.}
  \label{fig:comchain}
\end{figure}
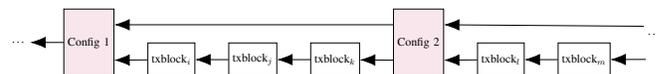

ComChain~\cite{comchain}, the ``Community Blockchain'', proposes the
\textit{configuration block}, a new block type that defines a subset of nodes
chosen to form the next consensus committee.  When a new consensus committee is
being proposed, a configuration block is sent to the current committee to be
agreed upon. Once agreed, the new committee is formed and begins participating
in consensus on the blocks containing transactions. The proposed configuration
must be validated and all pending transactions and blocks should be transferred
to the new committee to commit through the consensus. The structure is depicted
in Figure~\ref{fig:comchain}.

\subsection{New Structures}

Rather than proposing changes to block types, some proposals focus primarily on
the blockchain structure. The blockchain originated as a single, canonical
chain appending blocks sequentially. However, new proposals provide a number of
changes that allow for blocks, or transactions, to be committed in a variety of
ways. This presents opportunities for parallel block processing or new
communication patterns between nodes.

\subsubsection{Fruitchains}

\begin{figure}[h]
  \centering
  \resizebox{\columnwidth}{!}{
    \begin{tikzpicture}
      \node[block] (b1) at (0,0) {B$_i$};
      \node[left= of b1] (d1) {\dots};

      \node[block, right = of b1] (b2) {B$_j$};
      \node[block, right = of b2] (b3) {B$_k$};
      \node[block, right = of b3] (b4) {B$_l$};
      \node[block, right = of b4] (b5) {B$_m$};
      \node[right = of b5] (d2) {\dots};

      % Fruits
      \node[fruit, below =of b1] (f1) {F$_i$};
      \node[fruit, below =of b3] (f2) {F$_k$};
      \node[fruit, below = 2cm of b4] (f3) {F$_l$};
      \node[fruit, below = of b5] (f4) {F$_m$};

      \path[<-, >={Latex[scale=2.0]}, thick]
      (b1) edge (b2)
      (b2) edge (b3)
      (b3) edge (b4)
      (b4) edge (b5)
      (d1) edge (b1)
      (b5) edge (d2);

      % Fruits to blocks
      \path[-, thick]
      (f1) edge (b1)
      (f2) edge (b3)
      (f3) edge (b4);

      % Fruits to reference
      \draw[<-, >={Latex[scale=2.0]}, rounded corners] (d1) |- (f1);
      \draw[<-, >={Latex[scale=2.0]}, rounded corners] (b2.300) |- (f2);
      \draw[<-, >={Latex[scale=2.0]}, rounded corners] (b2.300) |- (f3);
      \draw[<-, >={Latex[scale=2.0]}, rounded corners] (b4.300) |- (f4);

    \end{tikzpicture}
  }
  \caption{Fruitchains Structure}
  \label{fig:fruitchains}
\end{figure}
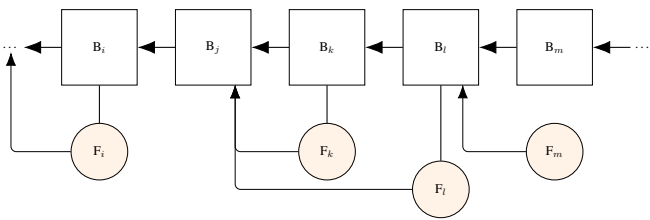

The Fruitchains~\cite{pass2017fruitchains} blockchain introduces an alternate
structure to the blockchain and decouples transaction processing from block
processing. Rather than transaction records being appended directly onto the
chain inside a block, they are packed into a \textit{fruit}, which is a batch
of transactions. To append a fruit to a chain, it \textit{hangs} from one
block, referencing a recent block. The referenced block must not be too far
from the block the fruit is attached to, as shown in
Figure~\ref{fig:fruitchains}. By decoupling the transaction batching inside a
block and having the fruits, transactions can be processed and committed to the
chain with reference to the state they were processed in. This provides the
ability to process more transactions while still ensuring the validity of the
state.

In the Bitcoin blockchain, miners creating orphaned blocks will not
get any reward, which is not fair to them as they also contributed
their computing power. Fruitchains provides a fairer reward
distribution, as the creator of fruits hanged from a recent block can
also get its mining reward.

\subsubsection{ByzCoin}

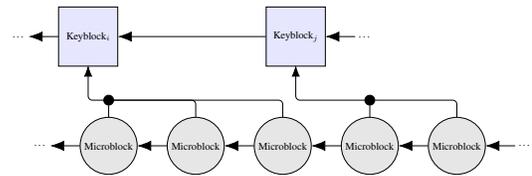
\begin{figure}[h]
  \begin{center}
    \resizebox{.8\columnwidth}{!}{
      \begin{tikzpicture}
      \node[keyblock] (K1) at (0,0) {Keyblock$_{i}$};
      \node[left = of K1] (d1) {\dots};
      \node[keyblock, right = 5cm of K1] (K2) {Keyblock$_{j}$};
      \node[right= of K2] (d2) {\dots};
      \node[microblock, below right= 2cm and -1cm of K1] (m1) {Microblock};
      \node[microblock, right= of m1] (m2) {Microblock};
      \node[left = of m1] (dm1) {\dots};
      \node[microblock, right= of m2] (m3) {Microblock};
      \node[microblock, right= of m3] (m4) {Microblock};
      \node[microblock, right= of m4] (m5) {Microblock};
      \node[right= of m5] (dm2) {\dots};
      \node[lineconnect, above = 0.4cm of m1] (lc1) {};
      \node[lineconnect, above = 0.4cm of m4] (lc2) {};

      % Keyblocks
      \path[<-, >={Latex[scale=2.0]}, thick]
        (d1) edge (K1)
        (K1) edge (K2)
        (K2) edge (d2);

      % Microblocks
      \path[<-, >={Latex[scale=2.0]}, thick]
      (dm1) edge (m1)
      (m1) edge (m2)
      (m2) edge (m3)
      (m3) edge (m4)
      (m4) edge (m5)
      (m5) edge (dm2);

      \path[-] (lc1) edge (m1);
      %\draw[smooth, rounded corners] (lc1) -- (m1);
      \draw[smooth, rounded corners] (lc1) -| (m2);
      \draw[smooth, rounded corners] (lc1) -| (m3);
      \draw[<-, >={Latex[scale=2.0]}, rounded corners] (K1) |- (lc1);

      \draw[smooth, rounded corners] (lc2) -- (m4);
      \draw[smooth, rounded corners] (lc2) -| (m5);
      \draw[<-, >={Latex[scale=2.0]}, rounded corners] (K2) |- (lc2);

      %\path
      %(K1) edge[<-, bend left=20] (m1)
      %(K1) edge[<-, bend left=20] (m2);

      %\draw[<-, smooth, rounded corners] (K1) -| (m3);

    \end{tikzpicture}
  }
  \end{center}
  \caption{ByzCoin Microblock and Keyblock structure \label{fig:byzcoin}}
\end{figure}

ByzCoin~\cite{Kokoris-KogiasJ16-ByzCoin}, inspired by
Bitcoin-NG~\cite{EyalGencer-etc-16-BitcoinNG}, provides a new structure for the
blockchain to improve transaction throughput by decoupling the transactions from
the block creation. Unlike Bitcoin-NG, ByzCoin separated the microblocks and
keyblocks into separate chains as shown in Figure~\ref{fig:byzcoin},
where two parallel chains store the information. The creation of Keyblocks is
utilized to form a consensus committee. The consensus committee participates
in BFT consensus to produce microblocks, which are appended to a second parallel
chain every few seconds.

\paragraph{RepuCoin}

Similar to ByzCoin, RepuCoin~\cite{Yu2018Repucoin} presents a similar
idea of decoupling transactions. RepuCoin utilizes the keyblocks for
leader election and the consensus committee. When a keyblock is
proposed, and the consensus committee agrees on this new block, it
becomes \textit{pinned}. From the moment the keyblock is pinned and
committed, the microblocks are produced from the leader and consensus
group. Like ByzCoin, RepuCoin follows a structure of parallel chains,
where keyblocks are mined and committed, referencing each predecessor
keyblock, whereas the microblocks run on a parallel chain referencing
predecessor microblocks but also reference the most recent keyblock.

% \CN{OLD:\\
% \textbf{ByzCoin} ByzCoin used the foundations of Bitcoin-NG to further
% improve the Bitcoin blockchain. Due to the possibility of forks in Bitcoin-NG,
% and to improve the security, ByzCoin utilises the PBFT consensus. Keyblocks are
% now selecting new consensus members, rather than a single leader. Microblocks
% are now signed by the entire consensus group and sent to the consensus leader
% before being accepted on to the chain.
%
% \JS{This paragraph did not mention anything about the structure. Both
%   ByzCoin and repucoin use a new structure, where two parallel chains
%   are used. One chain stores keyblocks, and the other chain records
%   microblocks. (See Figure 3, page 7,
%   https://arxiv.org/pdf/1602.06997.pdf, and Figure 1, page 5,
%   https://eprint.iacr.org/2018/239.pdf)}
% }

\subsubsection{HashGraph (Swirlds)}

\begin{figure}[h]
  \centering
  \resizebox{.6\columnwidth}{!}{
    \begin{tikzpicture}
      % Actors
      \node at (0,0) {Alice};
      \node at (2,0) {Bob};
      \node at (4,0) {Carol};
      \node at (6,0) {Dave};

      % Actor lines
      \draw (0, -1) -- (0, -7.5);
      \draw (2, -1) -- (2, -7.5);
      \draw (4, -1) -- (4, -7.5);
      \draw (6, -1) -- (6, -7.5);

      % Events
      \node[hgwitness] (a1) at (0, -1) {A}; %{A$_1$};
      \node[hgwitness] (b1) at (2, -1) {B}; %{B$_1$};
      \node[hgwitness] (c1) at (4, -1) {C}; %{C$_1$};
      \node[hgwitness] (d1) at (6, -1) {D}; %{D$_1$};

      % Event Transfers
      \node[hgevent] (bd1) at (6, -2) {};
      \node[hgevent] (db1) at (2, -3) {};
      \node[hgevent] (ba1) at (0, -4) {};
      \node[hgevent] (bd2) at (6, -4) {};
      \node[hgevent] (cb1) at (2, -4.5) {};
      \node[hgevent] (bc1) at (4, -6) {};
      \node[hgevent] (bd3) at (6, -6) {};
      \node[hgevent] (db2) at (2, -7) {};
      \node[hgevent] (ac1) at (4, -7) {};

      % Paths
      \path[-]
        (b1) edge (bd1)
        (bd1) edge (db1)
        (db1) edge (ba1)
        (db1) edge (bd2)
        (c1) edge (cb1)
        (cb1) edge (bc1)
        (cb1) edge (bd3)
        (ba1) edge (ac1)
        (bd3) edge (db2);

      % % End of round one
      % \path (-0.5, -7.5) edge[dashed] (6.5, -7.5);

      % % Event (R2)
      % \node[hgwitness] (d2) at (6, -8) {D$_2$};

      % \path [-]
      %   (ba1) edge (d2);

    \end{tikzpicture}
  }
  \caption{HashGraph Structure. An example round of communication and event
  gossiping.}
  \label{fig:hashgraph}
\end{figure}
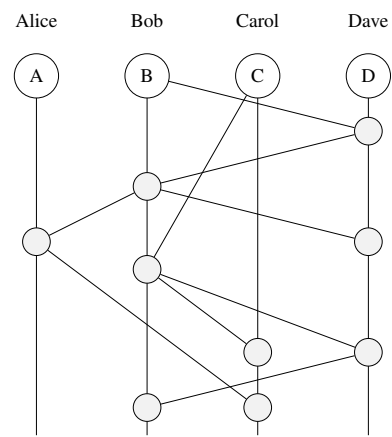

Swirlds~\cite{baird2016swirlds} modified the traditional blockchain structure
by proposing a graph structure where each ``block'' has multiple children or
parents.  Swirlds presents the \textit{HashGraph}, a directed acyclic graph in
which the ``blockchain'' utilizes all blocks that are mined. This removes the
concept of stale blocks and encourages forked chains. As seen in
Figure~\ref{fig:hashgraph}, the hashgraph contains vertices and edges, dubbed
events and relationships respectively.  An event is gossiped by a node can
contain transaction information and must require hashes referencing two past
events, which is then signed by the gossiping node.
Through gossiping, the receiving node learns about the context of the event and
all other information by previous nodes in the communication. The nodes can
then agree on the ancestors of the event through offline virtual voting, as all
needed votes from other nodes are contained in the received gossips.  Although
nodes may learn of events in differing places of their graph, an event is said
to be consistent if it has equivalent ancestors in all node's graphs.

\subsubsection{Beacon Chain}

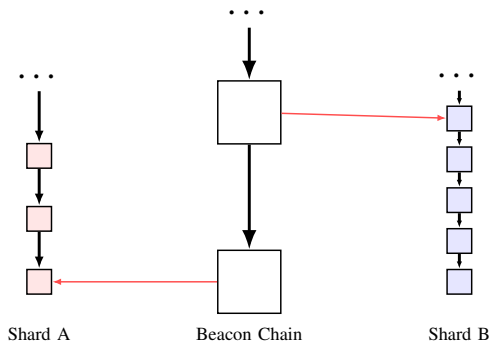
\begin{figure}[h]
  \begin{center}
    \resizebox{0.8\columnwidth}{!}{
      \begin{tikzpicture}
        % Side Chain 1
        \node (adesc) at (-2, -0.5) {\tiny Shard A};
        \node[beaconsideblocka] (A1) at (-2,0) {};
        \node[beaconsideblocka, above = 10pt of A1] (A2) {};
        \node[beaconsideblocka, above = 10pt of A2] (A3) {};
        \node[above = 14pt of A3] (A4) {\ldots};

        \path[<-, >={Latex[scale=0.5]}, thick]
          (A1) edge (A2)
          (A2) edge (A3)
          (A3) edge (A4);

        % Side chain 2
          \node (cdesc) at (2, -0.5) {\tiny Shard B};
        \node[beaconsideblockb] (C1) at (2, 0) {};
        \node[beaconsideblockb, above = 4pt of C1] (C2) {};
        \node[beaconsideblockb, above = 4pt of C2] (C3) {};
        \node[beaconsideblockb, above = 4pt of C3] (C4) {};
        \node[beaconsideblockb, above = 4pt of C4] (C5) {};
        \node[above = 4pt of C5] (C6) {\ldots};

        \path[<-, >={Latex[scale=0.3]}, thick]
          (C1) edge (C2)
          (C2) edge (C3)
          (C3) edge (C4)
          (C4) edge (C5)
          (C5) edge (C6);

        % Beacon Chain
          \node (bdesc) at (0, -0.5) {\tiny Beacon Chain};
        \node[beaconblock] (B1) at (0, 0) {};
        \node[beaconblock, above = 1cm of B1] (B2) {};
        \node[above = 0.5cm of B2] (B3) {\ldots};
        \path[<-, >={Latex[scale=0.8]}, thick]
          (B1) edge (B2)
          (B2) edge (B3);

        % Beacon Links
        \path[<-, >={Latex[scale=0.5]} , red!70!white]
          (A1) edge (B1)
          (C5) edge (B2);

      \end{tikzpicture}
    }
  \end{center}
  \caption{Beacon Chain as adopted by Ethereum Serenity \label{fig:beacon-chain}}
\end{figure}

Dfinity~\cite{dfinity-arxiv} envisioned a chain where a set of nodes
selected through a \textit{random beacon} would propose and notarize blocks.
Ethereum Serenity~\cite{eth2.0-specs, ethresearch-beaconchain}, proposed a
similar concept but adopted a sharded blockchain model.
The single, canonical beacon chain would serve as the backbone for the shards
operating in parallel. The shard blockchains handle state transitions and hold
transaction information, whereas the beacon chain acts as a point of
cross-shard communication and knowledge of shard states.
Figure~\ref{fig:beacon-chain} illustrates the concept of the beacon chain with
two shards.

\subsubsection{Tangle (\textsc{Iota})}

\begin{figure}[h]
  \centering
  \resizebox{\columnwidth}{!}{
    \begin{tikzpicture}

      \node[tangletx] (t0) at (0, 0) {tx$_0$};
      \node[tangletx] (t1) at (2, 0.5) {tx$_1$};
      \node[tangletx] (t2) at (2, -1) {tx$_2$};
      \node[tangletx] (t3) at (4, -0.5) {tx$_3$};
      \node[tangletx] (t4) at (6, -1.5) {tx$_4$};
      \node[tangletip] (t5) at (6, 1) {tx$_5$};
      \node[tangletx] (t6) at (7, 0) {tx$_6$};
      \node[tangletx] (t7) at (8, -1) {tx$_7$};
      \node[tangletip] (t8) at (10, -0.5) {tx$_8$};
      \node[tangletip] (t9) at (10, 0.5) {tx$_9$};

      % Paths
      \path[<-, >={Latex}, thick]
      (t0) edge (t1)
      (t0) edge (t2)
      (t1) edge (t5)
      (t1) edge (t6)
      (t2) edge (t3)
      (t2) edge (t4)
      (t3) edge (t7)
      (t4) edge (t7)
      (t6) edge (t9)
      (t6) edge (t8)
      (t7) edge (t8);

    \end{tikzpicture}
  }
  \caption{\textsc{Iota}'s tangle structure.}
  \label{fig:tangle}
\end{figure}
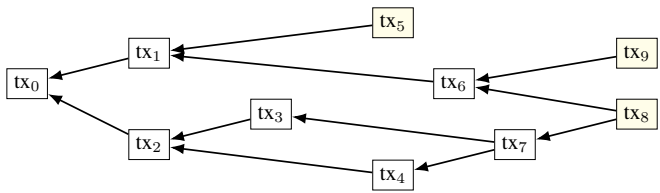

\textsc{Iota}~\cite{iota-whitepaper} presents the \textit{tangle}, a DAG which
removes blocks from the storage structure. The tangle DAG, as illustrated in
Figure~\ref{fig:tangle}, stores transactions as vertices and requires each
vertex to have an edge between two (or more) previous transactions. This forms
the directed structure and the edges constitute to the validity of the
transactions. The transactions are gossiped through the network and are placed
into the DAG of the node when received. Unconfirmed transactions, named
\textit{tips} are placed into the graph as leaves and are assigned a weight.
The weight of the tip is proportional to the transactions it references, with
respect to the recency and weight of referenced transactions.

For a tip to become validated, it must gain a high
\textit{confirmation confidence} by getting referenced by new
transactions over time as the DAG continues to grow. Although the DAG
may differ between nodes, the transaction ancestors will be common
amongst nodes and will therefore be passed through the consensus to be
agreed upon. Analysis~\cite{dagiot-trustcom19} shows that such structure
largely improves the scalability of blockchain.

\subsubsection{Block Lattice (RaiBlocks)}

\begin{figure}[th!]
  \centering
  \resizebox{.7\columnwidth}{!}{
    \begin{tikzpicture}
      % Actors
      \node at (0,0) {Alice};
      \node at (2,0) {Bob};
      \node at (4,0) {Carol};
      \node at (6,0) {Dave};

      % Actor lines
      \draw (0, -1) -- (0, -7);
      \draw (2, -1) -- (2, -7);
      \draw (4, -1) -- (4, -7);
      \draw (6, -1) -- (6, -7);

      % Transactions

      % Alice to Bob
      \node[blsend] (ab1s) at (0,-1) {tx$_1$ (S)};
      \node[blrec] (ab1r) at (2,-1.5) {tx$_1$ (R)};

      % Alice to Dave
      \node[blsend] (ad1s) at (0, -2.5) {tx$_2$ (S)};
      \node[blrec] (ad1r) at (6, -3) {tx$_2$ (R)};

      % Dave to Carol
      \node[blsend] (dc1s) at (6, -4) {tx$_3$ (S)};
      \node[blrec] (dc1r) at (4, -4.5) {tx$_3$ (R)};

      % Carol to Bob
      \node[blsend] (cb1s) at (2, -5.5) {tx$_4$ (S)};
      \node[blrec] (cb1r) at (4, -6) {tx$_4$ (R)};

      % Paths
      \path[-]
      (ab1s) edge (ab1r)
      (ad1s) edge (ad1r)
      (dc1s) edge (dc1r)
      (cb1s) edge (cb1r);

    \end{tikzpicture}
  }
  \caption{Block Lattice example execution and structure overview.}
  \label{fig:blocklattice}
\end{figure}
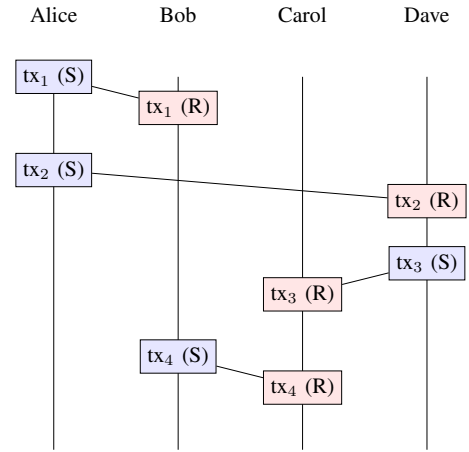

RaiBlocks~\cite{raiblocks-whitepaper} introduced the \textit{block lattice},
depicted in Figure~\ref{fig:blocklattice}. The distinguishing feature of the
block lattice is that each account maintains its own personal blockchain, the
DAG ledger consists of the global set of accounts and their relevant
blockchain. Each transaction sent from user to user requires two transactions,
a ``send'' transaction signed by the transaction origin owner, and a
``receive'' transaction, signed by the party receiving the transaction.
Similarly, each party must create a block on their respective chain, a ``send
block'' and ``receive block''. Figure~\ref{fig:blocklattice} depicts an example
communication between parties.  The pattern formed by the transactions forms a
lattice of send and receive blocks, forming the DAG structure. Any conflicting
transactions that are detected require a vote to be passed, in which the
majority vote wins.

\section{Attacks}\label{sec:attack}

Blockchains are susceptible to a number of attacks~\cite{conti2017survey}. Each
varying implementation requires assumptions which often leads to limitations
and vulnerabilities. The known attacks utilize these assumptions to empower the
adversary to double spend, or, deny the service of the blockchain to others.
Table~\ref{ref:attacksoverview-1}, Table~\ref{ref:attacksoverview-2} and
Table~\ref{ref:attacksoverview-3} present an overview of the attacks reviewed
in this section.

\subsection{Mining Power Attacks}

The mining power attacks target systems that utilize Proof-of-Work
membership selection combined with any consensus that suffers from forks. The
mining power attacks are focused on an adversary gaining sufficient amount of
mining power to double spend, or, take control of the chain. To allow for more
profitable mining, nodes often collude together to form mining pools, allowing
cost sharing of profits and higher probability of block rewards. The Bitcoin
network consists of a number of pools~\cite{blockchaininfo-pools}. In July
2014, the GHash.IO mining pool exceeded the 51\% threshold. At the time of
viewing, the largest pool is the BTC.com pool which controls $29.1\%$ of the
networks computational power. The second highest, AntPool, controls $13.3\%$,
closely followed by ViaBTC ($10.6\%$) and SlushPool ($10.1\%$). It can be seen
that if the top three pools collude, they control the majority of the network
and can therefore control the chain. Similarly, in the Ethereum
blockchain\footnote{Data available online:
  \url{https://etherscan.io/stat/miner?range=7&blocktype=blocks}. Accessed 25
Apr. 2018}, mining pools, such as Ethermine ($27.5\%$), f2pool\_2 ($16.6\%$), SparkPool ($15.7\%$)
and Nanopool ($14.1\%$), control a large portion of the network mining power.

\paragraph{51\%}

The most noted attack is the majority mining attack, also referred to as the
51\% Attack. The attack allows an adversary to control the blockchain,
validating invalid transactions and censoring the entire network as they wish.
If an adversary is able to gain the majority of the mining power, they have the
ability to produce valid blocks faster than other nodes, or pools, with high
probability, meaning they are able to create alternative chains with a false
history, allowing arbitrary double-spending.

This attack targets the adversary assumption that the honest nodes hold the
majority of the network's resources. Due to the distribution of computing power
in current Proof-of-Work blockchains, gaining a majority of the network's
resources can be seen as a practically infeasible task, as the cost would be
too significant unless the attack was carried out in early stages.

However, proposed techniques~\cite{Bonneau16-bribery-attacks,
natoli2017balance,HeilmanKZG15-Eclipse-Attacks, apostolaki2017hijacking} make
it plausible for an adversary to gain a 51\% of the resources
and exploit this assumption. It is also worth noting that large mining pools
develop over time and collusion amongst the large pools may provide enough
power to control the network.

\subsubsection{Bribery Attack (Flash Attacks)}

The Bribery Attack~\cite{Bonneau16-bribery-attacks}, also known as the flash
attack, consists of an adversary temporarily gaining the majority of the mining
power for malicious use. This attack allows for the adversary to gain temporary
control of the chain, allowing a window where they are able to double spend or
censor transactions. The adversary gains a majority of the mining power by
renting computing power. This can be achieved through out-of-band bribes,
in-band bribes or through renting computing power. The methodology follows an
attacker issuing a transaction in the chain to a merchant. Once the merchant is
satisfied with the block confirmations, the adversary utilizes the gained
computing power to create a conflicting transaction in a forked chain. The
adversary controls the majority of the network mining power, and is then able
to extend the forked chain quicker than the network and win the longest chain
in fork resolution, where the merchant will no longer gain the funds due to the
conflicting transaction. The adversary is not concerned about the long-term
health of the chain, so consequences of the bribery attack may be felt by the
network.

This attack can be facilitated by the use of cloud services providing easily
available computing power, or through collusion of mining pools. However, a
large investment or bribe is needed to gain the majority of the network power,
even for a short period of time.

\subsection{Strategic Mining}

Strategic mining attacks exploit the decentralized nature of the blockchain, as
each node contributing to the system is required to uphold their own local copy
of the chain. Strategic mining attacks consist of withholding information from
other nodes for later use to cheat the system.

\subsubsection{Selfish Mining attack}

This selfish mining attack was proposed where pools mine blocks in
secret~\cite{EyalSirer14-Selfish-attack}. The attack requires an adversary to
control a significant amount ($>$25\%) of mining power, such that it can keep
up with the chain progress. The attack methodology requires the pool to mine
blocks in secret, selfishly keeping the produced blocks from the other nodes in
the network. Selectively, the pool can choose to release the newly produced
blocks to waste the honest miners computational power on producing old blocks.

An extension to this work~\cite{Sapirshtein2016optimalselfishminging}
highlighted optimal strategies for selfish mining and showed that the selfish
mining had optimal and sub-optimal conditions for success. Such optimal
suggestions include conditions where the adversary should adopt network blocks,
override network blocks of their own, match the block or wait. These actions,
if selected effectively, could lead to a more optimal selfish mining strategy.

This attack is highly dependent on the pool's mining power. The pool must be
able to produce a blockchain branch that is longer than the public branch. When
they choose to release their branch, they will win the longest chain rule,
allowing them to falsely operate the chain.

\afterpage{
\onecolumn
\begin{landscape}
\begin{center}
  \begin{table*}[h]
    \vspace{-0.5cm}
  \caption{Attacks Overview: Part 1\label{ref:attacksoverview-1}}
%  \hspace{-8cm}
  \resizebox{1.35\textwidth}{!}{%
  \begin{tabular}{p{0.1\textwidth} p{0.11\textwidth}p{0.43\textwidth}p{0.45\textwidth}x{0.25\textwidth}}
    \toprule
    \textbf{Category} & \textbf{Attack} & \textbf{Overview} & \textbf{Attack Core} & \textbf{Cause}\\
    \midrule
        \multirow{2}{*}{\raisebox{-\heavyrulewidth}{Mining}}
        & 51\%

        & An adversary that controls the strict majority of the mining power
        has the ability to double spend coins and control the chain.

        & \begin{itemize}
        \item An adversary is able to obtain strictly greater than 50\% of the
          total network mining power.
          \end{itemize}

          & \hfill\vfill $f' > \frac{N'}{2}$
        \\\cmidrule{2-5}
        & Bribery/Flash

        & An adversary bribes others to gain a strict majority of the mining
        power to perform a 51\% attack in a short time period.

        & \begin{itemize}
            \item Mining power can be paid/bribed or rented.
            \item The overall cost of the attack is (a) less than purchasing
              resources to perform a 51\% attack, and (b) less than the overall
              reward gained from performing the attack.
          \end{itemize}
          & \hfill\vfill\hfill\vfill
          $f' > \frac{N'}{2}$
    \\\midrule
    \multirow{5}{0.5cm}{\raisebox{-\heavyrulewidth}{\parbox{0.05\textwidth}{Strategic Mining}}}
        & Selfish Mining

        & A mining pool mines a fork in secret. They then release the fork to
        the network using a smart strategy to waste the work of others and
        ensure their private fork is considered the longest chain.

        & \begin{itemize}
            \item Pool has significant amount of mining power, and, can mine
              quicker than network.
            \item No transaction or blocks leaked to network.
          \end{itemize}
          &
                %\item Synchrony
                %\item Favour Availability
                \hfill\vfill
                \hfill\vfill
                Probabilistic Consensus
        \\\cmidrule{2-5}
        % Selfish
        & Finney's

        & The adversary sends a transaction to a merchant for some irrevocable
        action, however, the adversary mines a separate conflicting transaction
        into a block. The merchant performs the action as soon as the
        transaction is sent into the network. At this point, the adversary
        releases the block containing the conflicting transaction and the
        original transaction to the merchant is invalidated.

        & \begin{itemize}
            \item Merchant trades with 0 block confirmations.
            \item Adversary's original transaction does not get mined into a
              block before releasing the conflicting block.
          \end{itemize}
          & \begin{itemize}
                \item Synchrony
                \item Favor Availability
                \item Probabilistic Consensus
          \end{itemize}
          \vspace{-0.4cm}
        \\\cmidrule{2-5}
        & Double Spending Fast Payments
        & An adversary connects as a peer to a merchant and issues a
        transaction directly for an irrevocable action. The merchant accepts the
        transaction without any block confirmations and performs the action.
        Simultaneously, the adversary sends a conflicting transaction to the
        other peers. With non-zero probability, the conflicting transaction propagates
        faster and is mined into a block.
        & \begin{itemize}
          \item A merchant accepts a transaction with 0 blocks.
          \item The adversary's conflicting transaction propagates the network quicker.
          \item The adversary's transaction is mined into a block before the merchants.
        \end{itemize}
          & \begin{itemize}
                \item Synchrony
                \item Probabilistic Consensus
          \end{itemize}
        \\\cmidrule{2-5}
        % Selfish
        & Rosenfeld's
        & An adversary solo mines a private longest branch in secret,
        withholding all information from other nodes. They then interact with a
        merchant, who requires some number of block confirmations before
        executing an irreversible action. Once the merchant is satisfied and
        performs the action, the adversary releases the solo mined branch which
        wins the longest chain and is accepted as the main chain, invalidating
        the transaction to the merchant.
        & \begin{itemize}
          \item An adversary can mine a branch in secret without any
            information leakage.
          \item The adversary can mine the longest branch.
          \item A merchant trades with a low block confirmation.
          \end{itemize}
          & \begin{itemize}
                \item Synchrony
                \item Probabilistic Consensus
          \end{itemize}
        \\\cmidrule{2-5}
        & Vector76
        & An adversary interacts with a merchant who performs an action as soon
        as the transaction is accepted into a block. The adversary quickly
        sends a self-mined block containing the transaction directly to the
        merchant and the action is performed, while simultaneously submitting a
        conflicting transaction to the network. Once the merchant learns of the
        network's chain, the adversary's block is invalidated and lost.
        % \vspace{1em} %this is for fixing space
        & \begin{itemize}
          \item A merchant trades with 1 block confirmation.
          \item The adversary can mine a block quicker than the network.
          \item The Merchant never learns of the network's block and the
            adversary can propagate the block directly to the merchant quicker
            than merchant learning of network block.
        \end{itemize}
          & \begin{itemize}
                \item Synchrony
                \item Probabilistic Consensus
          \end{itemize}
        \\\cmidrule{2-5}
        % Selfish
        & Withholding Mining Pools
        & An adversary joins a mining pool, but hides all valid Proof-of-Work
        they mine. This lowers the potential revenue for the pool as they miss
        out on a number of blocks.
        & \begin{itemize}
          \item Static pool membership; all members stay in the pool
            regardless of revenue profits.
          \item Membership rules are not strict and do not require nodes to
            submit certain amounts of work.
          \end{itemize}
       & \multicolumn{1}{p{0.25\textwidth}}{This attack is not on membership selection nor on
         consensus. Rather, it is an attack on the mining pool
         software and infrastructure, which is independent to the
         blockchain system.}
    \\\midrule
    Strategic \vfill Mining \vfill + \vfill Communication
        %\multirow{1}{*}{\raisebox{-\heavyrulewidth}{Network Selfish}}
        & Stubborn Mining
        & An adversary mines blocks with the honest network, only selfishly
        mining when an optimal strategy proves successful. By releasing blocks
        periodically (rather than the entire chain at once), the adversary is
        able to mine a private chain and gain profit for longest chain.
        & \begin{itemize}
          \item Adversary can mine blocks in a private chain and keep up with
            the network.
          \item No information leakage of the adversary's chain unless
            specifically by the adversary.
        \end{itemize}
          & \begin{itemize}
                \item Synchrony
                \item Probabilistic Consensus
          \end{itemize}
        % Selfish
    \\\bottomrule
  \end{tabular}
}
\end{table*}
\end{center}
\end{landscape}
\twocolumn
}

\subsubsection{Finney Attack}

The Finney Attack~\cite{Fin11} was proposed in the Bitcoin forums
as a way to double-spend against a merchant. The pre-requisite for this
attack's success is that a merchant delivers an irrevocable product or service
at the time of payment, meaning waiting for 0 block confirmations. The
adversary engages with the merchant and sends coins for the service or goods
that are irreversible. At this point, the merchant completes the trade.
Simultaneously, the adversary then creates a transaction to themselves and
mines a block, which includes this transaction. The adversary then quickly
broadcasts the block with the transaction to themselves. If quick enough, the
transaction to the merchant will conflict, and the transaction back to self
will be confirmed in a block and accepted by the network.

This attack leverages the merchant's assumption that transaction confirmations
do not require block confirmations. To remedy this attack, merchants are
advised to wait the recommended block confirmation, implementation specific,
which will give a higher probability of the transaction being confirmed into
the chain.

\subsubsection{Fast-Payment Double Spending}

Similar to Finney's attack, the Double Spending attacks on
Fast-Payments~\cite{KAC12fast} leverages a merchant's acceptance of a
transaction with 0 block confirmations. \textit{Fast Payments} are payments
where a transaction is considered accepted within a minute of it being proposed
in the network. Conversely, \textit{Slow Payments} require a threshold block
confirmation.

The adversary begins the attack by connecting to the merchant as a peer. The
adversary sends the transaction to the merchant for an irrevocable action.
After some small delta time, the adversary sends a conflicting transaction to
the other peers it is connected to. There is a high probability that the
conflicting transaction propagates through the network faster than the original
transaction to the merchant and is mined into a block. Once the transaction has
been mined into a block, the double spend has occurred.

This leverages the merchant accepting ``fast payments'' where a transaction is
accepted with no block confirmations in a small period of time. Secondly, it
requires the conflicting transaction to propagate the network and be mined into
a block quicker than the original merchant transaction.

\subsubsection{Rosenfeld's Attack}

Rosenfeld's proposed attack~\cite{Ros12} is a variant of the strategic mining
attack, where nodes withhold block production from other nodes. The adversary
solo mines a long branch in secret, keeping all information from other nodes.
The adversary then interacts with a merchant, issuing a transaction for an
irreversible action. Once the merchant has successfully completed the action,
and is satisfied with block confirmations, the adversary releases the longer,
solo-mined branch and conflicts with the main, public branch. Since the
adversary holds the longest branch, the adversarial chain is accepted and the
merchant's transaction is no longer in the chain.

This attack is highly dependent on the adversary controlling a major portion of
the computational power. If the adversary is unable to solo-mine a longer
branch, the attack will not be successful as the smaller branch will not
be accepted by other nodes in the network. Furthermore, if the merchant waits
for a large amount of confirmations, the adversary will be required to mine a
longer branch for longer, making the cost of the attack rise.

\subsubsection{Vector76 Attack}

The Vector76 attack~\cite{Vec11} was proposed on the BitcoinTalk forums by the
user ``vector76'', where an adversary wishes to double-spend against a
merchant.  The requirement for this attack's success is the merchant's
assumption that a transaction is confirmed within 1 block, meaning as soon as
the transaction is accepted into a block it is confirmed. The adversary
interacts with the merchant, and creates a transaction to pay for irreversible
goods or services. The adversary mines a block in secret, containing the
transaction to the merchant. When the adversary learns about a new block, they
quickly send the block they mined to the merchant directly. At this point, the
merchant can see the transaction in the block and performs the action, or
trades the goods. When the merchant learns of the forked chain, they will
select the other chain as correct, discarding the adversary's block, which
includes the adversary's transaction. The adversary double spends the coins
by including a conflicting transaction in the alternate chain.

This attack is dependent upon the merchant accepting the transaction as soon as
it has been included in a block. Furthermore, the adversary must be able to
provide a valid block faster than the network, and propagate it to the merchant
quicker than the merchant learning of the alternate chain. If the merchant
receives the network's block faster than the adversary's block, they may decide
to reject the adversary.

\subsubsection{Withholding Attack against Mining Pools}

The Block Withholding Attack~\cite{bag2017bitcoin, rosenfeld2011analysis} is an
attack targeted towards mining pools. An adversarial miner joins a mining pool
and begins the mining protocol. Mining in a pool generates both partial
Proof-of-Work and Full Proof-of-Work. The adversary can determine whether the
proof-of-work they have mined is valid or invalid. When the adversary mines a
valid hash, they withhold it from the pool, wasting the time of the pool and
having the potential for the pool to miss out on a number of blocks. The
motivation for this attack is to attack mining pools and lower their overall
block reward.

This attack heavily relies on the share model of the mining pool, but can be
mitigated by having strict conditions on the membership of the nodes in the
pool as well as adjusting the pool based on the profits made.

\subsubsection{Stubborn Mining}

Selfish mining attacks alone may prove to be too impractical for real-world
applications. The requirements for performing selfish mining lower the profit
margin. However, through applying optimal mining strategies, and potentially
combining with network-level attacks, the revenue may prove to be profitable.

The Stubborn Mining attack~\cite{NayakKMS16stubbornmining}, proposed by Nayak
et al., builds upon selfish mining by introducing mining strategies to increase
success and revenue, rather than selfishly mining and releasing the private
chain in one instance. The \textit{Lead Mining} strategy is where the adversary
leads by $k$ blocks and releases blocks the same height as the network-mined
blocks. Alternatively, the \textit{Equal Fork} strategy means the adversary
would conceal any new blocks even if they have no lead, and mine until they
lead. The \textit{Trail} strategy depicts a miner that continues to mine even
if they fall behind, until a certain threshold $j$ number of blocks behind. If
the miner is $j$ blocks behind, they accept the honest network chain and
continue to mine, however, if they succeed and catch up to the network's chain
they then introduce the fork resolution where they own the blocks that are in
the chain and can profit. By applying these strategies, a stubborn miner is
able to increase revenue and make their chain the chosen chain.

These strategies applied by a stubborn miner can be combined with network
attacks to further improve the revenue potential and percentage of success of
the attack. Examples such as the Eclipse
Attack~\cite{HeilmanKZG15-Eclipse-Attacks} can be used to isolate nodes and the
adversary can then censor or collude with the eclipsed miner to gain profit.

This attack requires an adversary to mine blocks to try and keep up with the
network. Furthermore, the strategies require no blocks to be leaked from the
adversarial chain.

\subsection{Communication Attacks}

Alternatively to mining attacks, the adversaries can utilize the
peer-to-peer network of the blockchain to attack the system. The
segregation and isolation of nodes can often be leveraged to perform
double-spending attacks.

\subsubsection{BGP Routing Attack}

BGP is the routing protocol responsible for connecting Autonomous Systems (AS)
together and routing IP packets through the internet. Autonomous Systems are
responsible for publishing IP prefixes, and Routers utilize this to route
packets. Since the blockchain requires nodes to connect to peers over the
internet, BGP is inherently used to route the packets to the other nodes.
However, in BGP the validity of any route advertisements are unchecked, as any
AS can advertise a given IP prefix and if they advertise a more specific prefix
(e.g. \textbackslash21 is more specific than \textbackslash20) they will become
the preferred route for the routing from other ASes. However,
monitors~\cite{yan2009bgpmon} and methods exist to
detect~\cite{schlamp2016heap} and provide resilience~\cite{zhang2007practical}
against BGP prefix hijacking.

The BGP Routing Attack~\cite{apostolaki2017hijacking} proposed against the
Bitcoin blockchain exploits the lack of verification in BGP to isolate a group
of nodes. Once the group of nodes has been successfully isolated, the adversary
can then delay block propagation or leave the nodes uninformed, utilizing their
mining power for their own benefit.

This attack relies on the ability of the adversary to isolate nodes in the
network through BGP route hijacking, and upholding the advertised route for a
long period of time.

\subsubsection{Eclipse Attack}

Hijacking BGP routes can be seen as a difficult task, as the adversary is
required to advertise a specific prefix and maintain the preferred route for
all isolated nodes. However, other methods exist to isolate specified node(s)
on the network for an adversary to perform an attack.

The blockchain is a decentralized, peer-to-peer network where all
nodes connect to others and form a gossip-based network. All nodes learn about
information broadcast from other nodes. Different implementations
allow nodes to connect to various numbers of other nodes, in Bitcoin it allows
for 125 peers to be connected, 117 incoming and 8 outgoing connections,
whereas Ethereum defaults to 25 peers.

The Eclipse attack~\cite{HeilmanKZG15-Eclipse-Attacks} describes an attack in
which a node can be eclipsed and isolated from the network by leveraging the
peer-to-peer connections of the blockchain. The adversary populates the peer
tables of the victim node, spoofing as many possible IPs into the victim node
as possible. When the victim node restarts, they will try to reconnect to the
nodes in the peer IP table. If the adversary has successfully populated a large
percentage of the IPs in the peer table, the node will connect all peers to the
adversary. The adversary can now control the information seen by the victim,
and is now able to execute double-spending attacks or delay attacks.

This attack relies on the adversary being able to insert a large number of IPs
into the victim's peer table. It also leveraged the eviction mechanism used by
the node to replace old IPs. Countermeasures have been suggested, and
implemented~\cite{bitcoind_0_10, HeilmanKZG15-Eclipse-Attacks}, to become
resilient to the attack.

\subsubsection{The Balance Attack}

Alternative to isolating a single node, the assumptions of the blockchain
consensus can be used to assist an adversary to execute double-spending.
Inspired by the Blockchain Anomaly~\cite{natoli2016blockchain}, the Balance
Attack~\cite{natoli2017balance} proposes an attack where the adversary
partitions the network into groups of nodes with balanced computational power.
Each subgraph of nodes will create a fork of the chain, which at some point,
will require a resolution. The adversary issues a transaction into the fork
containing the merchant. The merchant waits for the specified block
confirmations, and performs the action providing irrevocable goods or services
to the adversary.

At this point, the adversary then issues a conflicting transaction in another
subgraph, and provides additional computational power, assisting to create a
longer, or heavier, chain. The adversary slowly re-introduces the subgraphs
together, growing the number of nodes that accept the adversary fork containing
the conflicting transaction. When the merchant's subgraph learns of the fork,
they will accept the adversary fork and the transaction to the merchant will be
invalidated on the fork, executing the double-spending.

This attack heavily relies on the ability of the attacker to partition the
network into balanced subgroups where their computational power will provide a
significant benefit to the subgraph, and keep the network segregated until the
fork can be resolved.

\afterpage{%
\onecolumn
\begin{landscape}
  \vspace*{2.4cm}
\begin{center}
  \begin{table*}[h]
    \vspace{-0.5cm}
  \caption{Attacks Overview: Part 2\label{ref:attacksoverview-2}}
  \resizebox{1.35\textwidth}{!}{%
  \begin{tabular}{p{0.09\textwidth} p{0.11\textwidth}p{0.5\textwidth}p{0.45\textwidth}x{0.2\textwidth}}
    \toprule
    \textbf{Category} & \textbf{Attack} & \textbf{Overview} & \textbf{Attack Core} & \textbf{Cause}\\
    \midrule
        %\multirow{3}{*}{\raisebox{-\heavyrulewidth}{Communication}}
        Communication
        & BGP Routing
        & An adversary hijacks BGP routes and isolates groups of nodes. Once
        successful, and no information is being leaked, the adversary can control
        the information the nodes observe. This allows for potential double spending
        or utilizing the group's mining power for personal profit.
        & \begin{itemize}
            \item Adversary can advertise (and hold) BGP routes without
              autonomous systems being suspicious.
            \item No multi-homing for nodes.
            \item No information leakage across the isolated group(s).
        \end{itemize}
        & \hfill\vfill
        Synchrony
        \\\cmidrule{2-5}
        % Network
        & Eclipse Attack
        & An adversary eclipses a node to isolate them from the network. The
        adversary achieves this by populating the victim's peer tables, due to
        eviction mechanisms the adversary can control enough peers to
        completely isolate the node. Once successful, the adversary now
        controls information observed by the victim and is able to double
        spend.
        \vspace{5pt}
        & \begin{itemize}
            \item Adversary can spoof a high number of peers, and keep the
              connections alive.
            \item The peer eviction mechanism doesn't use ``try before replace''.
            \item There is a low number of buckets in the peer table.
        \end{itemize}
        & \hfill\vfill
        Synchrony
        \\\cmidrule{2-5}
        % Network
        & Balance Attack
        & The adversary partitions the network into balanced powered subgraphs
        of nodes. Once the subgraphs have forked, the adversary transacts to a
        merchant in one subgraph, and joins another subgraph to increase it's
        block production. Once the adversary's subgraph has a longer fork, they
        begin to resolve the partition across subgraphs, resolving the fork
        and double spending the coins to the merchant.
        & \begin{itemize}
            \item The adversary can partition and control the network.
            \item The adversary can transact in two subgraphs without any
              information leakage or nodes knowing of the segregation.
            \item Adversary can contribute enough block creation power
              that their chosen subgraph forms the longer chain.
            \end{itemize}
          & \begin{itemize}
                \item Synchrony
                \item Probabilistic Consensus
          \end{itemize}
        \\\cmidrule{2-5}
        % Network
        & MitM Attacks
        & The adversary hijacks routes (BGP) or spoofs (ARP) to
        Man-in-the-Middle the communication against nodes. Once the adversary
        is positioned correctly, they are able to manipulate the communication
        delay between nodes and perform other communication attacks.
        & \begin{itemize}
            \item Adversary can advertise (and hold) BGP routes without
              autonomous systems being suspicious.
            \item No multi-homing for nodes/mining pools.
            \item The adversary is the preferred route for communication and is situated
              in the middle.
            \item The adversary is able to manipulate communication delay between
              nodes.
        \end{itemize}
        & \hfill\vfill\hfill\vfill\hfill\vfill
        Synchrony
        \\\cmidrule{2-5}
        % Network
        & Attack of the Clones
        & The adversary partitions the Proof-of-Authority network and
        duplicates their node to be present in each partition with duplicated
        public-private key pairs. Proof-of-Authority will then adjust to ensure
        progress, and the adversary will be able to produce blocks on each
        partition and therefore make conflicting transactions or double spend.
        & \begin{itemize}
          \item Adversary can duplicate their instance without being detected
            as a duplicate.
          \item Adversary is able to partition the authority groups and place
            one clone in each partition.
          \item The block proposal/validation will still progress even with
            a significant number of authorities partitioned.
        \end{itemize}
        & \hfill\vfill\hfill\vfill\hfill\vfill
          Synchrony
        \\\bottomrule
  \end{tabular}
}
\end{table*}
\end{center}
\end{landscape}
\twocolumn
}

\subsubsection{Man-in-the-Middle Attacks}

The Man-in-the-Middle (MitM) attacks proposed against
Ethereum~\cite{ekparinya2018impact} illustrates the applicability of BGP route
hijacking to double spend on a public blockchain. By hijacking a BGP route, the
adversary can manipulate the communication delay between nodes and subgroups of
nodes. Once the adversary is in control of the communication delay, they can
leverage their position to perform other attacks, illustrated with the Balance
Attack~\cite{natoli2017balance} in the proposal.

Similarly, Ekparinya et al.\ also show the effectiveness of a MitM attack in a
private blockchain context by using ARP spoofing~\cite{ekparinya2018impact} to
double spend.

\subsubsection{Attack of the Clones}

The Attack of the Clones~\cite{ekparinya2019clones} presents an attack on the
Proof-of-Authority blockchains in Ethereum~\cite{wood2014ethereum, parityPoA,
EIP225}. This attack requires an adversarial authority, or a number of
adversarial authorities, to duplicate their node instances and require them to
have a exhibit the same public-private key pair. The adversary then partitions
the network, with one cloned instance in each partition. The network continues
to create blocks in separate partitions, and therefore double spending
and conflicting transactions can be proposed and accepted into the partitions.

The proposed mitigations to the attack include the use of a partially
synchronous consensus algorithm, such that message delays have lower impact on
the operation of the system. Similarly, requiring a threshold of the sealers
to sign the blocks helps to harden the protocol, as it requires the adversary to
have a larger pool of sealers to allow the block to be produced in the forks.

\subsection{Stake Attacks}

With the increasing popularity of Proof-of-Stake based approaches,
identified attacks allow an adversary to gain power and profit in the
chain by investigating minimal stake, or to make other selected nodes
(a.k.a. validators or consensus group members) loss their stake.
Proof-of-Stake, however, suffers from considerable validator
collusion. If the majority of the validators collude to perform
actions, then the Proof-of-Stake mechanism can fail in some cases.

\subsubsection{Nothing at Stake}

One of the first attacks on Proof-of-Stake is called ``Nothing at
Stake''~\cite{ethproblemswiki}, where rational validators aim at getting
rewards by not following the protocol's specification. Proof-of-Stake
based validation incentive means validators are encouraged to follow
protocol to earn rewards. In the case of a fork, the most optimal
strategy for validators is that they should validate on all possible
chains for the chance of most reward.

Adversaries are able to leverage this behavior, and with non zero probability
execute a double spending attack. The adversary creates a transaction to a
merchant, and simultaneously creates a fork in the chain with a conflicting
transaction. At this point, the validators will begin validating all chains due
to the economic incentive. With non-zero probability, the adversary's forked
chain may be selected as the correct chain to extend and the double spending
has occurred with the attacker staking nothing during the process.

A solution to the Nothing at Stake problem is for the validators to commit
value to sign and create valid blocks. This acts as a disincentive for the
validators creating and signing blocks on all forks, or they will lose some
money and be ultimately punished.

Alternatively, unforkable blockchains also present a solution to the nothing at
stake problem as the adversary will be unable to create a fork with the
conflicting transaction.

\subsubsection{Discouragement}

Alternatively, the adversary can discourage validator participation in the
validation of blocks. In some proof-of-stake schemes, the validator group is
punished if certain conditions are met and disagreement occurs. The
Discouragement Attack~\cite{discouragementAttack} outlines an attack where the
adversary can purposely act maliciously and broadcast their intention to other
validators. Due to validators possibility to lose money, validators withdraw
themselves from the committee. Eventually, the number of nodes in the validator
set will diminish, due to discouragement, and the adversary can now perform
a majority attack.

\subsubsection{Censorship}

Instead of validators working independently, adversarial collusion can lead to
validators dictating the operation of the blockchain. The Censorship
attack~\cite{censorship} can occur when validators collude to censor certain
transactions from being placed into blocks. The attack ramifications lead to
the destruction of the blockchain ecosystems and breaking the core values of
the blockchain. This attack relies on the collusion of the validators, as the
majority of the validators would have to agree on the censoring of specific
blocks or transactions. However, adversaries may leverage other methods to
lower the number of validators in the set until satisfied with the conditions.

Possible solutions to censorship arise with introduction of cost or mechanisms
to the consensus. An example of changing the reward structure to be
proportional to the number of validators that signed the block, for example if
$98$ of $100$ validators sign the block, then the validators that signed would
only be getting $98\%$ of the reward. Unless full collusion amongst all
validators occurs, then the validators censoring would not receive the full
bonus and may not be economically incentivized to uphold the censoring.
Rotation of validators for each block can also overcome censorship, as the
validators performing the censoring would only be able to in the time window.

\subsubsection{Grinding}

Implementations of Proof-of-Stake have mechanisms to select a validator from
the active set of validators to create and propose the next block. In some
cases, the selection of a validator follows the rule that a node has a
probability of being selected proportional to the amount of stake they have
contributed as a percentage of the total. The Grinding~\cite{ethPoSFaq}
attacks are attacks where a validator takes steps to increase the bias of their
selection as a validator, through some computation or other methods.

With PeerCoin, or PPCoin~\cite{king2012ppcoin}, the stake is based upon the
coin-age, and the validators probability of success is determined based on the
age they consume in the block creation process. The Grinding attack can
be applied in this scenario where the adversary performs computation and grind
through combinations of parameters in blocks to provide the best probability of
their coins creating a valid block.

Similarly, NXT~\cite{nxt} implements Proof-of-Stake where the randomness for
the next block is dependent on the signature of it's predecessor in the chain.
Due to the nature of the signatures, the validator can execute a grinding
attack and compute the randomness, selectively manipulating
randomness~\cite{vitalikrandomness} by skipping their turn to create a block.
The validator can then calculate the optimal strategy such that they gain an
above-average number of blocks to sign, leading to higher reward for the
validator.

Known solutions~\cite{king2012ppcoin, vitalikrandomness, ethPoSFaq}
exist to the stake grinding attack. In PPCoin, validators can be asked to stake
their coins well in advance of the block creation, such that the validator
cannot grind the best probability for their coins to win. The randomness can
also be improved through the use of secret sharing and threshold signatures, as
well as validators all collaboratively generating the randomness.

\afterpage{
\onecolumn
\begin{landscape}
  \vspace*{2.4cm}
\begin{center}
  \begin{table*}[h]
    \vspace{-0.5cm}
  \caption{Attacks Overview: Part 3\label{ref:attacksoverview-3}}
  \resizebox{1.35\textwidth}{!}{%
  \begin{tabular}{p{0.09\textwidth} p{0.11\textwidth}p{0.5\textwidth}p{0.45\textwidth}x{0.2\textwidth}}
    \toprule
    \textbf{Category} & \textbf{Attack} & \textbf{Overview} & \textbf{Attack Core} & \textbf{Cause}\\
    \midrule
          \multirow{6}{*}{\raisebox{-\heavyrulewidth}{Stake}}
        & Nothing at Stake
        & Members have incentives to validate blocks on all possible chains
        to gain the most profit. This means an adversary can fork the chain and
        have a high probability of successfully getting their chain accepted.
        & \begin{itemize}
          \item Members do not lose stake on validating an incorrect chain.
          \item Adversary does not lose money by forking the chain.
          \item The blockchain suffers from forks.
          \end{itemize}
        &
        \hfill\vfill
        Incentive Punishment
        \\\cmidrule{2-5}
        %\multirow{5}{*}{\raisebox{-\heavyrulewidth}{Stake}}
        & Discouragement
        & In Proof-of-Stake, members lose stake if agreement cannot be reached.
        An adversary threatens and discourages members from staying in the
        consensus committee by acting maliciously and purposely avoiding
        reaching agreement. As members leave the consensus committee, the
        adversary will gain a higher percentage of stake and can eventually
        perform a majority stake attack.
        & \begin{itemize}
          \item An adversary can sacrifice enough stake for members to leave.
          \item Honest validators do not join the validation and overthrow the
            adversary.
          \end{itemize}
        &
        \begin{itemize}
            \item Incentive Punishment
            \item $f > \frac{N}{2}$
        \end{itemize}
        \\\cmidrule{2-5}
        %\multirow{5}{*}{\raisebox{-\heavyrulewidth}{Stake}}
        & Censorship
        & Adversarial collusion between members can lead to censorship of
        blocks and transactions from being accepted to the chain.
        & \begin{itemize}
          \item Majority validator collusion.
          \item No cost model that hinders rewards.
          \item No adequate validator rotation.
          \end{itemize}
        &
        \hfill\vfill
        $f > \frac{N}{2}$
        \\\cmidrule{2-5}
        %\multirow{5}{*}{\raisebox{-\heavyrulewidth}{Stake}}
        & Grinding
        & An adversary grinds through parameters, or known computations, to
        find a strategy that will increase the number of blocks they validate,
        or, the amount of reward they get. The main aspect is to maximize the
        node's probability of selection by finding optimal parameters.
        \vspace{1em}
        & \begin{itemize}
          \item Validators don't require to stake coins in advance.
          \item Bad, or computable randomness.
          \end{itemize}
        &
        \hfill\vfill
        Incentive Punishment
        \\\cmidrule{2-5}
        %\multirow{5}{*}{\raisebox{-\heavyrulewidth}{Stake}}
        & Stake Bleeding
        & An adversary creates a hidden, forked chain. In the main chain, each
        time they are asked to validate they refuse and slow the block
        production. Eventually, the adversarial chain will grow and can win
        the longest chain if the main chain is delayed enough.
        & \begin{itemize}
          \item No checkpoint mechanism implemented.
          \end{itemize}
        & \begin{itemize}
              \item Synchrony
              \item Probabilistic Consensus
        \end{itemize}
        \\\cmidrule{2-5}
        & Long Range
        & An adversary interacts with a merchant and sends a transaction for an
        irrevocable action. Once the action has been performed, the Adversary
        starts a fork from a significant distance in the past (e.g.\ directly
        after the genesis block) and eventually produces a longer chain. The
        adversary can then produce the longest chain and win the fork resolution.
        & \begin{itemize}
          \item Timestamps are not used for authentication.
          \item Context sensitive transactions are not used.
          \item No checkpoint mechanism implemented.
        \end{itemize}
        & \begin{itemize}
              \item Synchrony
              \item Probabilistic Consensus
        \end{itemize}
    \\\bottomrule
  \end{tabular}
}
\end{table*}
\end{center}
\end{landscape}
\twocolumn
}

\subsubsection{Stake Bleeding}

Proof-of-Stake provides a number of improvements to the current blockchain,
however, it has been outlined that it suffers from a number of theoretical
attacks and vulnerabilities. To oppose the vulnerabilities, proposed security
measures have been implemented which provide mechanisms countering the possible
attacks. The introduction of checkpointing, a measure to solidify the
blockchain state and eliminate all forks that are not complying with the latest
checkpoint, allowed for a number of attacks to be mitigated.

For Proof-of-Stake blockchains that fail to implement checkpointing, as well as
employ transaction-fee reward-based incentives, can be vulnerable to the Stake
Bleeding attack~\cite{GKR18-Stakebleeding}, where an adversary can convince the
network to adopt a secondary chain. The attack methodology requires an
adversary that holds a moderate proportion of the stake in a blockchain that
adheres to the longest chain protocol of Nakamoto's Consensus. The adversary
creates a secondary chain, hidden from honest parties. Each time the adversary
is chosen to extend the valid chain, they refuse and slow down the process,
whereas they continue extending their private chain.

Over time, the adversary will lose stake in the honest chain, but will gain
rewards in the fake chain and slowly with the inclusion of valid transactions
will exceed the valid chain. At this point the adversary can release their
private chain and all honest nodes will accept it as the valid chain due to
it's superior length.

Proposed mitigations~\cite{GKR18-Stakebleeding, KiayiasKRDO16-Ouroboros,
lar-pos, BentovPS16a-SnowWhite, snowwhite-daian} to this attack consist of frequent
checkpointing, chain and block density, and context-sensitive transactions
which may include the hash of the most recent blocks in the transaction. Similarly,
PPCoin's coin-age~\cite{king2012ppcoin} also helps to prevent stake bleeding
by requiring the investment of time.

\subsection{Long Range}

The Long Range Attack~\cite{vb-long-range-attacks} was theorized during the
early development of Ethereum by Vitalik Buterin. The attack details a sequence
of events where an adversary is able to double spend. The adversary sends a
transaction to the merchant, the merchant then accepts the transaction after a
certain number of block confirmations. The adversary then creates a fork, but
starts it from a much earlier block, for example just after the genesis block.

In Proof-of-Work, the adversary would be required to hold a majority of the
network's computing power, constituting this attack as a majority mining
attack.  However, the methodology, where the attacker can utilize a smart
contract to perform hashes that are very easy to mine, producing a known value
for itself, but would be extremely costly to the other miners. For example, it
can be a loop for a specific value that the adversary knows, but others will
have to re-compute the value each time.  In this way, the attacker is able to
slow the progress of the honest chain. The adversary then continues to mine
until they create the longest chain and can overtake the main chain. This
attack requires a high amount of computing power, such that the adversary is
able to create the longest chain. In Proof-of-Stake, however, the attack
requirements are lower due to the validation process. An adversary with as low
as 1\% of all coins can perform the long range attack with more blocks
trivially.

Proposed mitigations~\cite{ethPoSFaq, vb-long-range-attacks, king2012ppcoin}
exist for the long range attack. One example mitigation is to ensure that
timestamps on the blocks are used as a rejection mechanism. Another alternative
is to ensure that a large portion of the coins, for example 30\%, are required
to sign every $N^{th}$ block.

\begin{table*}[!t]
  \caption{Membership Selection and Consensus}\label{tab:MS-C-overview}
  \begin{threeparttable}
  %\resizebox{\textwidth}{!}{
  \begin{xtabular}{
      p{0.15\textwidth}|% Chain
      p{0.07\textwidth} |% Year
      p{0.22\textwidth} % Membership
      p{0.11\textwidth} |% Category - Membership
      p{0.18\textwidth} % Consensus
      p{0.1\textwidth}  % Category - consensus
    }
    \toprule
    \textbf{Chain} & \textbf{Released} & \textbf{Membership} & \textit{Category} & \textbf{Consensus Mechanism} & \textit{Type} \\
    \midrule
    Bitcoin & 2008 & Proof-of-Work & Work &  Nakamoto's & Nakamoto \\
    PeerCoin & 2012 & Proof-of-Coin-Age & Stake & Nakamoto's (Highest Age) & Nakamoto \\
    Ethereum\tnote{*} & 2014/20XX & Proof-of-Work/Proof-of-Deposit & Work/Stake & Nakamoto's /(Casper) & Nakamoto/BFT \\
    Tendermint & 2014 & Proof-of-Lock & Stake & Tendermint BFT & BFT \\
    PeerCensus & 2014 & Proof-of-Work & Work & PeerCensus (PBFT Variant) & BFT \\
    Permacoin & 2014 & Proof-of-Retrievability & Capacity &  Nakamoto's Consensus & Nakamoto \\
    BurstCoin & 2014 & Proof-of-Space & Capacity & Nakamoto's & Nakamoto \\
    SpaceMint & 2015 & Proof-of-Space & Capacity &  Nakamoto's & Nakamoto \\
    ByzCoin & 2016 & Proof-of-Work & Work & ByzCoin (PBFT Variant) & BFT \\
    HoneyBadger BFT & 2016 & - & - & HoneyBadger BFT & BFT \\
    Solida & 2017 & Proof-of-Work & Work & Solida (PBFT Variant) & BFT \\
    Ouroboros & 2017 & Ouroboros PoS & Stake &  Ouroboros' Consensus & BFT \\
    AlgoRand & 2017 & AlgoRand & Stake &  AlgoRand BA$\star$ & BFT \\
    HyperLedger Sawtooth & 2017 & PoET & TEE & Nakamoto's\tnote{$\dagger$} & Nakamoto\\
    RepuCoin & 2018 & Proof-of-Work & Work/Reputation & RepuCoin BFT & BFT \\
    RedBelly Blockchain & 2018 & - & - & DBFT & BFT \\
    \bottomrule
  \end{xtabular}
  %}
  \begin{tablenotes}
  \item[*] The Ethereum blockchain is moving to Proof-of-Stake, currently using
    Proof-of-Work with the longest chain, but moving to Casper Proof-of-Stake
    with a BFT variant (named Casper) for the consensus. This consensus has not
    been included in the survey as it is still being developed and has a lack
    of formal documentation at this time.
  \item[$\dagger$] The HyperLedger Sawtooth
    documentation~\cite{sawtoothjournal} states that the PoET runs a measure of
    total time waiting. However, if a PBFT consensus is used, it will never
    fork, making the fork resolution and consensus in the system flexible.
  \end{tablenotes}
  \end{threeparttable}
\end{table*}

\section{Discussion}

 \label{sec:discussion}

 In this section, we discuss (a) how membership selection, consensus
 mechanism, and structure are combined in existing systems and their
 impact on each other; (b) other interesting areas of blockchains not
 explored in this survey; and (c) upcoming and future proposals
 integral to the improvement of the blockchain, but not mature enough
 for critical evaluations and inclusion in the survey.

 % \JS{Not sure if whitepaper would fit into this section, maybe in
 % intro?}

 % \CN{I think the discussion is meant for what we say about the whitepapers, the
 % introduction mentions it briefly but we should detail it more in the
 % discussion as it is a limitation?}

 % In this section, we present areas of blockchains not explored in this survey,
 % as well as detail the limitations. We explore alternative concepts as well as
 % detail upcoming and future proposals integral to the improvement of the
 % blockchain, but not mature enough for critical evaluation and inclusion in the
 % survey.

\subsection{Membership Selection and Consensus}

This section provides an overview on how existing systems make use of
different membership selection and consensus mechanisms. An overview
is presented in Table~\ref{tab:MS-C-overview}.

\paragraph{Role of Membership in Consensus Mechanisms}

Although consensus algorithms are powerful, there are some limitations
that impede the security and scalability to high node numbers.

For security, consensus algorithms rely on the membership selection to
provide a set of trustworthy nodes to run consensus, such that an
attacker cannot become a consensus group member easily. For example,
Bitcoin uses proof-of-work to select consensus group members, where an
attacker with a temporally high mining power can become a consensus
member easily. However, with proof-of-reputation in RepuCoin, it is
much more difficult for such an attacker to become consensus member,
thus it helps enforcing the security assumptions of consensus
algorithms.

For scalability, running optimally, without saturating network
bandwidth with messages, requires a smaller committee of members to
run the consensus. Membership selection fills the gap by providing the
consensus with a subset of nodes to form the small committee of
consensus members. The consensus also provides the membership
selection with the requirements and assumptions that it must fulfill.

Nakamoto's Consensus, for example, requires optimally one proposer of blocks
but places an assumption that the blocks are propagated to the entirety of the
network prior to any new proposals. This assumption is translated into the
difficulty of the Proof-of-Work puzzle and limits the speed of block proposals
and the number of nodes in the membership selection. This is also evident in
the \ghost consensus mechanism, as it requires an amount of time between
proposals so the heaviest sub-tree is observed by the nodes in the network.

Similarly, in AlgoRand as an example, the \BAstar consensus runs
optimally with a smaller subset of nodes, rather than all nodes
performing the consensus. The AlgoRand membership selection utilizes
VRFs to provide a subset of nodes as the committee to run the \BAstar
consensus.

\paragraph{Improvements on Bitcoin Membership Selection}

To save the wasted energy in PoW, many membership selection mechanisms have
been proposed. In particular, the proposed mechanisms either use a stake-based
lottery protocol (e.g. PoCA, PoD) for membership selection, or repurpose the
required work by replacing the to-be-wasted computational task with something
useful (e.g.\ distributed archive in Proof-of-Space). Since they are only
considering the issue of wasted energy, they keep using the Nakamoto's
consensus mechanism.

With proposals of the former type, stake-based lottery, new members are
selected at random having a probability proportional to the stake they have.
The focus on these proposals is designing the ``stake'' they rely on. For
example, PeerCoin makes use of coin age as the stake; and Proof-of-Activity
uses the balance of a participant and its online activity as the stake.

With proposals of the latter, new members are selected also at random, but the
probability is depending on the alternative useful work they perform. The focus
of these proposals has been on repurposing the wasted energy. For example,
Permacoin, proposes to use distributed archive of files as the work, and
BurstCoin and SpaceMint propose to use distributed storage space as a means to
selected members.

\paragraph{Improvements on Bitcoin Consensus}

As mentioned previously, Nakamoto's consensus only provides probabilistic
guarantees. This results in potential forks of the chain, and can be exploited
by an attacker to launch double spending attacks. Prior works try to remedy
this issue by replacing the Nakamoto's consensus with the classic consensus
mechanisms, i.e., Byzantine fault tolerant protocols that provide deterministic
guarantee.

To the best of our knowledge, PeerCensus~\cite{decker2016bitcoin} was
the first blockchain to propose Proof-of-Work for membership
selection, and a BFT-style protocol for reaching consensus. In
particular, the miners who successfully solved the recent
Proof-of-Work puzzles form the consensus committee to run the BFT
consensus on the proposed blocks. Similar to PeerCensus, a number of
other chains (e.g. ByzCoin, Solida, and RepuCoin) utilize
Proof-of-Work to select consensus committee members, and BFT-style
consensus for reaching consensus. The focus of ByzCoin is on scaling
the PBFT consensus schemes. Solida is focusing on providing a solution
for consensus committee reconfiguration. Hybrid
consensus~\cite{Pass-Shi-DISC17} aims at providing an efficient
bootstrapping with fast response. RepuCoin uses reputation-based
weighted BFT protocol for reaching consensus. It aims at exploring
ways to prevent bribery attack and flash attacks, where an attacker
bribes existing power miners or rents cloud mining power to obtain a
majority of the mining power quickly for a short time period.

\paragraph{\bf Hybrid systems}

Other systems have been working on combining a number of different techniques.
The Proof-of-Activity~\cite{bentov2014proof} combines both Proof-of-Work and
Proof-of-Stake to perform the membership selection. The Proof-of-Work aspect
produces the initial proposer of the empty block, then used to derive the
validators that will sign off the block. This is a combination of two
membership selection techniques to harmoniously utilize the guarantees of both.

Similarly, Thunderella~\cite{pass2017thunderella} proposes the use of the BFT
consensus during the optimal periods. In non-optimal conditions, or during the
cooldown phases, they switch from using the BFT to Nakamoto consensus. This
provides a hybrid nature between the BFT based model and the Proof-of-Work
longest chain model.

\subsection{Consensus Mechanism and Structure}

This section discusses the impact of the structure on the consensus mechanisms
of choice.

\paragraph{Role of Structure in Consensus}

The structure defines the way transactions and events are recorded in
blockchain systems. This definition places heavy requirements on the
consensus algorithm and what blocks are proposed and decided upon. The
traditional blockchain, as depicted in the original Bitcoin
paper~\cite{Nakamoto08-Bitcoin}, describes one canonical chain with
homogeneous block types. This gives the requirement to the consensus that a
single block should be decided for each height.

Bitcoin-NG~\cite{EyalGencer-etc-16-BitcoinNG} introduces different
block types, decoupling leader election (keyblocks) from transaction
serialization (microblocks). In particular, miners create keyblocks,
which do not contain any transaction, through PoW. The successful
miner in PoW is selected as a leader for a period of time until a new
keyblock is created. A leader includes transactions into microblocks,
and is allowed to validate microblocks directly without requiring
further proof-of-work. This greatly increases the scalability of the
consensus algorithms, in terms of the system throughput.

The further evolution of Blockchains introduced the use of other structures.
One such evolution is utilizing a Directed Acyclic Graph (DAG) as the
structure of the blockchain, where multiple events are able to occur
simultaneously but require new consensus approaches to provide finality on
multiple events. DAG structures apply different techniques to achieve
consensus on the events that have occurred in the graph.

Additionally, the integration of sharding techniques has exhibited similar
aspects such as parallel transaction processing and the benefit of less
storage space required. However, it also brings new challenges to achieving
consensus, such as transaction and state finality requiring the context of
the shards.

% The further evolution of Blockchains introduced using directed
% acyclic graph (DAG) as the structure of blockchains, where multiple
% events can occur simultaneously. This provides the consensus
% mechanisms with the requirements to provide finality on multiple
% events. DAG structures apply different techniques to achieve
% consensus on the events that have occurred in the graph.

\paragraph{Improving Bitcoin Structure}

The original bitcoin structure, the one canonical chain of homogeneous blocks,
provides the requirement for a distributed ledger of transactions for a
decentralized payment system. However, the evolution of blockchains involve a
number of functionality updates and improvements, which propose new structure
changes to provide the new functionality or improvements.

By decoupling the transactions from the block creation, this allows for a
higher throughput of transactions to be processed and appended to the ledger,
whilst still maintaining the guarantees of the original chain structure. DAG
structures on the other hand require a significantly different consensus and
membership selection model.

%%% Vincent's comment to remove
%The
%DAG structure, on the other hand, proposes another alternative means to provide
%higher throughput and more flexible scalability to the blockchain, but requires
%significantly different models for consensus and membership selection.

\subsection{Membership Selection and Structure}

The nature of the structure also imposes requirements on the membership
selection. The structure denotes how blocks are to be accepted and added,
imposing requirements on how the consensus is to be performed. This impacts
the way members are selected and the type of members that are selected.

\paragraph{Role of Structure in Membership Selection} The structure
may require different types of nodes to be selected in a variety of
ways. The original Bitcoin structure requires one set of membership
to be chosen for each block. The Microblock structure proposed by
ByzCoin and RepuCoin require groups of nodes to be selected in intervals,
differing from the original where optimally one node would propose
and decide on the next block and be agreed upon by everyone. The
introduction of DAG structures imposes a major change to the
membership selection, where different nodes control different paths
on the total graph, or, are in charge of managing their own ledger
of information.

The introduction of sharding will also impose a number of changes to the
way membership selection is performed. As each shard will operate
in different conditions, different memberships can be employed for each shard
and operate uniquely.

\subsection{Scalability \label{subsec:scalability}}

Scalability of blockchains can be encompassed by three main aspects; the number
of consensus nodes, the number of clients, and the number of transactions
processed per second (TPS).

Typically, a client is a node that is not participating in
the consensus and is not actively producing blocks, however is
actively transacting and receiving and gossiping information around
the network. The number of clients have minimal impact on the overall
performance of the blockchain, however, may impact the overall time
for message bandwidth and propagation in the network. All types of
blockchain consensus scale well
\cite{Vukolic15}.

With respect to the number of consensus nodes, non-BFT blockchain consensus
offers impeccable scalability due to the simplicity of its protocol. BFT
protocols, however, have limited scalability due to message complexity and the
types of communication patterns in the protocols. For example, Nakamoto's
consensus only requires nodes to choose the longest chain locally to reach an
agreement, so consensus is able to be reached even in the presence of thousands
of nodes.  PBFT, on the other hand, requires several rounds of communication
with all-to-all message passing involving all consensus members. So in
practice, PBFT can only scale to a few dozen nodes.

For the number of processed transactions per second, Nakamoto's consensus,
often seen paired with Proof-of-Work, performs poorly due to the requirement of
proposals needing to propagate through the network. In particular, with
Bitcoin, each block is 1Mb in size, and the size of each transaction is on
average approximately 256 bytes. This limits the block to contain only approximately $4,000$
transactions, which is proposed every 10 minutes on average leading to an
average of roughly 7 TPS. Increasing the size of a block is not an
effective solution to improve the performance. For example, Bitcoin
Cash, a hard fork of Bitcoin, increases the maximum block size from 1
MB to 8MB. Later, in 2018, Bitcoin
Cash\footnote{https://www.bitcoincash.org} has been split into two ---
one is called Bitcoin Adjustable Blocksize Cap (Bitcoin
ABC)\footnote{https://www.bitcoinabc.org} with maximum block size of
32 MB, and the other is called Bitcoin Satoshi's Vision (Bitcoin
SV)\footnote{https://bitcoinsv.io} which supports block size up to 128
MB. However, while providing a better number of TPS, the performance
is still limited. Additionally, since the PoW-based membership
selection requires competition in physical computing power, this may
put an extra constraint on the actual number of consensus nodes in
practice. For example, currently (as of Dec 2018), 14 mining pools
collectively have 80.5\% mining power of the entire network, which
shows the limit of scalability w.r.t. the number of consensus nodes in
practical \footnote{https://www.blockchain.com/pools}.

On the other hand, BFT-based blockchain consensus provides good
throughput~\cite{Vukolic15} leading to potentially tens of thousands of
transactions per second. However, as BFT protocols normally require a
pre-defined consensus group, they cannot be applied directly into blockchains.
This promotes systems, such as ByzCoin and RepuCoin, to combine different
membership selection algorithms with BFT consensus algorithms, to adapt BFT
protocols in Blockchain. The new combinations also impose changes on the
blockchain structure. For example, with the combination of PoW+BFT, systems
(such as ByzCoin and RepuCoin) deployed paralleled chains.

% Payment channels: lightning, raiden
\paragraph{Off-chain payments\label{payment-channel}}

Another proposal to improve blockchain scalability is the use of payment, or
state, channels. These channels allow for transactions to be performed
off-chain with the same security as on-chain payments, which results in overall
higher throughput. Techniques such as the Lightning
Network~\cite{poon2015bitcoin} and Raiden~\cite{raidenfaq} utilize off-chain
messages to perform transactions amongst parties. The messages and transactions
are signed off-chain by both parties and the on-chain transactions are used for
final settlement.

\subsection{Ancillary Components}

Although this paper provides insight into three blockchain
components, namely the membership selection, consensus and structure,
there are a number of other components that form the blockchain and
heavily contribute and influence the design and operation.

\subsubsection{Cryptography}

The blockchain heavily relies on cryptography for many aspects of it's
operation. The cryptographic properties integrate into the security of the
blockchain, allowing it to provide critical guarantees. The use of cryptography
falls into a number of aspects of the blockchain, each providing unique
properties and security.

\paragraph{Hash Functions}

Hash functions provide the basis for the mining performed on
blockchains as well as integrating with the coins and transactions. A
cryptographically secure hash function is a deterministic one-way
function, that is pre-image resistance and collision resistant. The
pre-image resistance property ensures that for any given hash value,
it should be difficult to find the original input message. The
collision resistance property guarantees that it is difficult to find
two different inputs, such that they share the same hash value.

During the mining process, the pre-image resistance property of hash
functions makes it difficult to find a number that fits the threshold
provided by the difficulty. The collision resistance property makes it
difficult\footnote{With the current computing technology and known
  mathematical principles.} to find a collision where the hash of an
entity leads to the same hash of another. This is critical during the
mining process, but also in the way transactions and blocks are
formed, as the hash should be unique to identify each transaction and
block uniquely. Furthermore, the collision resistant deterministic
one-way function also ties into the user accounts, as each user
account is addressed by a hash, which if found to be non-unique may
cause issues.

\paragraph{Authentication and Authorization}

The use of public key infrastructure forms the basis of wallets and user
accounts. Each account is given an associated private key and public key which
an address is then derived from. This ties a user to valid signatures when they
sign transactions, and is a large component of the blockchain. Each transaction
must be signed with a valid signature from a user, and can be verified using
the wallets associated public key. Recently, a number of hardware-based wallets
have been released, such as the Trezor~\cite{trezor}
and the Ledger~\cite{ledgernano},
providing hardware
private keys to increase the security for the users.

\paragraph{Transaction privacy}

A major component of the blockchain is transaction privacy.
Zcash~\cite{sasson2014zerocash} and Monero~\cite{monero-about} are
two blockchains that are heavily working on the privacy aspect of the
blockchain.

Ring Signatures~\cite{howtoleak-ringsig} allow a user in a group to
create a signature, in the way that a verifier can verify that the
signature is created by a member of the group, but does not know
exactly which member has created this signature. This allows the user
to stay anonymous when submitting transactions as part of the
group. This provides privacy for groups of users, however, it can lead
to double-spending of coins as a user can create two transactions
spending the same coin without being detected. Linkable Ring
Signature~\cite{DBLP:conf/acisp/LiuWW04} addresses this problem by
revealing the identity of the signer if the signing key is used more than
once. A variant of Linkable Ring Signature~\cite{Fujisaki:TRS} is
adopted by CryptoNote~\cite{cryptonote,Inscrypt18}. In CryptoNote, an
input of a transaction is a Traceable Ring Signature, where the ring
contains one coin to be spent, and several decoy inputs known as
``mix-ins''. An observer only knows that one of the coins in
the ring is spent, but does not know which coin. Monero
further improved the work on privacy with ring signatures, extending
CryptoNote's implementation by designing ``Ring Confidential
Transactions'' (RingCT)~\cite{eprint-ringCT}. RingCT provides
anonymity of the transaction protecting both the user and the
transaction amount, allowing for a higher level of privacy amongst
transactions.

However, attacks on the privacy of CryptoNote-style blockchains have
been identified. Two concurrent papers~\cite{KumarFTS17,MillerMLN17}
found that there exists ``zero-mix-in'' transitions, where a ring only
contains the real coin to be spent. This not only provides no privacy
guarantee on the transaction with zero-mix-in, but also reduces the
privacy of other transactions that use the input of the zero-mix-in
transitions. They also found that other attributes of a coin, such as
its age, leak information on the likelihood of this coin being spend
in a transaction. Other work~\cite{YUCSF19, closed-set-attack, Fork-attack} show that even assuming
all transactions contain mix-ins, and all attributes do not leak any
information, the different ways to choose mix-ins or having a hard fork
would also leak information about the trace of transactions. There are
also network level analysis \cite{cryptoeprint:2019:411} on the
topology of such blockchains, as knowing the network topology would
help compromising the security and privacy of a system.

ZCash, as well as a number of other blockchains~\cite{ethzksnarks},
are working on applying \textit{Zero Knowledge Proof of Knowledge} (ZKPK) to
the blockchain. ZKPK allows parties to prove their knowledge without
revealing any information about the knowledge. For example, it allows
users to prove possession of information without revealing the actual
information, retaining privacy. The use of Zero Knowledge Proof of
Knowledge on the blockchain is becoming increasingly popular, allowing
for the verification of transactions and the verifying of arbitrary
computations~\cite{ethzksnarks}.  The introduction of
\textit{zk-SNARKs}~\cite{zcash-zksnarks} in the
Zcash~\cite{sasson2014zerocash} blockchain sparked widespread
interest into the adoption of Zero Knowledge Proofs on the
blockchain. This will lead to an advancement of the cryptographic
techniques used as well as provide more insight into better privacy
and efficiency. Similar to the effort on security analysis of
CryptoNote-style blockchain, reducing the privacy guarantee of Zcash
is shown to be possible \cite{DBLP:conf/uss/KapposYMM18}.

\subsubsection{Internal Structures and Storage} Another component critical to the
operation of the blockchain is the use of internal structures. These structures
are integral to the operation of the blockchain and can help define what
functionality the blockchain can provide. The structures handle the storage,
retrieval and handling of data that is saved to the chain, or, used as state.

In Bitcoin variants, the most common structure is the in-memory map of
storing Unspent Transaction Outputs (UTXOs)~\cite{bitcoinutxo}, which
tracks any unspent transaction outputs, representing coins.

Ethereum uses prefix trees, or ``Tries''~\cite{ethereumpatricia}, to
handle state, storage, transactions and receipts. Ethereum updates
state through blocks, the use of the state trie helps to append new
state to the chain, updating new balances, changing the state and
storage of smart contracts.

Other blockchains use a number of internal structures to handle state,
balances, account and transaction information, and more. The choice of
structures can impact on the chain capabilities, differing with each
implementation but still a core component to consider.

\subsubsection{Virtual Machines and Platforms}

The Ethereum blockchain utilizes the Ethereum Virtual Machine
(EVM)~\cite{ethereum-dev-tutorial, wood2014ethereum} to run smart contracts and
handle state. Recently there have been efforts to move away from the EVM and
move into eWASM~\cite{ethereum-ewasm-compendium}, a
WebAssembly~\cite{webassembly}
%\footnote{https://webassembly.org/}
based virtual machine that will
then handle the state. The shift from the specialized Ethereum virtual machine
to eWASM will provide new support for different languages and provide a more
optimal execution.

\subsection{Whitepaper Information}

The rising interest of blockchains and its applications promote an influx of
proposals. These proposals encompass a number of components, including
membership selection, consensus, structure, cryptography or others. Due to the
nature of the surrounding community, many of the proposals are detailed
informally by means of whitepapers~\cite{slimcoin, cryptonote, blackcoin, nxt,
burstcoinwhitepaper, iota-whitepaper, raiblocks-whitepaper,
avalanchewhitepaper, cosmos, Nakamoto08-Bitcoin}, which in some cases lack peer
review and are often shallow in terms of technical details.

Many whitepapers are written as public-facing documents to support the future
developments, or provide content for potential investors to understand. This
often leads to the lack of formal models and technical details that form
critical components of the new proposals. Other whitepapers are presented in a
formal manner, clearly addressing the technical details and formalizing models
consistent with the literature. The proposals conforming to this format are
often able to be critically evaluated and discussed thoroughly, however they
make up a minority of the proposed whitepapers. Although this survey aims to
provide an aggregation of proposals, it is not exhaustive.

\subsection{Deployment Issues}

Currently, all mainstream blockchains are PoW-based. The transition from
Proof-of-Work to Proof-of-Stake can be seen as a deployment challenge for
employing the energy efficient solutions. In particular, to constitute
fairness for existing miners, the transition period must reward the miners in
such a way that they are incentivized to drop mining. Most Proof-of-Stake
mechanisms planned today, for example Casper for
Ethereum~\cite{casperfriendlyghost, ethPoSFaq}, increase the block difficulty
to slowly phase out mining by making it impractical. Other new systems, such
as the Proof-of-Activity~\cite{bentov2014proof}, integrated the
Proof-of-Stake with the existing PoW-based system.

\subsection{Future Outlook}

% A number of new proposals not included in the survey detail new
% proposals to the blockchain, including new BFT
% varients~\cite{% BentovPS16a-SnowWhite,
%   avalanchewhitepaper}, formal verification~\cite{cbccons-zamfir}, as
% well as significant structure changes and
% others~\cite{cbccasperresources}.

One major upcoming release is the adoption of sharding of the
blockchain. The issue of scalability has been a factor on improving
the blockchain. Pruning of state~\cite{ethstatetrieprune}, light
clients~\cite{ethstatetrieprune, vblightclient, ethwikilight,
  paritylgihtethereum, gethlight}, block sizes~\cite{bchfaq} and
others.  However, the issue of the size of the blockchain remains
unanswered, as the blockchain is growing over time, issues of storage
space and availability to smaller nodes arise. The proposal of having
a sharded blockchain~\cite{ethereumshardingfaq} allows for a reduction
in the size of the chain by splitting it amongst
participants. However, the challenge of applying this while
maintaining the blockchain guarantees is currently in work.

% Side-chains: plasma, ``Pegged sidechains''

Similar to payment channels, there have been proposals~\cite{plasma,
back2014enabling} to have side-chains operating alongside the main
blockchain, enabling the use of ``tokens'' or other added functionality.

% Cross-chain: polkadot

Recently, a number of cross-chain techniques have been devised~\cite{polkadot,
interledger, cosmos} to address the issue of governance and allow for
transactions across multiple blockchains, providing an opportunity to diversify
the blockchain model and open possibilities to offer different guarantees for
users.

% Sentence to finish off the Section.
As the blockchain matures, new mechanisms will be proposed to improve the
blockchain. The future developments will impact performance, scalability or
security of the blockchain in a variety of ways.

\balance
\section{Conclusion \label{sec:conclusion}}

Although in its infancy, the blockchain paradigm is generating
widespread interest. As with all successful concepts, the current
blockchain ecosystem quickly became quite complex and difficult to
understand, from both an introductory view and an in-depth dive into
the state of the art.

To address this problem, we deconstructed the blockchain into three main
critical components: membership selection, consensus mechanism and
structure. We introduced an evaluation framework to get insight into system
models, desired properties and analysis criteria, using the decoupled
components as parameters.  We tested our approach by analyzing the relevant
state of the art, and performing a categorization of how systems fit in the
different criteria outlined.

% The work presented is non-exhaustive, as the number of proposals for
% blockchains increase, a number of new proposals can be added to the
% framework.  Limited by available information, improvements to the
% framework may include the addition of quantitative examination of
% proposals.

The number of proposals for blockchains increases by the month. However, we
trust our framework to be comprehensive enough that proposals can be added and
classified in the future. Improvements to the framework may include the
addition of quantitative examination of proposals.

This survey provided clear insight into the blockchain proposals landscape,
through a novel perspective --- deconstructing the blockchain complex into few
simple but complete critical components.  We expect this work to be useful to
the community, providing direction and helping to simplify future designs, such
as inspiring innovative coherent combinations revealed by our decomposition and
categorization.

\ifCLASSOPTIONcompsoc
\section*{Acknowledgments}
\else
  \section*{Acknowledgment}
\fi
This research is in part supported under Australian Research Council Discovery
Projects funding scheme (project number 180104030) entitled ``Taipan: A
Blockchain with Democratic Consensus and Validated Contracts'' and Australian
Research Council Future Fellowship funding scheme (project number 180100496)
entitled ``The Red Belly Blockchain: A Scalable Blockchain for Internet of
Things''.

This work is in part supported by the Fonds National de la Recherche
Luxembourg through PEARL grant FNR/P14/8149128 and by Intel through the ICRI
Institute for Collaborative Autonomous \& Resilient Systems (ICRI-CARS).

\bibliographystyle{IEEEtranS}
\bibliography{IEEEabrv,reference}
\end{document}